\RequirePackage{snapshot}
\documentclass[twocolumn,letterpaper,accepted=2021-07-25]{quantumarticle}
\pdfoutput=1
\usepackage[numbers,sort&compress]{natbib}
\usepackage{amsmath, amsthm, amssymb,graphicx,color,bbm,listings}
\usepackage{algorithm}
\usepackage{algorithmic}
\begin{document}



\newcommand{\rrangle}{\rangle\!\rangle}
\newcommand{\llangle}{\langle\!\langle}
\newcommand{\rrpipe}{|}
\newcommand{\llpipe}{|}

\newcommand{\cM}{\mathcal{M}}
\newcommand{\cG}{\mathcal{G}}
\newcommand{\cD}{\mathcal{D}}
\newcommand{\cE}{\mathcal{E}}
\newcommand{\cL}{\mathcal{L}}
\newcommand{\cU}{\mathcal{U}}
\newcommand{\cH}{\mathcal{H}}

\newcommand{\reals}{\mathbb{R}}

\newcommand{\diff}{\mathrm{d}\!}
\newcommand{\pdiff}[2]{\frac{\partial #1}{\partial #2}}
\newcommand{\pdiffsq}[3]{\frac{\partial^2 #1}{\partial #2\partial #3}}
\newcommand{\expect}[1]{\ensuremath{\left\langle#1\right\rangle}}

\newcommand{\qhat}{\hat{q}}
\newcommand{\rhohat}{\hat{\rho}}
\newcommand{\rhoMLE}{\hat{\rho}_{\mathrm{MLE}}}
\newcommand{\rhotrue}{\rho_{\mathrm{true}}}
\newcommand{\Rhat}{\hat{\mathcal{R}}}
\newcommand{\Vbar}{\overline{V}}

\newcommand{\ket}[1]{\ensuremath{\left|#1\right\rangle}}
\newcommand{\bra}[1]{\ensuremath{\left\langle#1\right|}}
\newcommand{\braket}[2]{\ensuremath{\left\langle#1|#2\right\rangle}}

\newcommand{\ketbra}[2]{\ket{#1}\!\!\bra{#2}}
\newcommand{\braopket}[3]{\ensuremath{\bra{#1}#2\ket{#3}}}
\newcommand{\proj}[1]{\ketbra{#1}{#1}}

\newcommand{\sket}[1]{\ensuremath{\llpipe#1\rrangle}}
\newcommand{\sbra}[1]{\ensuremath{\llangle#1\rrpipe}}
\newcommand{\sbraket}[2]{\ensuremath{\llangle#1|#2\rrangle}}
\newcommand{\sketbra}[2]{\sket{#1}\!\sbra{#2}}
\newcommand{\sbraopket}[3]{\ensuremath{\sbra{#1}#2\sket{#3}}}
\newcommand{\sproj}[1]{\sketbra{#1}{#1}}

\newcommand{\logL}{\log\mathcal{L}}
\newcommand{\seqaction}{\tau}

\def\Id{1\!\mathrm{l}}
\newcommand{\Tr}{\mathrm{Tr}}
\newcommand{\Nparams}{N_{\mathrm{params}}}
\newcommand{\todo}[1]{\textcolor{red}{#1}}

\newlength\myindent
\setlength\myindent{2em}
\newcommand\bindent{%
  \begingroup
  \setlength{\itemindent}{\myindent}
  \addtolength{\algorithmicindent}{\myindent}
}
\newcommand\eindent{\endgroup}

\title{Gate Set Tomography}

\author{Erik Nielsen}
\affiliation{Quantum Performance Laboratory, Sandia National Laboratories}

\author{John King Gamble}
\affiliation{Microsoft Research}

\author{Kenneth Rudinger}
\affiliation{Quantum Performance Laboratory, Sandia National Laboratories}

\author{Travis Scholten}
\affiliation{IBM Quantum, IBM T.J. Watson Research Center}

\author{Kevin Young}
\affiliation{Quantum Performance Laboratory, Sandia National Laboratories}

\author{Robin Blume-Kohout}
\affiliation{Quantum Performance Laboratory, Sandia National Laboratories}

\maketitle

\begin{abstract}
  Gate set tomography (GST) is a protocol for detailed, predictive characterization of logic operations (gates) on quantum computing processors. Early versions of GST emerged around 2012-13, and since then it has been refined, demonstrated, and used in a large number of experiments.  This paper presents the foundations of GST in comprehensive detail.  The most important feature of GST, compared to older state and process tomography protocols, is that it is \emph{calibration-free}.  GST does not rely on pre-calibrated state preparations and measurements. Instead, it characterizes all the operations in a \emph{gate set} simultaneously and self-consistently, relative to each other.  Long sequence GST can estimate gates with very high precision and efficiency, achieving Heisenberg scaling in regimes of practical interest.  In this paper, we cover GST’s intellectual history, the techniques and experiments used to achieve its intended purpose, data analysis, gauge freedom and fixing, error bars, and the interpretation of gauge-fixed estimates of gate sets.  Our focus is fundamental mathematical aspects of GST, rather than implementation details, but we touch on some of the foundational algorithmic tricks used in the \texttt{pyGSTi} implementation.  
\end{abstract}

\tableofcontents

\section{Introduction}

Getting to useful quantum computation will require engineering quantum bits (qubits) that support high-fidelity quantum logic operations, and then assembling them into increasingly large quantum processors, with many coupled qubits, that can run quantum circuits.  But qubits are susceptible to noise.  High-fidelity logic operations are hard to perform, and failure rates as low as $10^{-4}$ are rare and remarkable.  Useful quantum computation may require fault tolerant quantum error correction, which distills a few high-fidelity logical qubits from many noisy ones.  Threshold theorems \cite{knill:2005,Aliferis:2006,aliferis:2007,Aliferis:2009} prove that this is possible in realistic quantum processors -- \emph{if} the processor's errors satisfy certain conditions.  Each threshold theorem assumes different conditions on the errors, but in general they must (1) be only of certain types, and (2) occur at a rate below a certain value (generally around $10^{-3}$ -- $10^{-4}$ for the most common types).  So it is not an exaggeration to say that the entire program of practical quantum computing depends on understanding and characterizing the errors that occur in as-built processors, controlling their form, and reducing their rates.

Assessing how qubits, logic operations, and entire processors behave is the central task of \emph{quantum device characterization}, a.k.a. ``quantum characterization, verification, and validation'' (QCVV).  There are many protocols for this task, all of which share the same basic structure: a set of experiments described by \emph{quantum circuits} are performed on the processor, yielding data that is processed according to some algorithm, to produce an estimate of an aspect of the processor's behavior.  Some protocols produce highly detailed predictive models of each logic operation (tomography \cite{Gale1968-mk, Vogel1989-vt, Hradil1997-fv, Munro2001-jn, James2001-hz, Artiles2005-ts, Blume-Kohout2010-vv, Blume-Kohout2010-hb, Christandl2012-am, Smolin2012-yv, Banaszek2013-zx, Granade2016-qy, Haah2017-jw, Poyatos1997-mz, Chuang1997-vf, Fiurasek2001-mp, DAriano2001-fx, Childs2001-en, Altepeter2003-fb, OBrien2004-tr, Weinstein2004-vn, Mohseni2006-cp, Riebe2006-tc, Bendersky2008-xa, Lobino2008-ud, Bialczak2010-et, Kimmel2014-nx, Kim2014-ix, GST2013, Greenbaum15, GST2015, GST2016, Blume-Kohout2017-kn}).  Others produce semi-predictive partial characterizations (component benchmarking \cite{EmersonScience2007, Knill2008, wallman_randomized_2014, Magesan2011-ra, Magesan2012-bo, Gaebler2012-vq, Magesan2012-sg, Gambetta2012-yu, Corcoles2013-zs, Barends2014-ap, Wallman2015-pa, Carignan-Dugas2015-pz, Wallman2015-uq, Chasseur2015-zz, Sheldon2016-sj, Wallman2016-kx, Proctor2017-ru, Harper2017-oa, Wallman2018-wy, Huang2019-zj, McKay2019-kf, Proctor2019-ma, Erhard2019-ig, Ekert2002-ma, Levi2007-rb, Toth2010-xi, Flammia2011-rw, Da_Silva2011-jv, Moussa2012-kd, Reich2013-oj, Kimmel2015-tj, Rudinger2017-vy, Aaronson2018-tj, Mayer2018-zt, Helsen2019-fi, Huang2020-ll}).  And some assess the performance of entire processors (holistic benchmarking \cite{Bishop2017-ub, Linke2017-ht, Michielsen2017-ds, Neill2018-nn, Boixo2018-wy, Hempel2018-ip, Yeter-Aydeniz2019-uz, Blume-Kohout2019-uv, Hamilton2019-xo, Cross2019-el, Arute2019-fl, McCaskey2019-kd, Wright2019-mx}).

All of these methods are valuable parts of the QCVV toolbox, serving distinct (though overlapping) purposes.  Some tasks demand simple holistic summaries of overall performance, while others are best served by partial characterization.  But for debugging, rigorous device modeling, and reliable prediction of complex circuit performance (including the circuits that perform quantum error correction) there is as yet no substitute for detailed tomography of individual logic operations, a.k.a. quantum gates.  The most powerful and comprehensive technique for this task is currently gate set tomography (GST).  Developed around 2012-13, GST has been used in a variety of experiments \cite{GST2013,GST2015, GST2016, Blume-Kohout2017-kn, Ware2018-cq, Proctor2019-oi, Song2019-fg,Hong2020-vc, Joshi2020-wo, Zhang2020-ux}, discussed in the literature \cite{White2019-ls, OBrien2017-fw, Scholten2019-jp, Cerfontaine2019-qh, Rol2017-wn, Mavadia2018-al, Puchala2019-tl, Lin2019-ck, Rudnicki2018-bn, Rudinger2017-vy, Kimmel2015-tj, Endo2018-kb}, and implemented in a large open-source software project \cite{pygsti,Nielsen2020-lu}.  But no comprehensive explanation of the theory behind GST has appeared in the literature.  This paper fills that gap.

\subsection{Gate set tomography}
The term ``gate-set tomography'' first appeared in a 2012 paper from IBM Research \cite{MerkelPRA13} that introduced several of GST's core concepts, but the group did not pursue GST further.  Around the same time, our group at Sandia National Labs started pursuing an independent approach to the same problem \cite{GST2013,GST2015}, which we have developed extensively since then \cite{GST2016, Blume-Kohout2017-kn, Proctor2019-oi, Nielsen2020-lu}.  This approach to GST, which has been used in a number of additional experiments over the past 5 years \cite{Rol2017-wn, Mavadia2018-al, Ware2018-cq, Song2019-fg, Hong2020-vc, Joshi2020-wo, Zhang2020-ux, cincio2020machine}, is the one that we discuss and present here.  We discuss its relationship to IBM's original approach where appropriate.

``Tomography'' means the reconstruction of a comprehensive model (of something) from many partial cross-sections or slices, each of which provides only a limited view that may be useless by itself.  Tomographic techniques are distinguished by their aspiration to construct a \emph{comprehensive} model for a system or component, by \emph{fitting} that model to the data from many independent experiments.  Quantum tomographic methods include state tomography \cite{Gale1968-mk, Vogel1989-vt, Hradil1997-fv, Munro2001-jn, James2001-hz, Artiles2005-ts, Blume-Kohout2010-vv, Blume-Kohout2010-hb, Christandl2012-am, Smolin2012-yv, Banaszek2013-zx, Granade2016-qy, Haah2017-jw}, process tomography \cite{Poyatos1997-mz, Chuang1997-vf, Fiurasek2001-mp, DAriano2001-fx, Childs2001-en, Altepeter2003-fb, OBrien2004-tr, Weinstein2004-vn, Mohseni2006-cp, Riebe2006-tc, Bendersky2008-xa, Lobino2008-ud, Bialczak2010-et, Kimmel2014-nx, Kim2014-ix}, detector or measurement tomography \cite{Luis1999-ek, Fiurasek2001-zc, Lundeen2009-qs, Blumoff2016-uk, Chen2019-zs}, Hamiltonian tomography \cite{Schirmer2004-uv, Cole2005-aq, Di_Franco2009-cr, Granade2012-xq, Ferrie2013-yi, Zhang2014-rl, Wang2015-md, Cole2015-wo, Rudinger_HamReconstruct_PRA2015, Wang2017-nh, Krastanov2019-xq, Hou2019-mp} and GST.

Like process tomography, GST is intended to characterize \emph{how logic operations (e.g. gates) impact their target qubits}. Those gates’ target subsystem and state space needs to be specified up front, and GST reconstructs the gate-driven evolution of those targets only. GST is not designed to probe the holistic performance of many-qubit processors, and as most commonly used, it does not capture general crosstalk. (Although GST can easily be adapted to focus on \emph{specific} types of crosstalk, and characterize them in detail.)

GST differs from state and process tomography, which we discuss later in Section \ref{sec:StateAndProcessTomography}, in two fundamental ways.  First, it is almost entirely \emph{calibration-free} (or ``self-calibrating'').  When GST reconstructs a model of a quantum system, it does not depend on a prior description of the measurements used (as does state tomography) or the states that can be prepared (as does process tomography).  Second, GST reconstructs or estimates not a single logic operation (e.g., state preparation or logic gate), but an entire \emph{set} of logic operations -- a \emph{gate set} including one or more state preparations, measurements, and logic gates.  These two features are inextricably linked, and we discuss this connection later in this paper.

GST's independence from calibration is its original \emph{raison d'etre}.  As IBM pointed out \cite{MerkelPRA13}, state and process tomography become systematically unreliable when the ``reference frame'' operations on which they rely (pre-calibrated measurements for state tomography, pre-calibrated state preparations \emph{and} measurements for process tomography) are either unknown, or misidentified.  This limits their practical utility and makes them unsuitable for rigorous characterization.

GST is not the only calibration-free method.  Today, new characterization protocols are more or less expected to be calibration-free.  The oldest calibration-free protocol is randomized benchmarking (RB) \cite{EmersonScience2007,Knill2008}, which predates GST by almost 10 years.  Detailed comparisons between RB and GST have been given elsewhere \cite{GST2016,Blume-Kohout2017-kn}, and they are complementary tools addressing different needs.


This paper is not a GST ``how-to'' manual.  We do provide some incidental discussion of how to perform GST in practice, and how it is implemented in our \texttt{pyGSTi} software \cite{pygsti}, but only inasmuch as it illustrates aspects of the theory.  Separately, we recently published a concise guide to applying GST to characterize hardware \cite{Nielsen2020-lu}.  In contrast, this article is intended as a theory paper presenting GST's mathematical background, justifications, and derivations.

\subsection{Section guide \& expert summary}
Because this paper attempts to be comprehensive, it is rather long.  So we begin here with a short expert-level summary of the major ideas and results, which may help readers decide what parts to read.  A reader who understands \emph{everything} in this section may not need to read any further at all!  Other readers may wish to skip to the first part that discusses a statement (from this summary) that is not obvious.

We begin in Section \ref{sec:background} by establishing mathematical and conceptual foundations.  We introduce quantum logic operations (\ref{sec:circuits}); the mathematical representations of states and measurements (\ref{sec:HSSpace}); and superoperators as a model of noisy quantum logic gates, including transfer matrix and Choi matrix representations, and introduce super-Dirac (superbra / superket) notation (\ref{sec:Qgates}).  We introduce gate sets, their representation as a collection of superoperators and operators, and gauge freedom (\ref{sec:GateSets}).  Finally, we establish notation and conventions for quantum circuits and the superoperator $\seqaction(g)$ caused by a quantum circuit $g$ (\ref{sec:GateSequences}).

In the second part of Section \ref{sec:background}, we review standard forms of tomography for states,  processes, and measurements (\ref{sec:StateAndProcessTomography}).  We conclude this introduction by explaining the role of calibration in standard tomography (\ref{sec:calibration}) to set the stage for GST.

Section \ref{sec:LGST} is devoted to linear inversion GST (LGST).  We start with the history of GST, describing how LGST solved some of the problems with the initial versions of GST (\ref{sec:LGSTbackground}). We then derive LGST, emphasizing parallels with process tomography (\ref{sec:LGSTalgorithm}), and introducing techniques required for long-sequence GST.  Readers inclined to skip over the mathematics in Section \ref{sec:LGSTalgorithm} should feel free to do so, as these derivations are not prerequisites for subsequent material.  In the last parts of Section \ref{sec:LGST}, we address two issues left unanswered in the derivation; how to prepare fiducial states and measurements (\ref{sec:LGSTfiducials}), and how to fit LGST data better with maximum likelihood estimation (\ref{sec:LGSTMLE}).

Section \ref{sec:LongSequenceGST} introduces long-sequence GST, the standard ``GST'' protocol.  (LGST came first, and is easier to analyze theoretically, but is rarely used alone in practice).  We introduce the two main motivations for long-sequence GST, which are (1) much higher accuracy, and (2) the ability to impose constraints such as complete positivity.  After discussing early and obsolete versions of long-sequence GST (\ref{sec:HistoricalLSGST}), we explain how to construct circuits that amplify errors and enable Heisenberg-limited accuracy (\ref{sec:ExperimentSelection}).  Then, we define the GST likelihood function and show how to maximize it to find the maximum likelihood estimate of a gate set model (\ref{sec:ParameterEstimation}).

Section \ref{sec:AdvancedLongSequenceGST}'s two subsections cover two ``advanced'' aspects of long-sequence GST.  The first (\ref{sec:Parameterization}) conceptualizes gate sets as parameterized models, introducing the term ``gate set model'', and discusses how to impose constraints like trace preservation or complete positivity.  The second (\ref{sec:FPR}) describes ``fiducial pair reduction'', a method for eliminating redundancy among the standard set of GST circuits and thereby reducing the number of circuits prescribed.

Section \ref{sec:Interpretation} addresses the question ``What can you do with a GST estimate once you have it?''  We start with the critical step of assessing goodness-of-fit and detecting \emph{model violation} or non-Markovianity (\ref{sec:GoodnessOfFit}).  We also explain what we mean by ``Markovian'' in this context, and why model violation is associated with ``non-Markovian'' behavior.  We then turn to the important problem of extracting simplified metrics (e.g., fidelity) from a gate set, explain why choosing a gauge is necessary, and explain how to do so judiciously using a process we call ``gauge optimization'' (\ref{sec:GaugeOpt}).  Finally, we discuss how to place error bars around that point estimate, including how to deal with the complications induced by gauge freedom (\ref{sec:ErrorBars}).

Appendices address some of the most technical and tangential points.  These include the gauge (\ref{sec:Gauge}), the historically-important eLGST algorithm (\ref{sec:eLGST}), how to deal with overcomplete data in LGST (\ref{sec:LGSTovercomplete}), specific implementation choices made in \texttt{pyGSTi} (\ref{sec:PyGSTiImplementation}), numerical studies that support claims made in the main text (\ref{sec:NumericalVerification}), and the bias we find in a particular estimator (\ref{sec:chi2bias}).  Appendices are referenced from the text where applicable.

\section{Background\label{sec:background}}

\subsection{Mathematical Preliminaries\label{sec:math}}

We begin with a concise introduction to the mathematics, formalism, and notation used to describe qubits, multiqubit processors, their logic operations, and the programs (quantum circuits) that they perform.

\subsubsection{Quantum processors and quantum circuits}\label{sec:circuits}

\begin{figure}
  \begin{center}
    \includegraphics[width=3in]{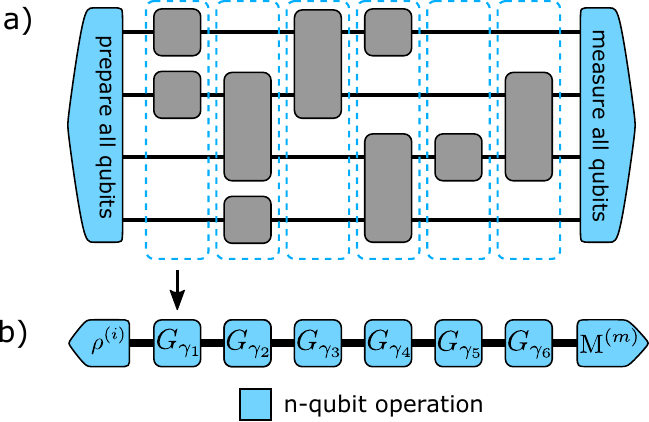}
    \caption[CaptionForLOF]{\textbf{(a)} A fixed-input classical-output (FI/CO) quantum circuit begins by preparing all the qubits and ends by measuring all the qubits.  Individual 1- and 2-qubit \emph{elementary operations}\footnote{Often called ``gates'' but not here since we use the term ``gate'' to mean a $n$-qubit operation.} (gray boxes) that occur at the same time are grouped into circuit layers (dashed rectangles) which by construction act on all the qubits (horizontal black lines).  \textbf{(b)} In this work, a \emph{gate} refers to a $n$-qubit operation and thus corresponds to a \emph{circuit layer} (downward arrow).  FI/CO circuits are, for the purposes of this paper, composed of: a $n$-qubit state preparation, a sequence of $n$-qubit gates (circuit layers), and a $n$-qubit measurement.  The symbols $\rho$, $G_{\gamma}$ and $M = \{E_j\}_j$ label these operations within a gate set (see Eq.~\ref{eq:GatesetDef}).  The particlar circuit shown begins with preparation in the $i$-th available state, proceeds with execution of gates indexed by $\gamma_i, \ldots \gamma_6$, and concludes with performance of the $m$-th available measurement.\label{fig:circuit}}
  \end{center}
\end{figure}

An idealized quantum information processor comprises one or more qubits that can be (i) initialized, (ii) transformed by quantum logic gates, and (iii) measured or ``read out'' to provide tangible classical data \cite{DiVincenzo2000-iw}.  Using such a processor is usually described by physicists as running experiments, and by computer scientists as running programs.  These are the same thing.  Experiments/programs are described by \emph{quantum circuits} (see Figure \ref{fig:circuit}a), which specify a schedule of logic operations.

In this paper, the experiments we consider correspond to circuits that begin with initialization of all the qubits, conclude with measurement of all the qubits, and apply zero or more logic gates in the middle.

If a processor contains just one qubit, then only one logic gate can be performed at a time, and every circuit corresponds to a state preparation, a sequence of logic gates, and a measurement.  Circuits on many qubits typically divide each clock cycle into operations on just one or two qubits.  We refer to these as \emph{elementary operations}, as we use the more common term ``gate'' to denote a $n$-qubit operation (see below).  The elementary operations of a single clock cycle constitute a \emph{circuit layer}, and so each circuit corresponds to a preparation, sequence of layers, and measurement.   Models of many-qubit circuits must then have a means of describing the way a circuit layer propagates a $n$-qubit quantum state based on the one- and two-qubit elementary operations within that layer.  Such modeling, when the layers are composed of imperfect operations, is nontrivial \cite{Sarovar2019-xc}.



The implementation of GST described in this paper utilizes simple models where each circuit layer is an independent operation.  That is, the only models we consider describe only $n$-qubit operations that never occur in parallel.  To simplify our language, we refer to these $n$-qubit operations as \emph{gates}, and never dissect them further.  This means that \emph{in this paper, the term ``gate'' is synonymous with circuit layer}, as every circuit layer is modeled as an independent operation (a ``gate'') that acts on all the qubits.  For example, a GST model assumes no \emph{a priori} relationship between the behavior of the following three layers (labeled by $G$ for ``gate''):
\begin{itemize}
\item ``$G_{XI}$'' -- An $X$ gate on qubit 1, in parallel with an idle gate on qubit 2.
\item ``$G_{IX}$'' -- An idle gate on qubit 1, in parallel with an $X$ gate on qubit 2.
\item ``$G_{XX}$'' -- An $X$ gate on qubit 1, in parallel with an $X$ gate on qubit 2.
\end{itemize}
This approach rapidly becomes unwieldy as the number of qubits grows, since exponentially many distinct layers are generally possible.  While this limits the scalability of the GST protocol (which was developed in the context of one- and two-qubit systems where modeling parallel gates wasn't an issue), many of the ideas and theory surrounding it can be extended to many-qubit models \cite{MultiQubitGSTInPrep}.

QCVV protocols, including tomography, seek to deduce properties of a processor's elementary logic operations from the results of running circuits.  Doing so requires a model for all three kinds of logic operations (initialization, gates, and measurements) that has adjustable parameters, and can be used to predict circuit outcome probabilities.  The most common model, and the one used by GST, is based on \emph{Hilbert-Schmidt space}.

\subsubsection{States, measurements, and Hilbert-Schmidt space\label{sec:HSSpace}}

A $n$-qubit processor is initialized into a particular \emph{quantum state}.  Knowledges of a quantum state enables predicting probabilities of outcomes of any measurements that could be made on the system.  After initialization, the state evolves as gates are applied.  The final state, just before the processor is measured, determines the probabilities of all possible measurement outcomes.  So we model the three kinds of quantum logic operations as follows:  initialization is modeled by a quantum state; logic gates are modeled by dynamical maps that transform quantum states; measurements are described by linear functionals that map quantum states to probability distributions.

A quantum system is described with the help of a $d$-dimensional \emph{Hilbert space} $\mathcal{H} = \mathbb{C}^d$, where $d$ is the largest number of distinct outcomes of a repeatable measurement on that system (and is therefore somewhat subjective).  For a single qubit, $d=2$ by definition, and for a system of $n$ qubits, $d=2^n$.  \footnote{Formal quantum theory is complicated by the possibility of infinite-dimensional Hilbert spaces, but these are rarely required for quantum computing, so we will assume finite $d$.}

A quantum state is described by a $d\times d$ positive semidefinite, trace-1 matrix $\rho$ that acts on its Hilbert space ($\rho: \mathcal{H} \to \mathcal{H}$) \cite{NielsenChuangQCBook}.  This is called a \emph{density matrix}.  A matrix must be positive semidefinite and have trace 1 in order to represent a physically valid state.  Such density matrices form a convex set, and it is very convenient to embed this set in the complex $d^2$-dimensional vector space of $d\times d$ matrices.  This vector space is called \emph{Hilbert-Schmidt space} and denoted by $\mathcal{B}(\mathcal{H})$.  \emph{Hermitian} (self-adjoint) matrices form a real $d^2$-dimensional subspace of Hilbert-Schmidt space.  In this work, we only consider density matrices that are Hermitian, but disregard the trace and positivity constraints, using the symbol $\rho$ to describe any element of real Hilbert-Schmidt subspace.  We state explicitly if/when the trace and positivity constraints are assumed.

Hilbert-Schmidt space is equipped with an inner product $\langle A, B \rangle \equiv \Tr(A^\dag B)$ that, as we will see, is very useful in quantum tomography.  In a generalization of Dirac notation on a Hilbert space we represent an element $B$ of Hilbert-Schmidt space by a column vector denoted $\sket{B}$, and an element of its (isomorphic) dual space by a row vector denoted $\sbra{A}$, so that $\sbraket{A}{B} = \Tr(A^\dag B)$.  We refer to $\sket{B}$ as a \emph{superket} and to $\sbra{A}$ as a \emph{superbra}, since operations on $\mathcal{B}(\mathcal{H})$ are called \emph{superoperators} (see the next section).

Measuring a quantum system yields an outcome or ``result'' which we assume is drawn from a discrete set of $k$ alternatives.  Each outcome's probability is a linear function of the state $\rho$.  Therefore, the $i$th outcome can be represented by a dual vector $\sbra{E_i}$, so that $\mathrm{Pr}(i|\rho) = \sbraket{E_i}{\rho} = \Tr(E_i\rho)$.  The laws of probability require that $E_i\geq0$ and $\sum_i{E_i}=\Id$.  The $E_i$ are called \emph{effects}.  The set $\{E_i\}$ completely describes the measurement and is called a \emph{positive operator-valued measure (POVM)}.  As with states, we sometimes relax positivity constraints on POVM effects for convenience.

We find it useful to define a basis for Hilbert-Schmidt space, $\left\{B_i\right\}$, with the following properties:
\begin{enumerate}
\item Hermiticity:  $B_i = B_i^\dagger$,
\item Orthonormality:  $\Tr(B_i B_j) = \delta_{ij}$,
\item Traceless for $i>0$:  $B_0 = \Id/\sqrt{d}$ and $\Tr(B_i)=0\ \forall\ i>0$.
\end{enumerate}
Here, $\Id$ is the $d$-dimensional identity map.  For a single qubit, the normalized Pauli matrices $\left\{\Id/\sqrt{2}, \sigma_x/\sqrt{2}, \sigma_y/\sqrt{2}, \sigma_z/\sqrt{2} \right\}$ constitute just such a basis.  For $n$ qubits, the $n$-fold tensor products of the Pauli operators satisfy these conditions.  

Since states and effects are both Hermitian, choosing a Hermitian basis makes it easy to represent states and effects in the $d^2$-dimensional \emph{real} subspace of $\mathcal{B}(\mathcal{H})$ mentioned above containing Hermitian operators.  We abuse notation slightly and hereafter use ``Hilbert-Schmidt space'' and $\mathcal{B}(\mathcal{H})$ to denote this real vector space, because we never need its complex extension.

We conclude with a brief example to illustrate these concepts more concretely.  Consider the single-qubit density matrix
\begin{equation}
  \rho = \left( \begin{array}{cc}
    3/4 & (1+i)/8 \\
    (1-i)/8 & 1/4 \end{array} \right).
\end{equation}
This matrix can be written as a linear combination of the normalized Pauli matrices,
\begin{equation}
  \rho = \frac{1}{\sqrt{2}} \left(\frac{\Id}{\sqrt{2}}\right) + \frac{1}{4\sqrt{2}} \left(\frac{\sigma_x}{\sqrt{2}}\right) -  \frac{1}{4\sqrt{2}} \left(\frac{\sigma_y}{\sqrt{2}}\right) + \frac{1}{2\sqrt{2}} \left(\frac{\sigma_z}{\sqrt{2}}\right),
  \end{equation}
where each coefficient can be found by taking the inner product $\Tr(B_i^\dag \rho)$ (these coefficients will always be real).  We then represent $\rho$ as $\sket{\rho} \in \mathcal{B}(\mathcal{H})$ by the real column vector
\begin{equation}
  \sket{\rho} = \left( \begin{array}{c}
    1/\sqrt{2} \\
    1/(4\sqrt{2}) \\
    -1/(4\sqrt{2}) \\
    1/(2\sqrt{2})\end{array} \right).
\end{equation}
Similarly, the projectors onto the $\ket{0}$ and $\ket{1}$ states, given by the complex-valued matrices
\begin{equation}
  E_0 = \left( \begin{array}{cc}
    1 & 0 \\
    0 & 0 \end{array} \right)
  \quad \mbox{and} \quad
  E_1 = \left( \begin{array}{cc}
    0 & 0 \\
    0 & 1 \end{array} \right)
\end{equation}
may be represented as the row (dual) vectors
\begin{eqnarray}
  \sbra{E_0} &=& \left(\begin{array}{cccc}1/\sqrt{2} & 0 & 0 & 1/\sqrt{2}\end{array}\right) \,\mbox{and}\\
  \sbra{E_1} &=& \left(\begin{array}{cccc}1/\sqrt{2} & 0 & 0 & -1/\sqrt{2}\end{array}\right).
\end{eqnarray}
A measurement with the POVM $\{E_0, E_1\}$ gives the probability 0 or 1 via the Born rule,
\begin{eqnarray}
  p_0 &=& \sbraket{E_0}{\rho} = \sbra{E_0} \cdot \sket{\rho} = \Tr(E_0 \rho) = \frac{3}{4} \\
  p_1 &=& \sbraket{E_1}{\rho} = \sbra{E_1} \cdot \sket{\rho} = \Tr(E_1 \rho) = \frac{1}{4},
\end{eqnarray}
where $\cdot$ is the standard vector dot product.  This example illustrates the convenience of representing states and measurements as real vectors in Hilbert-Schmidt space, and the similarity with standard Dirac notation.
  

\subsubsection{Quantum logic gates\label{sec:Qgates}}

What happens between the beginning of a quantum circuit (initialization into a state $\rho$) and its end (a measurement $\{E_i\}$), is defined by a sequence of circuit layers (Figure \ref{fig:circuit}a).  Because we treat each circuit layer as a unique and independent ``gate'' (see Section \ref{sec:circuits}), the middle portion of a circuit can be seen as a sequence of \emph{quantum logic gates} (Figure \ref{fig:circuit}b).  Gates are dynamical transformations on the set of quantum states.  An \emph{ideal} quantum logic gate is reversible and corresponds to a unitary transformation of $\mathcal{H}$.  Such a gate would transform $\rho$ as $\rho \to U\rho U^\dagger$ for some $d\times d$ unitary matrix $U$.  This is a linear transformation from $\mathcal{B}(\mathcal{H})$ to itself, and is called a \emph{superoperator}.  The superoperator describing an ideal (unitary) gate is \emph{orthogonal}.  \footnote{To be clear, the $d\times d$ unitary matrix $U$ is not a superoperator -- it's an operator.  The linear transformation $\rho \to U\rho U^\dagger$ acting on Hilbert-Schmidt space is the superoperator.}

Real logic gates are noisy, and not perfectly reversible.  They still act linearly on states, so they can be represented by superoperators on $\mathcal{B}(\mathcal{H})$.  But the superoperators for noisy gates are not orthogonal, and are generally contractive.  These superoperators are known as \emph{quantum processes} or \emph{quantum channels}.  Given any superoperator $\Lambda$, we can represent it as a $d^2 \times d^2$ matrix that acts associatively (by left multiplication) on $\sket{\rho}\in\mathcal{B}(\mathcal{H})$, by just choosing a basis for $\mathcal{B}(\mathcal{H})$.  This representation is often called the \emph{transfer matrix} of $\Lambda$.  We adopt this terminology, and denote it by $\mathsf{S}_\Lambda$.  Thus, we write
\begin{equation}
  \Lambda : \sket{\rho} \mapsto \mathsf{S}_\Lambda\sket{\rho},\label{eq:transfermxaction}
\end{equation}
where the column vector $\mathsf{S}_\Lambda\sket{\rho}$ is obtained by multiplying the column vector $\sket{\rho}$ from the left by the matrix $\mathsf{S}_\Lambda$.  If $\rho$ is prepared, and then $\Lambda$ is performed, and finally $\{E_i\}$ is measured, then the probability of outcome $i$ is
\begin{equation}
p_i = \sbraopket{E_i}{\mathsf{S}_\Lambda}{\rho} = \Tr(E_i \mathsf{S}_\Lambda \rho).
\end{equation}

Not every linear superoperator describes a physically allowed operation.  For example, the summed-up probability of all measurement outcomes is given by $\Tr(\rho)$, so $\Tr(\rho)$ must equal 1.  A superoperator $\Lambda$ that changes $\Tr(\rho)$ violates the law of total probability.  So only \emph{trace-preserving} superoperators are physically allowed.  Furthermore, if $\rho$ is not positive semidefinite, then some outcome of some measurement will have negative probability.  So states must satisfy $\rho \geq 0$.  If there exists any $\rho\geq0$ such that $\Lambda(\rho)$ is \emph{not} $\geq0$, then that $\Lambda$ is not physically possible.  A superoperator that preserves $\rho\geq0$ is called \emph{positive}.

However, a stronger condition is actually required:  when $\Lambda$ acts on \emph{part} of a larger system, it must preserve the positivity of that extended system.  This corresponds to requiring $\Lambda \otimes \Id_{\mathcal{A}}$ to be positive for any \emph{auxiliary} state space $\mathcal{A}$, and is called \emph{complete positivity} \cite{Choi1975-hx,Kraus1983-gg}.  It is stronger than simple positivity; some positive maps aren't completely positive.  Superoperators representing physically possible operations must be \emph{completely positive and trace-preserving} (CPTP).  The CPTP constraint turns out to be both necessary \emph{and} sufficient; any CPTP superoperator can be physically implemented given appropriate resources by the \emph{Stinespring dilation} construction \cite{NielsenChuangQCBook}.

The TP (trace-preserving) condition is easy to describe and impose in the transfer matrix representation; it corresponds to $\sbra{\Id}\mathsf{S}_\Lambda = \sbra{\Id}$.  If $\mathsf{S}_\Lambda$ is written in a basis whose elements are traceless except for the first one (as required earlier), then $\Lambda$ is TP if and only if the first row of $\mathsf{S}_\Lambda$ equals $[1, 0 \ldots 0]$.

The CP condition is more easily described in a different representation, the \emph{operator sum representation} \cite{Choi1975-hx,Kraus1983-gg}:
\begin{equation}
\Lambda : \rho \mapsto \sum_{ij} \chi^\Lambda_{ij} B_i \rho B_j^\dag\label{eq:choiRep}.
\end{equation}
Here, $\left\{B_i\right\}$ is a basis for $\mathcal{B}(\mathcal{H})$, and $\chi^\Lambda_{ij}$ is a matrix of coefficients called the ``Choi process matrix'' that represents $\Lambda$.  This \emph{Choi representation} of $\Lambda$ provides the same information as the transfer matrix $S_\Lambda$.  The mapping between them is known as the Choi-Jamiolkowski isomorphism \cite{Choi1975-hx,ChoiJamiolkowskiIsomorphism}:
\begin{equation}
  \chi^\Lambda = d(\mathsf{S}_\Lambda\otimes\Id)\sket{\Pi_{\mathrm{EPR}}},\label{eq:ChoiJIso}
\end{equation}
where $\Pi_{\mathrm{EPR}} = |\Psi_{\mathrm{EPR}}\rangle\langle\Psi_{\mathrm{EPR}}|$ and $|\Psi_{\mathrm{EPR}}\rangle$ is a maximally entangled state on a space that is the tensor product of $\mathcal{B}(\mathcal{H})$ and a reference auxiliary space of equal dimension.  Note that Eq.~\ref{eq:ChoiJIso} is technically problematic: the right hand side is a super-ket, but the left hand side is a matrix.  Equality here means that, in a consistent basis, these two objects are element-wise equal.  This element-wise equality is the Choi-Jamiolkowski isomorphism.  \emph{Krauss operators} correspond to the eigenvectors of $\chi^\Lambda$, and provide an intuitive picture of many common superoperators\cite{NielsenChuangQCBook}.

The CP condition is elegant and simple in the Choi representation:  $\Lambda$ is CP if and only if $\chi$ is positive semidefinite.  This condition is independent of the basis $\{B_i\}$ used to define $\chi$.

Hereafter, we use the transfer matrix representation almost exclusively, and so we use the term ``superoperator'' to also refer to the superoperator's transfer matrix.  Likewise, when we refer to a superoperator's matrix representation, this should be understood to be the transfer matrix.  When we use the Choi matrix representation (only to test/impose complete positivity), we will state it explicitly.  We use the term ``quantum operation'' interchangeably with ``gate'' to refer to \emph{general} quantum operations (not necessarily CP or TP), and will explicitly state when CP and/or TP conditions are assumed.

\subsubsection{Gate sets\label{sec:GateSets}}

Prior to GST, tomographic protocols sought to reconstruct individual logic operations -- states, gates, or measurements -- in isolation.  But in real qubits and quantum processors, this isolation isn't possible.  Every processor has initial states, gates, \emph{and} measurements, and characterizing any one of them requires making use of the others.  Underlying GST, RB, robust phase estimation \cite{Kimmel2015-tj}, and other modern QCVV protocols is the simultaneous representation of \emph{all} a processor's operations as a single entity -- a \emph{gate set}.

Most processors provide just one native state preparation and one native measurement.  Gates are used to prepare other states and perform other measurements.  For completeness, we consider the most general situation here.  Consider a processor that can perform $N_{\mathrm{G}}$ distinct gates, $N_\rho$ distinct state preparations, and $N_{\mathrm{M}}$ distinct measurements, and where the $m$-th measurement has $N^{(m)}_{\mathrm{E}}$ distinct outcomes.  We define the following representations for those operations:
\begin{eqnarray}
  G_i: \mathcal{B}(\mathcal{H}) \to \mathcal{B}(\mathcal{H}) &\,\mbox{for}\,& i=1\ldots N_{\mathrm{G}},\nonumber\\
  \sket{\rho^{(i)}} \in \mathcal{B}(\mathcal{H}) &\,\mbox{for}\,& i=1\ldots N_\rho, \mbox{ and}\label{eq:gateset_matvecs}\\
  \sbra{E^{(m)}_i} \in \mathcal{B}(\mathcal{H})^* &\,\mbox{for}\,& m=1\ldots N_{\mathrm{M}}, \, i=1\ldots N^{(m)}_{\mathrm{E}}\nonumber.
\end{eqnarray}
The collective set of symbols $G_i$, $\sket{\rho^{(i)}}$, and \sbra{E^{(m)}_i} serve two distinct roles.  First, by simply naming a set of operations, they provide a specification of the quantum processor's capabilities.  Secondly, each symbol represents a mathematical \emph{model} for the behavior (ideal, estimated, or unknown, depending on context) of the physical hardware device when the corresponding operation is performed in a quantum circuit.  The notation given above is used throughout this paper.  We refer to all these matrices and vectors collectively as a \emph{gate set}, denoted by $\mathcal{G}$, and defined as
\begin{equation}
\mathcal{G} = \left\{ \left\{\sket{\rho^{(i)}}\right\}_{i=1}^{N_\rho}; \left\{G_i\right\}_{i=1}^{N_{\mathrm{G}}}; \left\{\sbra{E^{(m)}_i}\right\}_{m=1,i=1}^{N_{\mathrm{M}},N^{(m)}_{\mathrm{E}}} \right\}\label{eq:GatesetDef}.
\end{equation}
A gate set is a complete model and description of a quantum processor's behavior when running quantum circuits.

A gate set is more than just a convenient way to package together independent gate, preparation, and measurement operations.  The operations on a quantum processor that a gate set describes are \emph{relational and inter-dependent}.  As a consequence, the representation given above -- where gates correspond to $d^2 \times d^2$ superoperators, state preparations to $d^2$-element column vectors, and POVM effects to $d^2$-element row vectors -- is an \emph{over-specification} of the physical gate set.  To see this, consider a transformation of the gate set that acts as
\begin{eqnarray}
\sbra{E^{(m)}_i} &\to& \sbra{E^{(m)}_i} M^{-1} \nonumber \\
\sket{\rho^{(i)}} &\to& M \sket{\rho^{(i)}} \nonumber \\
G_i &\to& M G_i M^{-1} \label{eq:GaugeTransform},
\end{eqnarray}
where $M$ is any invertible superoperator.  This transformation changes the concrete representation of the gate set, but does not change any observable probability (see Eq.\ref{eq:generalFreqToProb}).  Since absolutely nothing observable has changed, these are equivalent descriptions or representations of the same physical system.  

The transformations defined by Eq.~\ref{eq:GaugeTransform} form a continuous group, so the standard representation of gate sets described above contains entire ``orbits'' of equivalent gate sets. This degeneracy, referred to generically as ``gauge freedom'', is a perennial irritant and obstacle to characterization of quantum processors \cite{Blume-Kohout2017-kn,Proctor2017-ru,Lin_2019,matteo2020operational}.  More importantly, it shows that a gate set is not just the sum of its parts, and that tomography on a gate set is not just tomography on its components.  GST is tomography of a novel entity.

\subsubsection{Circuits\label{sec:GateSequences}}

The term ``quantum circuit'' is used in many contexts to mean subtly different things.  For clarity, we find it helpful to distinguish two related but distinct types of quantum circuits: fixed-input, classical-output (FI/CO) and quantum-input, quantum-output (QI/QO) circuits.

Data for standard QCVV protocols -- e.g., state/process tomography, RB, or GST -- consists of the counted outcomes of experiments.  Each experiment is described by a quantum circuit that begins by initializing and ends by measuring of all the qubits.  Because the input to such a circuit is ``fixed'' by the state preparation and the output of the circuit is classical measurement data, we call this type of circuit a \emph{fixed-input classical-output circuit}.  A FI/CO circuit represents and describes a probability distribution over classical bit strings, and a more-or-less concrete procedure for generating it with a quantum processor.

A quantum circuit can \emph{also} describe an arrangement of unitary logic gates, with no explicit initialization or measurement, that could be inserted into a larger circuit as a subroutine.  We call this kind of circuit a \emph{quantum-input quantum-output circuit}.  It is a sequence of circuit layer operations, none of which are preparation or measurement layers.  A QI/QO circuit represents and describes a quantum channel that maps an input quantum state to an output quantum state.

Thus, each FI/CO circuit comprises (1) one of the $N_\rho$ possible state preparations, (2) a QI/QO circuit (a sequence of layers), and (3) one of the $N_m$ measurements, which generates a specific outcome.  On a processor that has just one state preparation and one measurement operation, every FI/CO circuit can be described completely by the QI/QO circuit composed of all its circuit layers.

Let $\gamma$ be an index (or label) for available circuit layers (gates).  When we define a QI/QO circuit $S$ as $S = (\gamma_1, \gamma_2, ... \gamma_L)$, it means ``do layer $\gamma_1$, then do gate $\gamma_2$, etc.''  When the layer indexed (labeled) by $\gamma$ is performed, it transforms the quantum processor's state.  This transformation is represented by a superoperator $G_{\gamma}$.  Performing an entire QI/QO circuit, e.g. $S$, also applies a superoperator.  We denote the transfer matrix for QI/QO circuit $S$ by $\seqaction(S)$.  It denotes the map from $\mathcal{B}(\mathcal{H}) \to \mathcal{B}(\mathcal{H})$ formed by composing the elements of $S$, i.e., $\seqaction(S) = G_{\gamma_L} \cdots G_{\gamma_2} G_{\gamma_1}$.  Because it is a common source of confusion, we emphasize that gates appear in chronological order in layer sequences ($S$), but in reverse chronological order in matrix products $\seqaction(S)$, because that's how matrix multiplication works.  For later reference, the general action of $\seqaction$ on a layer sequence can be written
\begin{equation}
  \seqaction\left( (\gamma_1, \gamma_2, ... \gamma_L) \right) = G_{\gamma_L} \cdots G_{\gamma_2} G_{\gamma_1},\label{eq:sigmaDefinition}
\end{equation}
where $1 \le \gamma_i \le N_{\mathrm{G}}$ is an integer index into the set $\left\{G_i\right\}_{i=1}^{N_{\mathrm{G}}}$ of corresponding gates (quantum processes).  We use integer exponentiation of a QI/QO circuit to denote repetition.  So, if $S = (\gamma_1, \gamma_2)$ then $S^2 = SS = (\gamma_1,\gamma_2,\gamma_1,\gamma_2)$.  It follows that
\begin{equation}
  \seqaction(S^n) = \seqaction(S)^n.
\end{equation}

Data sets are generated by defining a set of FI/CO circuits, repeating each one $N$ times, and recording the results.  Results from each specific circuit are summarized by observed frequencies, $f_k = n_k/N$, where $n_k$ is the number of times the $k$th measurement outcome was observed, for $k=1\ldots N^{(m)}_{\mathrm{E}}$.  These frequencies are usually close to their corresponding probabilities,
\begin{equation}
f_k \approx \sbraopket{E^{(m)}_k}{G_{\gamma_L} \cdots G_{\gamma_2} G_{\gamma_1}}{\rho^{(i)}} \quad \forall k, \label{eq:generalFreqToProb}
\end{equation}
and can be used to estimate them.  More sophisticated estimators exist, but Eq.~\ref{eq:generalFreqToProb} motivates the idea that some or all of $\sket{\rho^{(i)}}$, $\sbra{E^{(m)}_k}$ and $G_j$ can be inferred from the observed frequencies.  Doing so is tomography.  Each kind of tomography treats some operations as known, and uses them as a reference frame to estimate the others.

It is usually apparent from context whether a FI/CO or QI/QO circuit is being referenced.  In the remainder of this paper, we only specify the type of circuit in cases when it may be ambiguous.

\subsection{State, Process, and Measurement Tomography\label{sec:StateAndProcessTomography}}

In this section we review standard state and process tomography.  We also briefly describe the (straightforward) generalization to measurement tomography.  These methods establish the context for GST, but they also introduce the mathematical machinery and notation that we will use for GST.

\begin{figure}
  \begin{center}
    \includegraphics[width=2.7in]{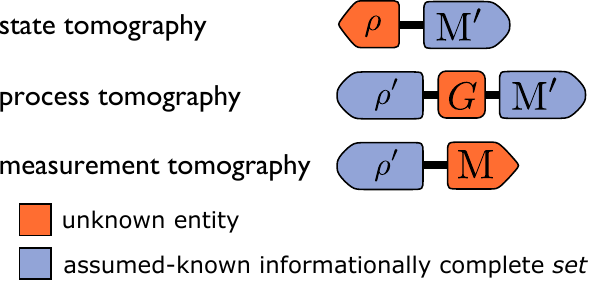}
    \caption{Structure of the circuits required for state, process, and measurement tomography.  Each of these protocols reconstructs an unknown entity (a state $\rho$, process $G$, or measurement $\mathrm{M}$) by placing that entity in circuits with an assumed-known reference frame formed by an informationally complete set of state preparations or measurements (or both).  Primed symbols ($\rho'$ and $\mathrm{M}'$) are meant to connote \emph{effective} state preparations and measurements, which are often implemented by applying gate operations after or before a native state preparation or measurement.  A critical problem with these standard tomographic techniques is that $\rho'$ and $\mathrm{M}'$ are in practice never known exactly.\label{fig:StdTomographyCircuits}}
  \end{center}
\end{figure}

\subsubsection{Quantum state tomography\label{sec:QuantumStateTomography}}

The goal of quantum state tomography \cite{Gale1968-mk, Vogel1989-vt, Hradil1997-fv, Munro2001-jn, James2001-hz, Artiles2005-ts, Blume-Kohout2010-vv, Blume-Kohout2010-hb, Christandl2012-am, Smolin2012-yv, Banaszek2013-zx, Granade2016-qy, Haah2017-jw} is to construct a full description of a quantum state $\rho$, from experimental data.  It is assumed that some quantum system can be repeatedly prepared in the unknown state $\rho$, that various \emph{fiducial measurements} can be performed on it, that those measurements are collectively \emph{informationally complete}, and that the POVMs representing those measurements are \emph{known} (see Figure \ref{fig:StdTomographyCircuits}).

``Fiducial'' means ``accepted as a fixed basis of reference'', and the measurements used in state tomography constitute a frame of reference.  A set of measurements $\{E^{(m)}_i\}_{m,i}$ is informationally complete if and only if, for any state $\rho$, the probabilities
\begin{displaymath}
  p^{(m)}_i(\rho) = \Tr\left[\rho E^{(m)}_i\right]
\end{displaymath}
\emph{uniquely} identify $\rho$ -- i.e., there is no other $\rho$ consistent with them.  This implies a simple linear-algebraic condition: the $\{E^{(m)}_i\}$ must span the entire space of effects, forming a complete dual basis for states.

To perform state tomography, many copies of $\rho$ are made available, and divided into $M$ pools.  The $m$th fiducial measurement is applied to all the copies in the $m$th pool.  The observed frequencies $f^{(m)}_i$ are recorded, and used to estimate the corresponding probabilities
\begin{displaymath}
  p^{(m)}_i(\rho) = \Tr\left[\rho E^{(m)}_i\right].
\end{displaymath}
Then $\rho$ is deduced from these probabilities.

Practical state tomography is more complicated (and interesting) because only finitely many copies can be measured, so each observed frequency isn't generally equal to the corresponding probability.  Those probabilities have to be \emph{estimated} from the data.  One option is to estimate $\hat{p}_i = f_i$ (where $\hat{\cdot}$ over a variable indicates an estimate).  This is called \emph{linear inversion}. It's not very accurate, and often yields an estimate $\hat\rho$ that isn't positive, but despite these practical drawbacks it's a theoretical cornerstone of tomography.  Linear inversion underlies the concept of informational completeness, which captures whether an experiment design (here, a set of measurements) is sufficient for estimation.  These ideas also underly GST, and are used directly in linear GST.  So we now present a concise overview of linear inversion state tomography and informational completeness.

Let the Hilbert-Schmidt space vector $\sket{\rho}$ denote an unknown quantum state that we want to reconstruct.  Let $\{E^{(m)}_i\}_{m,i}$ represent the set of $M$ \emph{known} fiducial measurements, indexed by $m=1\ldots M$, with $i=1\ldots N^{(m)}$ indexing the outcomes of the $m$th measurement.  We represent each of these effects as a dual vector $\sbra{E^{(m)}_i}$ in Hilbert-Schmidt space.  Only the list of effects matters, not which measurement $m$ a given effect $E^{(m)}_i$ belongs to, so we simply list all the effects as $\{E_j:\ j=1\ldots N_{f1}\}$, where $N_{f1}$ is the total number of distinct measurement outcomes for \emph{all} measurements performed. 

Now, let us assume that we have learned the exact probabilities of each measurement outcome in state $\rho$ (ignoring the entire process of performing measurements on samples, collating the results, and estimating probabilities from frequencies).  These probabilities are given by Born's rule,
\begin{equation}
p_j = \Tr\left[E_j \rho\right].
\end{equation}
We can we write this as a Hilbert-Schmidt space inner product,
\begin{equation}
p_j = \sbraket{E_j}{\rho},
\end{equation}
and then stack all the row vectors $\sbra{E_j}$ atop each other to form an $N_{f1}\times d^2$ matrix
\begin{equation}
A = \left(\begin{array}{c} \sbra{E_{1}} \\  \sbra{E_{2}} \\  \vdots \\ \sbra{E_{N_{f1}}} \\ \end{array} \right).\label{eq:defineA}
\end{equation}
Now, $\vec{p} = A\sket{\rho}$ is a vector whose $j$th element is $p_j$ (the probability of $E_j$, as defined above).  In the context of state tomography, all the measurement effects $E_j$ are assumed to be known \emph{a priori}, so the entire matrix $A$ is known.  $A$ also has full column rank ($d^2$), because the $\{\sbra{E^{(m)}_i}\}$ are informationally complete and therefore span Hilbert-Schmidt space. Therefore, we can reconstruct $\rho$ using linear algebra.  If $A$ is square, it has an inverse, and $\sket{\rho} = A^{-1} \vec{p}$.  If $N_{f1}$ is greater than $d^2$, then the $\{\sbra{E^{(m)}_i}\}$ form an overcomplete basis and $A$ is not square.  In this case, we can solve for $\sket{\rho}$ using a pseudo-inverse:
\begin{equation}
\sket{\rho} = (A^T A)^{-1}A^T \vec{p}.
\end{equation}

\subsubsection{Quantum process tomography\label{sec:QuantumProcessTomography}}

The goal of quantum process tomography \cite{Poyatos1997-mz, Chuang1997-vf, Fiurasek2001-mp, DAriano2001-fx, Childs2001-en, Altepeter2003-fb, OBrien2004-tr, Weinstein2004-vn, Mohseni2006-cp, Riebe2006-tc, Bendersky2008-xa, Lobino2008-ud, Bialczak2010-et, Kimmel2014-nx, Kim2014-ix} is to construct a full description of a quantum process (e.g., a gate) from experimental data.  This is done by constructing an informationally complete set of known \emph{fiducial states}, preparing many copies of each of them, passing those copies through the process of interest, and performing state tomography on the output states (see Figure \ref{fig:StdTomographyCircuits}).  All the same caveats and complications mentioned for state tomography apply here.  As above, we ignore them and focus on the underlying math problem of reconstructing an unknown process from exactly measured probabilities.

Let $G$ be the superoperator (transfer matrix) representing the process we want to reconstruct.  Let $\{\sbra{E_j}\}$ be the list of POVM effects representing all the outcomes of all the fiducial measurements, as in state tomography.  And let $\{\sket{\rho_i}\}$ be a list of $N_{f2}$ fiducial quantum states that will be repeatedly prepared as inputs to $G$.  

If state $\rho_i$ is prepared, $G$ is applied, and a measurement is performed with possible outcomes $\{E_j\}$, then the probability of observing outcome $E_j$ is
\begin{eqnarray}
  P_{j,i} &=& \Tr\left(E_j G[\rho_i]\right)\\
	&=& \sbraopket{E_j}{G}{\rho_i}.
\end{eqnarray}
In addition to the $A$ matrix defined previously, we define a new $d^2 \times N_{f2}$ matrix $B$ from the column vectors representing the fiducial states $\sket{\rho_i}$:
\begin{equation}
B = \left(\begin{array}{cccc} \sket{\rho_1} &  \sket{\rho_2} &  \cdots \sket{\rho_{N_{f2}}} \\ \end{array} \right).\label{eq:defineB}
\end{equation}
Now, we can write the $N_{f1}\times N_{f2}$ matrix $P$ whose elements are $P_{j,i}$ as
\begin{equation}
P = AGB.
\end{equation}
In the context of process tomography, both the fiducial states and the fiducial measurement effects are known, so all the elements of both $A$ and $B$ are known.  $A$ must have full column rank and $B$ must have full row rank, because the $\{\rho_i\}$ and $\{E_j\}$ were assumed to be informationally complete.  If they are square ($N_{f1} = N_{f2} = d^2$), then they are invertible, and we can immediately reconstruct the original process as
\begin{equation}
G = A^{-1}PB^{-1}.
\end{equation}
If there are more than $d^2$ input states and/or measurement outcomes, then (as before) we use a pseudo-inverse to obtain a least-squares solution:
\begin{equation}
G = (A^T A)^{-1}A^T P B^T (B B^T)^{-1}.\label{eq:ProcessTomo_O}
\end{equation}
More sophisticated estimators -- i.e., maps from $\{p_{i,j}\}$ data to estimates of $G$ -- exist.  They are important when finite-sample errors in the estimated probabilities are significant.  But the linear inversion tomography described here is the essence of process tomography.

\subsubsection{Measurement tomography}

Using tomographic techniques to characterize measurements gets much less attention than state and process tomography -- partly, perhaps, because it's a straightforward generalization of state tomography.

Measurement tomography \cite{Luis1999-ek, Fiurasek2001-zc, Lundeen2009-qs, Blumoff2016-uk, Chen2019-zs} uses pre-calibrated fiducial states to reconstruct the POVM effects for an unknown measurement (see Figure \ref{fig:StdTomographyCircuits}).  This is easy to represent using the framework already developed for process tomography.  If $\{\rho_i\}$ are known fiducial states, and $\{E_j\}$ is the unknown POVM, then we can define one vector of observable probabilities $\vec{p_j}$ for each unknown effect:
\begin{eqnarray}
  [p_j]_i &=& \Tr[\rho_i E_j] \\
	&=& \sbraket{E_j}{\rho_i}.
\end{eqnarray}
Using the $B$ matrix defined in the preceding discussion of process tomography, this is simply
\begin{equation}
\vec{p_j}^T = \sbra{E_j}B,
\end{equation}
which means that each effect $E_j$ can be reconstructed by linear inversion as
\begin{equation}
\sbra{E_j} = \vec{p_j}^T B^{-1},
\end{equation}
where $B^{-1}$ is a matrix inverse or pseudo-inverse, as appropriate.

\subsection{The role of calibration in tomography\label{sec:calibration}}

State, process, and measurement tomography are conceptually straightforward, but rely on a critical (and unrealistic) assumption.  They reconstruct the unknown operation \emph{relative} to some reference operations -- fiducial measurements and/or states -- that are assumed to be perfectly known.  Often, the reference operations are also assumed to be noiseless.

These assumptions are never satisfied in practice.  Identifying the exact POVMs that describe the fiducial measurements for state tomography -- i.e., \emph{calibrating} them -- would require perfectly known states.  But identifying those states, by state tomography, relies on known measurements.  This leads to an endless loop of self-referentiality.  In the same way, process tomography relies on fiducial states and measurements that are almost always produced by applying quantum logic gates (specific quantum processes) to just a few native states and measurements \cite{TakahashiPRA13}.  A process tomographer's knowledge of those fiducial objects can be no more precise than their knowledge of the gates used to prepare them -- which would have to be characterized with process tomography.

Ref.~\cite{MerkelPRA13} articulated this problem clearly.  It also demonstrated that this concern is real, and has practical consequences.  In realistic scenarios, errors in state preparation and measurement (SPAM) dominate inaccuracy in process tomography.

Several forms of ``calibration-free'' tomography \cite{BranczykNJP12,MogilevtsevNJP12,GST2013,MedfordNatureNano13,QuesadaPRA13,QuesadaQIM13,QuesadaFO13,StraupePRA13,StarkPRA14,Kimmel2014-nx,MohammadiPRA14,SchwemmerPRL15,JacksonPRA15,HouJOSAB16,ChapmanPRL16,Blume-Kohout2017-kn,McCormickPRA17,KeithPRA18} were proposed either independently from, or in response to, Ref.~\cite{MerkelPRA13}. Each sought to characterize quantum gates, state preparations, and/or measurements self-consistently \emph{without} any prior calibration.  Gate set tomography has emerged as the most widely adopted of these, and is now a \emph{de facto} standard for performing calibration-free tomography.  However, it comes with certain costs, which have motivated the continued use of state and process tomography in some experiments.  While we recognize the costs and drawbacks of GST, the risks of traditional state and process tomography are so clear that they should not be used, except under remarkable circumstances (e.g., where there are very good reasons to believe the reference operations are known to higher precision than is needed in the final estimate).

\section{Linear gate set tomography (LGST)\label{sec:LGST}}

GST offers two benefits.  It resolves the circularity inherent to state and process tomography, and achieves higher accuracy with lower experimental cost than process tomography.  These features are only loosely linked.  Linear GST (LGST) resolves the pre-calibration problem, but has the same accuracy/effort scaling as process tomography.  In this section, we use LGST to introduce the basic concepts and methods of GST.

\subsection{Introduction to LGST\label{sec:LGSTbackground}}

Linear-inversion gate set tomography (LGST) is basically a calibration-free amalgamation of state, process, and measurement tomography.  It constructs a low-precision estimate of a gate set $\mathcal{G}$ using the data from a specific set of short circuits.  LGST demonstrates the feasibility of avoiding pre-calibrated reference frames, and also why doing so creates gauge freedom.  

LGST was developed roughly contemporaneously with IBM's ``overkill'' approach to gate set tomography \cite{MerkelPRA13}.  IBM identified the pre-calibration problem, recognized that circuit outcome probabilities are given by expressions of the form $p_S = \sbraopket{E}{G\ldots}{\rho}$, and observed that a sufficiently large set of such probabilities should be sufficient to reconstruct the gate set.  They proposed performing \emph{all} circuits of 3 or fewer gates to generate data, then fitting a gate set to that data by numerical maximization of the likelihood function (maximum likelihood estimation, or MLE)
\begin{equation}
\mathcal{L}(\mathcal{G}) = \mathrm{Pr}(\mathrm{data}|\mathcal{G}).
\end{equation}

But this likelihood function is not well-behaved.  It is not quasi-convex (i.e., its level sets are not convex), because (1) the probabilities for the experimental observations are not linear functions of the parameters, and (2) the existence of the gauge makes the likelihood's maximum very degenerate.  Plotting the likelihood would reveal, instead of a unimodal ``hill'', an assortment of winding ``ridges'' with perfectly level crests.   Standard optimization methods \cite{ConvexOptimization} often fail to find global maxima on such functions, getting stuck in local maxima.  The IBM ``overkill'' algorithm therefore relied on starting with a good initial guess for the gate set, that would lie with high probability in a local neighborhood of the likelihood's maximum.

LGST avoids this problem by using a very different, structured, experiment design.  Instead of an unstructured set of short circuits, LGST prescribes a specific set of them.  The resulting data can be analyzed using only linear algebra, and the analysis is very similar to that used for process tomography.

As LGST looks very much like process tomography, it's important to recognize that it is doing something significantly different.  If LGST \emph{was} trying to reconstruct transfer matrices for individual gates, it would fail, because reconstructing individual gates without a pre-calibrated reference frame is not possible.  LGST reconstructs gates \emph{up to gauge freedom}.  Actually, it does more -- it reconstructs the entire gate set up to the \emph{global} gauge freedom given in Eq.~\ref{eq:GaugeTransform}, recapitulated below:
\begin{eqnarray}
\sbra{E^{(m)}_i} &\to& \sbra{E^{(m)}_i} M^{-1} \nonumber\\
\sket{\rho^{(i)}} &\to& M \sket{\rho^{(i)}}  \\
G_i &\to& M G_i M^{-1}.\nonumber
\end{eqnarray}
Such transformations change the elements of the gate set, but not any observable probability.  So it's not possible to distinguish between gauge-equivalent gate sets, and reconstructing a gate set up to arbitrary $M$ constitutes success.

Since a gate set comprises states, gates, and measurements, it's tempting to say that LGST characterizes all of them simultaneously.  But this is not quite right.  A gate set is not a collection of \emph{unrelated} quantum operations.  Quantum operations are usually described relative to an implicit and absolute reference frame.  But in most experiments, no such reference frame is available.  So GST characterizes all these operations \emph{relative to each other}, and estimates every property of a gate set that \emph{can} be measured without a reference frame.  But some properties of gate sets can't be measured, even in principle, and they correspond to gauge degrees of freedom.

Gauge freedom makes some familiar properties of gates unmeasurable.  Other properties of gates turn out to be not associated with a single operation, but purely \emph{relational} properties -- i.e., they are properties of the gate set, but not of any individual gate within it.  This awkwardness is the unavoidable price of avoiding pre-calibrated reference frames.  GST outputs a self-consistent representation of the available states, processes, and measurements, \emph{but that representation is generally not unique}. 
If finite-sample errors did not exist, LGST would be a perfect estimator of the gate set, and this paper would be much shorter.  But real experiments always suffer from finite sample error.  $N$ trials of an event with probability $p$ does \emph{not} generally yield exactly $pN$ successes, so estimating $p$ from data generally yields $\hat{p} = p \pm O(1/\sqrt{N})$.  As a result, LGST's estimation error scales as $O(1/\sqrt{N})$, just like process tomography.  So estimating a gate set to within $\pm10^{-5}$ with LGST would require repeating each circuit $N\approx10^{10}$ times, which is impractical.  The ``long-sequence GST'' protocol described in Section \ref{sec:LongSequenceGST} is much more efficient, and makes standalone LGST preferable only when there are severe resource constraints.  But LGST remains important both pedagogically and as a key first step in long-sequence GST's analysis pipeline.

\subsection{The LGST algorithm\label{sec:LGSTalgorithm}}

In this section, we present the core LGST algorithm.  We focus on (1) what LGST seeks to estimate, (2) what data is required for LGST, and (3) how to transform that data into an estimate under ideal circumstances.  At first, we make some idealized simplifying assumptions in order to maximize clarity.  After presenting the core algorithm, we return to these assumptions, relax them, and show how to make this algorithm practical and robust.  The simplifying assumptions are:
\begin{enumerate}
\item We assume the ability to create informationally complete sets of fiducial states $\left\{\sket{\rho'_j}\right\}$ and measurement effects $\left\{\sbra{E'_i}\right\}$ (see Section \ref{sec:LGSTfiducials}).
\item We ignore finite sample error in estimated probabilities, and its effects (see Section \ref{sec:LGSTMLE}).
\item We assume that the fiducial states and effects are exactly informationally complete, not overcomplete, so that $N_{f1} = N_{f2} = d^2$ (see Appendix \ref{sec:LGSTovercomplete}).
\end{enumerate}

We use the notation of Eq.~\ref{eq:GatesetDef} to denote the contents of a generic gate set $\mathcal{G}$:
\begin{equation}
\mathcal{G} = \left\{ \left\{\sket{\rho^{(i)}}\right\}_{i=1}^{N_\rho}; \left\{G_i\right\}_{i=1}^{N_{\mathrm{G}}}; \left\{\sbra{E^{(m)}_i}\right\}_{m=1,i=1}^{N_{\mathrm{M}},N^{(m)}_{\mathrm{E}}} \right\}.\label{eq:GatesetDef2}
\end{equation}
As in Section \ref{sec:GateSets}, $N_\rho$, $N_{\mathrm{G}}$ and $N_{\mathrm{M}}$ are the number of distinct state preparations, gates, and measurements, and $N^{(m)}_{\mathrm{E}}$ is the number of possible outcomes for the $m$-th distinct measurement.

\begin{figure}
  \begin{center}
    \includegraphics[width=3in]{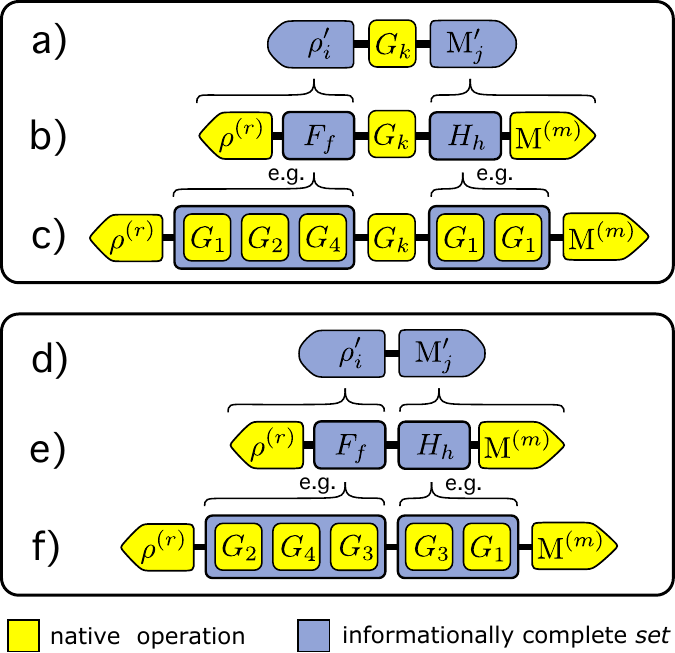}
    \caption{Structures of the two types of circuits required by the LGST algorithm.  \textbf{Upper panel:} Each native gate, $G_k$, is sandwiched between the elements of informationally complete sets of \emph{effective state preparations}, $\{\rho'_i\}$, and of \emph{effective measurements}, $\{\mathrm{M}'_j\}$.  These are the same circuits that process tomography requires to characterize $G_k$.  Line \textbf{(a)} shows these circuits in their simplest form, with each informationally complete set displayed as a unit.  Line \textbf{(b)} depicts the common case when the set of effective preparations (measurements) is implemented by following (preceding) a single native preparation (measurement) operation with a \emph{fiducial circuit} $F_f$ ($H_h$), see Eqs.~\ref{eq:effectiveEffects} and \ref{eq:effectivePreps}.  Line \textbf{(c)} exemplifies that the fiducial circuits are composed of native gates, and gives the circuit entirely in terms of native operations.  \textbf{Lower panel:} Because LGST does not assume knowledge of the $\rho'_i$ and $\mathrm{M}'_j$, it requires circuits that sandwich nothing between pairs of fiducials in order to be self-calibrating.  The circuit diagrams in lines \textbf{(d)}, \textbf{(e)}, and \textbf{(f)} parallel those in \textbf{(a)}-\textbf{(c)}.  LGST also requires the circuits that perform state (measurement) tomography on $\rho$ ($\mathrm{M}$), but these are not explicitly shown. They are similar to \textbf{(d)}-\textbf{(f)} (replacing $\rho'$ with $\rho$ or $\mathrm{M}'$ with $\mathrm{M}$), and are actually included as a subset of these circuits when the gate set contains only a single native state preparation (measurement) and one of the preparation (measurement) fiducial circuits is the empty (do-nothing) cirucit.  \label{fig:LGSTcircuits}}
  \end{center}
\end{figure}

To perform process tomography on an operation $G$ in Section \ref{sec:QuantumProcessTomography}, we constructed a matrix $P$ of observable probabilities using informationally complete fiducial states and effects.  For LGST, we will do the same thing.  But although we assume the existence and implementability of fiducial sets, we do \emph{not} assume that we know them.  They still form a reference frame, but we don't know what that frame is.

To reconstruct a \emph{set} of gates (processes) $\{G_k\}$, we will need one such matrix $P_k$ for each gate:
\begin{equation}
  [P_k]_{i,j} = \sbraopket{E'_i}{G_k}{\rho'_j}\label{eq:LGSTPk}.
\end{equation}
These probabilities are directly measurable -- we don't know what $\rho'_j$ and $E'_i$ are, but we can prepare/measure them.  The first line of Figure \ref{fig:LGSTcircuits} illustrates the circuit corresponding to Eq.~\ref{eq:LGSTPk}.

We can construct $A$ and $B$ matrices from the fiducial vectors exactly as for process tomography.  The difference, of course, is that although those matrices exist, their entries are not known to the tomographer.  Just as before, we can write
\begin{equation}
P_k = AG_kB. \label{eq:PeqAGB}
\end{equation}
Since we do not know $A$ or $B$, we cannot solve this equation for $G_k$.  Instead, to compensate for our ignorance about $\rho'_j$ and $E'_i$, we measure some additional probabilities that correspond to process tomography on the null operation.  We arrange these into a \emph{Gram matrix} for the fiducial states/effects:
\begin{equation}
\tilde{\Id}_{i,j} = \sbraket{E'_i}{\rho'_j} \label{eq:Itilde}.
\end{equation}
Figure \ref{fig:LGSTcircuits}d-f depicts the circuits whose outcome probabilities make up $\tilde{\Id}$.  This matrix can also be written in terms of the $A$ and $B$ matrices, as
\begin{equation}
\tilde{\Id} = AB.
\end{equation}
We assumed that these matrices are square ($N_{f1} = N_{f2} = d^2$) and invertible (which follows from informational completeness).  So we can invert the Gram matrix to get $\tilde{\Id}^{-1} = B^{-1}A^{-1},$ multiply Eq.~\ref{eq:PeqAGB} on the left by it,
\begin{equation}
\tilde{\Id}^{-1}P_k = B^{-1}G_k B,
\end{equation}
and solve for $G_k$ to get
\begin{equation}
G_k = B\tilde{\Id}^{-1}P_k B^{-1}.
\end{equation}

This may not look like the solution -- there's still an unknown $B$ involved -- but it is.  We've reconstructed $G_k$ \emph{up to a similarity transformation} by the unknown $B$.  Moreover, we can do this in exactly the same way for \emph{all} the gates $G_k$, and get estimates of them all up to the \emph{same} $B$.

We also need to reconstruct the native states $\rho^{(l)}$ and measurement effects $\{E_l^{(m)}\}$ in the gate set.  To do so, we construct the following vectors (denoted by $\vec{\cdot}$) of observable probabilities:
\begin{eqnarray}
  \left[\vec{R}^{(l)}\right]_j &=& \sbraket{E'_j}{\rho^{(l)}}\label{eq:LGST_R}\\
  \left[\vec{Q}^{(m)}_l\right]_j &=& \sbraket{E^{(m)}_l}{\rho'_j}\label{eq:LGST_Q}.
\end{eqnarray}
Measuring these probabilities corresponds to performing state tomography on each native state $\rho^{(l)}$, and measurement tomography on every native effect $\{E_l^{(m)}\}$ -- in the unknown frame defined by the fiducial effects and states.  They can be written using $A$ and $B$ as
\begin{eqnarray}
  \vec{R}^{(l)} &=& A \sket{\rho^{(l)}} \label{eq:LGST_R2} \\
  \vec{Q}_l^{(m)T} &=& \sbra{E^{(m)}_l} B, \label{eq:LGST_Q2}
\end{eqnarray}
and by using the identity $\tilde{\Id} = AB$ in Eq.~\ref{eq:LGST_R2}, we get the following equations for \emph{all} the elements of the gate set:
\begin{eqnarray}
  G_k &=& B\tilde{\Id}^{-1}P_k B^{-1} \label{eq:lgstG}\\
  \sket{\rho^{(l)}} &=& B \tilde{\Id}^{-1} \vec{R}^{(l)} \label{eq:lgstRho}\\
  \sbra{E^{(m)}_l} &=& \vec{Q}^{(m)T}_l B^{-1}.\label{eq:lgstE}
\end{eqnarray}
Perhaps surprisingly, this is the answer -- we've recovered the original gate set \emph{up to a gauge}.  The unknown $B$ is now a gauge transformation (see Eq. \ref{eq:GaugeTransform}).  It's not just unknown, but unknowable, and irrelevant.  The tomographer can set $B$ equal to \emph{any} invertible matrix and get back the original gate set, up to gauge freedom.  Equivalently, $B$ indexes different (but equivalent) representations of the gate set.  The simplest choice is $B=\Id$, but this almost always selects a very different gauge from that of the target gate set, resulting in the right hand sides of Eqs.~\ref{eq:LGST_R}-\ref{eq:lgstE} being significantly differentfrom their ideal values even when there is little or no error.  The best \emph{a priori} choice is to choose a $B$ corresponding to the tomographer's best \emph{a priori} guess for the fiducial states (see Eq.~\ref{eq:defineB}), because that (implicitly) defines the gauge in which the tomographer is expecting to work.  But although this is the best \emph{a priori} choice, it is almost never satisfactory for computing gauge-dependent quantities like the process fidelity.  Reliable analysis of the estimate requires \emph{a posteriori} gauge-fixing, which we discuss in Section \ref{sec:GaugeOpt}.

\subsection{Preparing the fiducial vectors \label{sec:LGSTfiducials}}

In Section \ref{sec:LGSTalgorithm} above, we assumed by fiat that informationally complete sets of fiducial states $\{\rho'_j\}$ and effects $\{E'_i\}$ could be prepared.  But most processors admit just \emph{one} native state preparation, and one measurement.  So fiducial states and measurements are implemented using gates from the gate set itself (as shown in Figure \ref{fig:LGSTcircuits} (b-c,e-f)).  

To do this, we define two small sets of QI/QO ``fiducial circuits''.  Each fiducial state is prepared by applying one of the \emph{preparation fiducial circuits} to a native state preparation.  Each fiducial measurement is performed by performing one of the \emph{measurement fiducial circuits} before a native measurement.  In the common case where the native state preparation and measurement are unique, the $\{\sket{\rho'_j}\}$ and $\{\sbra{E'_i}\}$ are entirely determined by the corresponding fiducial circuits:
\begin{eqnarray}
  \sbra{E'_{i}} &=& \sbra{E^{(0)}_{\lfloor i / M \rfloor }}\seqaction\left(H_{i\mod M }\right),\label{eq:effectiveEffectsSimple} \\
  \sket{\rho'_j} &=& \seqaction\left(F_{j}\right)\sket{\rho^{0}}.\label{eq:effectivePrepsSimple}
\end{eqnarray}
Equation \ref{eq:effectiveEffectsSimple} uses $M \equiv N^{(0)}_{\mathrm{E}}$ as shorthand to avoid too cumbersome a notation, and uses $\lfloor\cdot\rfloor$ and $\mathrm{mod}$ to denote the floor function and modular division.

In the more general case of multiple state preparations and measurements, these expressions become:
\begin{eqnarray}
  \sbra{E'_{i}} &=& \sbra{E^{(m(i))}_{t(i)}}\seqaction\left(H_{h(i)}\right),\label{eq:effectiveEffects} \\
  \sket{\rho'_j} &=& \seqaction\left(F_{f(j)}\right)\sket{\rho^{r(j)}},\label{eq:effectivePreps}
\end{eqnarray}
They must include simple functions that map each effective-item index ($i$ or $j$) to the native preparation index ($r(j)$), measurement index ($m(i)$), or fiducial index ($f(j)$ or $h(i)$)) corresponding to that item. Equations \ref{eq:effectiveEffects} and \ref{eq:effectivePreps} (or \ref{eq:effectiveEffectsSimple} and \ref{eq:effectivePrepsSimple}) introduce and define measurement fiducial circuits, $\left\{H_k\right\}$, and preparation fiducial circuits, $\left\{F_k\right\}$ that result in the sets $\{\sket{\rho'_j}\}$ and $\{\sbra{E'_i}\}$ being informationally complete.  Their essential function is describe how effective preparations and measurements are constructed from native operations, and are referred for this purpose in remainder of this work.   This is shown pictorially in Figure \ref{fig:LGSTcircuits}.

This construction has no consequences for the linear inversion analysis described above -- how the fiducial states/effects are produced doesn't matter, as long as they are consistent.  But it does have other consequences.

First, it ensures that every ``observable probability'' required by LGST can be obtained by running a specific circuit, composed from operations in the gate set itself.  

Second, it subtly reduces the number of free parameters in the model, because (e.g.) $\rho'_1$ and $\rho'_2$ are \emph{not} entirely independent, but are actually generated by the same set of operations applied in a different order.  The LGST linear inversion algorithm is blind to this correlation, but more careful statistical analysis can extract it and provide a somewhat more accurate estimate (see below).

Third, this construction places the burden of ensuring informational completeness of the fiducial states/effects onto the choice of fiducial circuits.  This choice is not ``canonical'' for GST -- it depends on the gates available in the gate set.  In practice, ``fiducial selection'' is usually done by (1) modeling the gates as error-free, and then (2) using that model to find a sets of fiducial circuits that produce informationally complete states/effects.  The goal is to have a Gram matrix $\tilde{\Id}$ that is well-conditioned, so that its inverse doesn't inflate finite-sample errors.  Optimal condition number is achieved when the fiducial states/effects form a 2-design\cite{Scott_tightmeasurements_2006}.

Of course, the gates are not necessarily error-free, and fiducial selection may fail if they are sufficiently erroneous.  Fortunately, post-hoc validation is straightforward, by computing the singular values of the measured $\tilde{\Id}$ before inverting it.  If the $d^2$ largest singular values are not all sufficiently large, then different (or additional) fiducials should be added (and more data taken).  If \emph{no} fiducials are found to produce informationally complete sets, then the operations in the gate set are not capable of exploring the system's nominal state space, and tomography must be limited to a subspace thereof\cite{bengtsson2017discrete}.

\subsection{Improving LGST with maximum likelihood estimation \label{sec:LGSTMLE}}

In the preceding sections, we have described the goal of tomography (including GST, and specifically LGST) as ``reconstructing'' states, processes, or gate sets.  This language has a long history in the tomography literature.  But admitting the existence of finite sample errors -- observed \emph{frequencies} of circuit outcomes will not exactly match their underlying \emph{probabilities} -- disrupts this paradigm.  The objects that we ``reconstruct'' are not quite the true ones.  Worse yet, in general there \emph{is} no ``true'' state or process!  Quantum states and quantum processes are mathematical \emph{models} for real physical systems.  Those models rely on assumptions (e.g., Markovianity), which are never quite true in practice.

``Reconstructing'' is a flawed term, and at this point we leave it behind.  It is more accurate to say that GST \emph{estimates} the gate set.  The goal of GST is to fit a gate set to data -- i.e., to find the gate set that fits the observed data best.  This description is more consistent with the least-squares generalization of linear inversion derived above.  But, having rephrased the problem this way, it is very reasonable to ask ``Is least-squares the \emph{right} way to fit the data?''  The LGST algorithms derived above minimize a sum of squared differences between predicted probabilities and observed frequencies:
\begin{equation}
f(\mathrm{gate\ set}) = \sum_j{\left(p_j^{(\mathrm{gate\ set})}-f_j^{(\mathrm{observed})}\right)^2}. \label{eq:SumSqError}
\end{equation}
But while minimizing this sum of squared residuals is very convenient, it's not especially well-motivated.  Furthermore, the LGST algorithms given above don't \emph{actually} minimize the average squared error over the entire dataset -- i.e., all the circuits performed to obtain LGST data -- because the projector $\Pi$ is chosen specifically to capture the support of the Gram matrix $\tilde{\Id}$, without regard for the other directly observed probabilities (e.g., those in the $P_k$).

The best way to fit the observed data optimally is to set aside traditional tomography and linear algebra, and focus directly on (1) the actual data (outcome frequencies of quantum circuits), and (2) the model being fit to it (gate sets).  The set of all possible gate sets can be viewed as a parameterized manifold.  Each point on that manifold is a gate set that predicts specific, computable probabilities for each quantum circuit.  It is therefore possible to write the objective function from Eq.~\ref{eq:SumSqError} precisely -- with $j$ ranging over every circuit performed -- as a function of the gate set, and minimize it numerically to find a gate set that really \emph{does} minimize the total squared error.

There's no particularly compelling reason to minimize the total squared error, though.  An objective function that \emph{is} very well motivated in statistics is the \emph{likelihood function}:
\begin{equation}
\mathcal{L}(\mathrm{gate\ set}) = \mathrm{Pr}(\mathrm{data}|\mathrm{gate\ set}).
\end{equation}
Although not immediately obvious, the sum-of-squares objective function in Eq.~\ref{eq:SumSqError} can be seen as an approximation to the likelihood.  Maximizing the likelihood function is the apotheosis of linear inversion, and yields a more accurate estimate when the likelihood function is nonlinear, as it is here.  If we numerically vary over gate sets to find the one that maximizes $\mathcal{L}$, that is \emph{maximum likelihood estimation} (MLE), and it is a strictly better and more accurate way of analyzing LGST data than the linear inversion methods defined above.

In the rest of this paper, we will adopt this approach.  We will view the data as a list of circuits with observed outcome frequencies.  We will view gate sets as statistical models that predict circuit probabilities.  And we will find good \emph{estimates} of real-world gate sets by fitting that model's parameters to observed data, using MLE or approximations to it.

A natural question, then, is ``Why develop the whole linear-inversion LGST framework, if having done so we immediately move beyond it?''  There are several good reasons.  The most practical is that the linear-inversion LGST algorithm developed above is actually a critical \emph{first} step in any numerical MLE algorithm!  Because circuit probabilities are nonlinear functions of the gate set parameters, both Eq.~\ref{eq:SumSqError} and the likelihood function itself have unpleasant \emph{global} behavior, including local extrema.  Linear-inversion LGST is a very fast algorithm to obtain a \emph{pretty good} estimate, which can serve as a starting point for (and be refined by) numerical MLE.

But LGST also serves as a critical \emph{conceptual} foundation.  Regardless of what estimator we use (linear inversion, MLE, or something else), we can't estimate the gate set's parameters unless the experiments that generated the data are sensitive to them.  LGST constructs a set of experiments guaranteed to be sensitive to all the parameters in the gate set.  Deeper circuits, introduced in the next section, \emph{amplify} those parameters and allow greater sensitivity -- but the experiments that we use in long-sequence GST are clearly derivative of the ones in LGST, and rely on the same structure to guarantee sensitivity to all amplified parameters.

\newpage

\section{Long-sequence Gate Set Tomography\label{sec:LongSequenceGST}}

Traditional process tomography on a gate $G_k$ uses circuits in which $G_k$ appears just once.  Observed probabilities depend linearly on $G_k$, making it very easy to invert Born's rule to estimate $G_k$.  LGST bends this rule -- a given $G_k$ may appear both in the middle of the circuit \emph{and} in the fiducial circuits -- but the circuits are short, and each gate appears only $O(1)$ times.  This enables the relatively straightforward analysis presented in the previous section, but it also limits precision.  If $A$ and $B$ (Eqs.~\ref{eq:defineA} and \ref{eq:defineB}) are well-conditioned [condition number $O(1)$] then each matrix element of $G_k$ is very close to a linear combination of observed probabilities.

If each circuit is performed $N$ times, then finite sample fluctuations cause estimation errors in each probability,
\begin{equation}
  \hat{p} = p \pm \frac{O(1)}{\sqrt{N}},
\end{equation}
and the accuracy of $\hat{G}_k$ is also limited to $\pm O(1)/\sqrt{N}$.
Under certain circumstances, some of these probabilities (with $p\approx 0$) can be estimated to within $O(1/N)$ \cite{MahlerPRL_2013}, but the resulting improvement in overall accuracy is limited by SPAM noise and by the fact that only a few of the circuits can be chosen to have $p\approx 0$.

A completely different way to break the $1/\sqrt{N}$ boundary is to incorporate data from \emph{deep} circuits, in which a gate appears many times.  The circuits must be designed so that their outcome probabilities depend more sensitively on the elements of $G_k$, with sensitivity proportional to the number of times $G_k$ appears in the circuit.  For example, the outcome probabilities for
\begin{equation}
  \Pr = \sbraopket{E}{G_k G_k G_k G_k}{\rho},
\end{equation}
may be four times as sensitive to changes in $G_k$ as the outcome probabilities for
\begin{equation}
\Pr = \sbraopket{E}{G_k}{\rho},
\end{equation}
and therefore allow four times the precision in estimating some aspects of $G_k$.  Long-sequence GST turns this simple idea into a practical algorithm for characterizing gate sets.  

The original motivation for long-sequence GST was precision and accuracy, but another advantage emerged later. Unlike traditional tomography and LGST, long-sequence GST is easily adaptable to a wide range of noise models.  In process tomography and LGST, each noisy gate is specifically described by an arbitrary transfer matrix.  Even small extensions -- e.g., focusing on \emph{low rank} transfer matrices -- require conceptual and practical modifications of tomography \cite{Gross2010-ck, Shabani2011-jk, Flammia2012-sw, Heinosaari2013-ti}.  But long-sequence GST only requires a parameterized model for the noisy gates that can be \emph{embedded} into transfer matrices.  The gate set of Eq.~\ref{eq:GatesetDef} is such a model, with one parameter per element of $\sket{\rho^{(i)}}$, $G_i$, and $\sbra{E^{(m)}_i}$ expressed in a chosen basis.  But a gate set can, more generally, be described by any mapping between a parameter space and the aforementioned operations.  Long-sequence GST can be used to estimate a gate set that is parameterized in any reasonable way. This flexibility can be used to model various features and constraints on the gate set, including complete positivity (CP) and trace preservation (TP).  We defer a discussion of this flexibility until Section \ref{sec:Parameterization}, as these details are not essential to understanding long-sequence GST.  Presently, it is sufficient to note that a gate set is a mapping from a parameter space (e.g., the direct sum of the vector spaces for each operator) to a set of operations capable of predicting circuit outcomes.

The core long-sequence GST protocol used by our research group has remained largely unchanged since about 2017.  But between 2013 and 2017, it morphed repeatedly as we discovered that certain things didn't work, or found better ways to solve certain problems.  Its current stable form is the result of a nontrivial evolution process. Like most products of evolution, it contains both historical artifacts and necessary solutions to non-obvious problems.  Rather than present the state of the art as a \emph{fait accompli}, we introduce it by outlining the historical path that led to it (Section \ref{sec:HistoricalLSGST}), emphasizing some of the techniques that \emph{didn't} work.  Next, we explain how to construct a good long-sequence GST \emph{experiment design}, a set of circuits of depth at most $O(L)$ that enables estimating gate set parameters to precision $O(1/L)$ (Section \ref{sec:ExperimentSelection}).  We conclude this presentation of long-sequence GST by showing how to estimate those parameters by efficiently maximizing the likelihood function, and discuss a variety of technical insights that make the numerical methods reliable in practice (Section \ref{sec:ParameterEstimation}).

\subsection{Historical background\label{sec:HistoricalLSGST}}

Our first attempt to achieve higher accuracy involved performing random, unstructured circuits and fitting gate set models to the resulting count data \cite{GST2013}.  We augmented the set of circuits prescribed by LGST with a variety of random circuits (much like those used in direct randomized benchmarking \cite{Proctor2019-ma}) of various depths, and then used numerical MLE to fit a gate set to the resulting data.  This approach produces a likelihood function that is generically messy -- it is not a quasi-convex function of the gate set, usually has local maxima, and has no particular structure.  However, it is straightforward to evaluate, and its derivative can be computed efficiently from the data.  Moreover, a subset of the data enable LGST, which provides a ``pretty close'' starting point for local gradient maximization of $\cL(\mathcal{G})$. 

Investigation of this approach revealed two problems.
\begin{enumerate}
\item Random circuits provided surprisingly low precision.  Although estimation error decreased with $L$, it declined as $O(1/\sqrt{L})$, not $O(1/L)$.  
\item The lack of structure in the likelihood function made numerical MLE problematic.  Local gradient optimization worked only unreliably, and was highly dependent on starting location.  We achieved reasonable success in simulations by starting with LGST, and then refining this estimate by adding successively deeper circuits to the likelihood function.  However, this technique proved less reliable in experiments, where the underlying model was less valid. Running the optimizer repeatedly with different starting conditions, and incorporating more global-optimization techniques, often revealed better local maxima of the likelihood.  This suggested that even the best estimates we found might not be global maxima.
\end{enumerate}
We concluded that (1) as the IBM group had observed earlier, MLE on unstructured GST data was not a satisfactory solution, and (2) we needed to choose circuits more cleverly (as with LGST) to make the analysis easier and more reliable.

We developed an approach called \emph{extended LGST} (eLGST).  It relied on two critical modifications to the original ``unstructured MLE'' approach, which impose additional structure to the experiment design and data analysis (respectively).

First, instead of performing random circuits, we constructed a set of circuits corresponding to performing LGST -- not on the gates $G_k$ themselves, but on a small set of ``base'' circuits, $\{S_l\}$.  We took each $S_l$ and sandwiched it between fiducial circuits (just as we did with each $G_k$ for LGST).  Originally, the base circuits were simple repetitions of individual gates, e.g. $G_k^p$ for $p=1,2,4,8,\ldots$.  We found that this amplified some, but not all, parameters of the gate set.

This extended-LGST (eLGST), experiment design eventually evolved into one where the base circuits were chosen to be a set of short ``germ'' circuits ($g_i$) repeated $p$ times ($g_i^p$).  This spawned the ``germ selection'' procedure used today in long-sequence GST, which we discuss in detail in Section \ref{sec:ExperimentSelection}.  This structure ensures that each base circuit amplifies \emph{some} set of deviations from the target gates, so that these deviations change the observed probabilities for the circuits based on that base circuit by $O(L)$.  The germs are chosen so that they amplify different deviations, and so that collectively they amplify \emph{all} deviations.  This careful, non-random experiment design allowed eLGST to achieve consistent, reliable, and predictable accuracy that scales as $1/L$.

Second, instead of using MLE to fit a gate set directly to the data, eLGST fits the gates indirectly via a 2-step process.  In the first step, we estimated the transfer matrix for each base circuit, $\widehat{\seqaction(g_i^p)}$.  Then, those estimates were used to back out a transfer matrix for each gate.
The eLGST protocol is a precursor to the long-sequence GST we now describe.  Appendix \ref{sec:eLGST} gives a full description of eLGST.

\begin{figure}
  \begin{center}
    \includegraphics[width=3.4in]{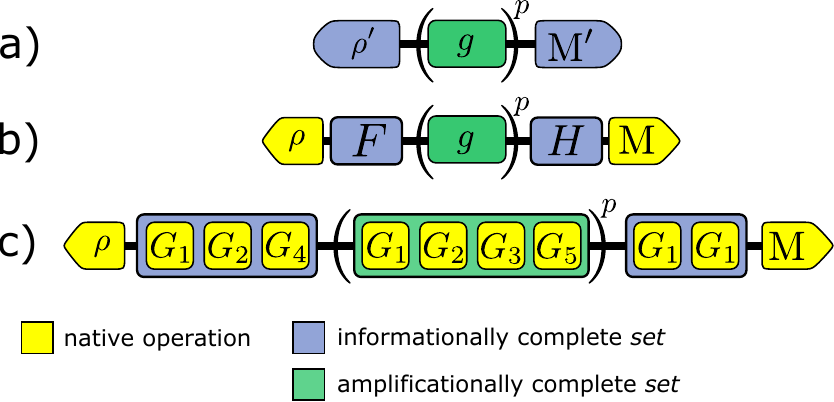}
    \caption{The structure of circuits in the standard GST experiment design, shown in increasing detail.  \textbf{(a)} Each GST circuit consists of an effective state preparation $\rho'$ (Eq.~\ref{eq:effectivePreps}), followed by a \emph{germ} circuit $g$ repeated $p$ times, followed by an effective measurement $\mathrm{M}'= \{E'_{i}\}$ (Eq.~\ref{eq:effectiveEffects}).  \textbf{(b)} Effective preparations are often implemented by a native state preparation $\rho$ followed by a \emph{preparation fiducial circuit} $F$, and similarly effective measurements are often implemented by \emph{measurement fiducial circuit} $H$ followed by a native measurement $\mathrm{M}$. \textbf{(c)} Writing the fiducials and germ in terms of native gate operations reveals how the native operations of a gate set compose to form a GST circuit.\label{fig:GSTcircuit}}
  \end{center}
\end{figure}

\subsection{Experiment design for long-sequence GST\label{sec:ExperimentSelection}}

A long-sequence GST experiment is designed to enable high accuracy with minimal experimental effort.  LGST can estimate a gate set to arbitrary accuracy, but because uncertainty in the estimated parameters scales as $O(1/\sqrt{N})$ if each circuit is repeated $N$ times, achieving precision $\epsilon$ requires $O(1/\epsilon^2)$ repetitions.  This makes precisions of $\epsilon \approx 10^{-5}$, as demonstrated in \cite{Blume-Kohout2017-kn}, practically impossible to reach.

Long-sequence GST overcomes this barrier by specifying a different \emph{experiment design} -- a set of quantum circuits to be run -- containing circuits that amplify errors in the gate set.  This experiment design retains the basic structure of LGST:  each of a list of ``operations of interest'' is probed by constructing circuits that sandwich it between pre- and post-operation fiducial circuits.  But instead of a single gate, the middle of each sandwich is a more complicated base circuit that amplifies certain errors so they can be measured more precisely by tomography.  In this section, we present the long-sequence GST experiment design.  We use the term \emph{experiment} to mean an experiment design along with the data obtained by repeating each of its circuits many times.  

Each long-sequence GST circuit constitutes three consecutive parts (see Figure \ref{fig:GSTcircuit}):
\begin{enumerate}
\item Prepare a particular state, $\sket{\rho'_k}$ by performing a native preparation followed by a fiducial circuit.
\item Perform $p$ repetitions of a short circuit $g$ that we call a \emph{germ}.
\item Perform a particular measurement $\{\sbra{E'^{(m)}_{i}}\}$ (Eq.~\ref{eq:effectiveEffects}) by performing a fiducial circuit and then a native POVM measurement.
\end{enumerate}
The outcome statistics from repeating such a circuit many times allow estimating probabilities like
\begin{equation}
p = \sbraopket{E'^{(m)}_{i}}{\seqaction(g_j^p)}{\rho'_{k}}.
\end{equation}
If the states and measurements are each informationally complete, these probabilities are sufficient for tomography on $\seqaction(g_j^p)$.  All GST experiments are derived from this basic structure, and every circuit performed for GST is specified by an $(j,k,p,m)$ tuple.

We call the short circuit $g$ a \emph{germ} because, like a germ of wheat or the germ of an idea, it serves as the template for something larger -- here, an arbitrarily deep circuit built by repetition.  We refer to $g^p$, the object of tomography, as a \emph{base circuit} and $p$ as a \emph{germ power}.  Repeating $g$ amplifies errors in $g$.  For example, if $g$ is a single unitary gate $G$, and $G$ over-rotates by angle $\theta$, then $\seqaction(g^p)$ will over-rotate by $p\theta$.  By varying the input state and final measurement ($k$ and $m$) over informationally complete sets, we probe the operation between them, just as in process tomography or LGST.  If this tomography on $g^p$ measures $p\theta$ to precision $\pm\epsilon$, then we've measured $\theta$ itself to precision $\pm\epsilon/p$.  

\begin{figure}[H]
  \begin{center}
    \includegraphics[width=3.4in]{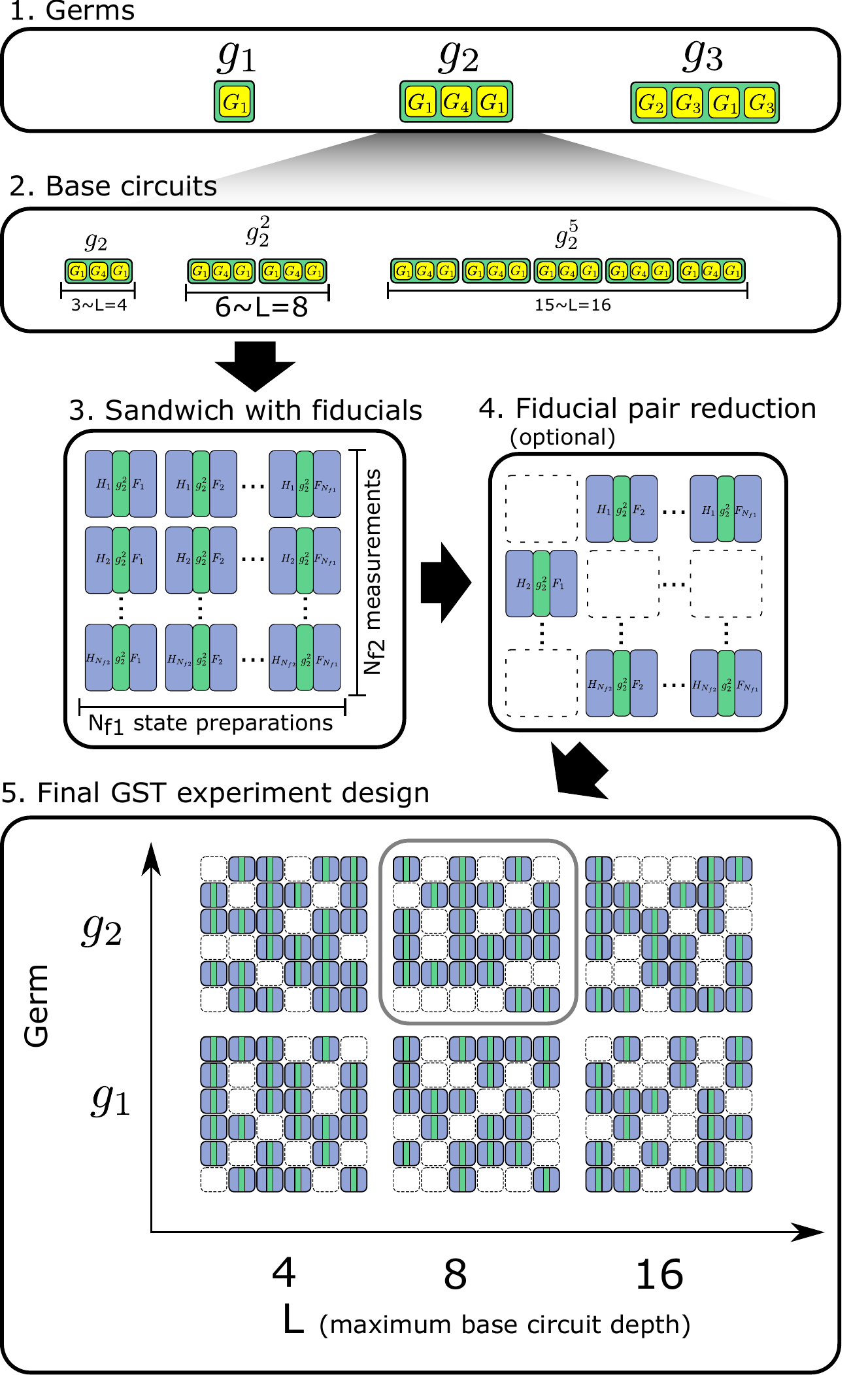}
    \caption{Overview of how to design a GST experiment. \textbf{Step 1.} Choose \emph{germ} circuits that amplify \emph{all} gauge-invariant parameters in the processor's gate set. \textbf{Step 2.} Define \emph{base circuits} by raising each germ to powers chosen so that the base circuits' depths are as close as possible to being logarithmically spaced. \textbf{Step 3.}  Sandwich each base circuit between each of $N_{f1} \times N_{f2}$ \emph{fiducial pairs} defined by $N_{f1}$ preparation and $N_{f2}$ measurement fiducial circuits. \textbf{Step 4.} Optionally apply \emph{fiducial pair reduction} (FPR) to eliminate redundant circuits from the experiment design, leaving just enough to ensure sensitivity to each base circuit's amplified parameters. \textbf{Final result 5.} A GST experiment design can be visualized as a grid of plaquettes.  The plaquettes correspond to base circuits, are arranged by germ and $L$, the maximum depth used to construct the base circuit from the germ.  Within a plaquette, squares indicate different fiducial pairs, and the plaquette will be ``sparse'' when FPR has been applied.\label{fig:GSTExperimentDesign}}
  \end{center}
\end{figure}

Simple germs comprising a single gate $G$ are not sufficient to amplify every error.  Some errors can only be amplified by first constructing a multi-gate circuit, e.g., $g = G_1G_2$, and then repeating \emph{it}.  Repeating $g$ many times amplifies errors that \emph{commute} with $\seqaction(g)$.  In the example above, an over-rotation error in $G$ is an error that commutes with $G$, so it is amplified.  But suppose that $G$ rotates by the correct angle, but around the wrong axis, e.g., instead of a rotation around $\hat{z}$, $G$ performs a rotation around $\hat{z}\cos\phi + \hat{x}\sin\phi$.  This \emph{tilt} error is not amplified by repeating $G$, but it can be amplified by a more complex germ that includes $G$ together with other gates.


To achieve high precision estimates efficiently, we need to amplify every parameter in the gate set that \emph{can} be amplified.  Two kinds of parameters cannot be amplified.  Gauge parameters cannot be measured at all, and properties of the SPAM operations cannot be amplified because SPAM operations only appear once in each circuit.  So germs amplify errors in gates.  Therefore, as we discuss selection of germs, we will focus exclusively on the gates and ignore SPAM operations.

In the following sections we describe how to select a set of base circuits -- germs and powers to raise them to that amplify all of the amplifiable parameters of a given ``target'' gate set.  This target gate set is typically taken to be the ideal gates a device is designed to implement.  More details may be found in Appendix \ref{sec:PyGSTiExperimentSelection}.

\subsubsection{Selecting germs ($g$)\label{sec:SelectingGerms}}  

To estimate a gate set efficiently and accurately, every variation in the gate set, except those corresponding to SPAM and gauge directions, must be amplified by some germ.  We call a set of germs that achieves this goal \emph{amplificationally complete}, in direct analogy to ``informationally complete'' sets of states or measurements.

To evaluate (and ensure) amplificational completeness, we model errors in the gate set as small perturbations to the target gate set.  Each germ $g$, when repeated, will amplify certain errors.  Specifically, any small perturbation to a gate set's parameters that causes a change in $\seqaction(g)$ that \emph{commutes} with $\seqaction(g)$ will be amplified by $g$.  The set of all such perturbations is easily shown to be closed under linear combination, so each germ $g$ amplifies error vectors that lie in a \emph{subspace} of the gate set's parameter space.  This subspace can be easily constructed from the target gate set and the germ.

It is also straightforward to construct the tangent subspace of \emph{gauge} variations.  It contains all small perturbations around the target gate set that do not change any observable probability at all.  Its complement is the subspace of physical errors that need to be amplified.  Now we can define amplificational completeness precisely:  \emph{a set of germs $\{g_j\}$ is amplificationally complete for a target gate set $\mathcal{G}$ if and only if the union of the error subspaces amplified by each $g_j$ span the complement of the subspace of gauge variations.}

Amplificational completeness defines a concrete condition that germs must satisfy for GST.  However, it depends on the target gate set, so each new target gate set requires finding a new set of germs.  Furthermore, different desiderata may motivate different amplificationally complete sets of germs (e.g., it is reasonable to prioritize the \emph{smallest} set of germs, or the \emph{shortest} possible germs, or the set that achieves the highest precision for a given maximum depth).  In the \texttt{pyGSTi} implementation of GST \cite{pygsti,Nielsen2020-lu}, we use a particular search algorithm to evaluate sets of short germs and find the smallest amplificationally complete set.  Algorithmic and mathematical details can be found in Appendix \ref{sec:PyGSTiExperimentSelection}.  We require that a chosen germ set always includes each gate $G_i$ in the gate set as an independent germ.

\subsubsection{Defining base circuits\label{sec:SelectingGermPowers}}

Once a set of germs is selected, a set of \emph{base circuits} is constructed by raising each germ to several powers $p$.  We begin by selecting $p=1$.  Since each gate $G_i$ is also germ, this ensures that each individual gate appears as a base circuit.  What remains is to choose all the \emph{other} values of $p$ that will appear in the experiment.

Large values of $p$ amplify errors more.  But including \emph{only} large powers $p$ would create an aliasing problem.  If $\seqaction(g)$ over-rotates by $\theta = \pi/16$, then repeating it $p=32$ times creates an over-rotation by $2\pi$, which appears to be no error at all!  So, exactly as in phase estimation, smaller powers $p$ are needed too.  Since this is essentially a binary search, logarithmically spaced values of $p$ are optimal.

More generally, a base circuit's $p$ is less relevant than its overall \emph{depth}.  If germ $g_1$ has depth 1, while $g_2$ has depth 5, then $g_1^{10}$ and $g_2^2$ both have depth 10.  So in describing GST experiments, we typically organize base circuits by their approximate depth $l$ rather than $p$.  If we denote the depth of germ $g$ by $|g|$, then the depth-$l$ base circuit based on $g$ is $g^{\lfloor l/|g| \rfloor}$.  Our intuition to use logarithmically spaced $p$ above leads us to choose logarithmically spaced $l$ -- i.e., $l = 1,m,m^2,m^3,\ldots$ -- to choose the base circuits.

Making $m$ larger reduces the total number of circuits to be run, but requires higher precision (more repetitions) at each value of $l$.  Empirically, we have found $m=2$ to work reliably -- i.e., $l = 1,2,4,8,16\ldots$ -- but other choices are possible.  Germs of depth $d>1$ do not have realizations for depths $l<d$, so a germ of depth 5 would appear first at $l=8$ (with 1 repetition), then $l=16$ (with 3 repetitions), etc.  Thus, the depth of a ``depth $l$'' base circuit is not necessarily equal to $l$, but is always in the interval $(l/m,l]$.

Any given experiment can only include finitely many circuits, so for each germ there must be a maximum depth $L$ at which it appears.  This is configurable.  Selecting $L$ should be guided by three facts.  First, increasing $L$ amplifies errors more and yields more precision (all else being equal).  Second, increasing $L$ makes the experiment larger (more circuits), which requires more time to obtain and analyze.  Third, there is essentially no benefit to increasing $L$ beyond the point where decoherence and stochastic errors dominate.  If the rate of decoherence in each gate is $\eta$, then circuits of depth $\gg1/\eta$ generally all produce the same equilibrium state, and little or nothing will be learned from circuits of depth $L > O(1)/\eta$.  If different gates have very different rates of stochastic errors, then it is useful to let $L$ vary from germ to germ.

\subsubsection{Putting it all together}

The circuits for a full GST experiment are constructed as follows.  First a set of amplificationally complete germs is selected (see \ref{sec:SelectingGerms}, Fig.~\ref{fig:GSTExperimentDesign} step 1).  Next a set of base circuits is chosen (see \ref{sec:SelectingGermPowers}, Fig.~\ref{fig:GSTExperimentDesign} step 2).  Let these be given by $\mathcal{O} = \left\{g_i^{p_{i,j}}\right\}_{i,j}$, where $i$ indexes a germ and $p_{i,j}$ is the $j$-th power that we apply to the $i$-th germ.  As explained above, to amplify the desired errors it is sufficient that the GST circuits estimate the probabilities
\begin{equation}
  p_{abij} = \sbraopket{E'_a}{\seqaction(g_i)^{p_{i,j}}}{\rho'_b},\label{eq:probabilities4GSTcondensed}
\end{equation}
where $a=1\ldots N_{f1}$ and $b=1\ldots N_{f2}$ range over the indices of the effective preparations and effects (Fig.~\ref{fig:GSTExperimentDesign} step 3 and 5).  Writing the effective state preparations and POVM effects in terms of their constituent fiducial circuits (Eqs.~\ref{eq:effectiveEffects} and \ref{eq:effectivePreps}) gives the probabilities entirely in terms of native operations, i.e.~elements of the gate set:
\begin{equation}
  p_{abij} = \sbra{E^{m(a)}_{t(a)}}\seqaction\left(H_{h(a)}\right)  \seqaction\left(g_i\right)^{p_{i,j}} \seqaction\left(F_{f(b)}\right)\sket{\rho^{r(b)}}.\label{eq:probabilities4GST}
\end{equation}
The corresponding circuit (which can be read off directly, from right to left because the matrix-ordering of Eq.~\ref{eq:probabilities4GST} is reversed from time-ordering) is repeated $N$ times to approximate $p_{abij}$.  The steps of this circuit are:
\begin{enumerate}
\item prepare the $r(b)$-th state
\item perform the (QI/QO) circuit $F_{f(b)} g_i^{p_{i,j}} H_{h(a)}$
\item measure using the $m(a)$-th type of measurement
\end{enumerate}
The frequency $f_{t(a)} = n_{t(a)}/N$, where $n_{t(a)}$ is the number of times the $t(a)$th outcome was observed, estimates $p_{abij}$. In this way, a circuit's outcome data can estimate all of the $p_{abij}$ that differ only in their $t(a)$ (POVM effect) index.

Letting $a$, $b$, $i$ and $j$ range over their allowed values in the steps above defines all the circuits needed to run long-sequence GST.  These circuits constitute a GST experiment design. Their structure is shown schematically in Figure \ref{fig:GSTcircuit}, where again we omit the many indices (see Eq.~\ref{eq:probabilities4GST}) for clarity.  For later analysis of the circuits' outcome data, it is helpful to separately keep track of the fiducial and germ circuits, and the germ powers.  In Appendix \ref{sec:NumericalVerification} we show that the circuits of a GST experiment design (using the particular \texttt{pyGSTi} algorithms discussed in Appendix \ref{sec:PyGSTiExperimentSelection}) result in an accuracy that scales as the inverse of the circuit depth and total number of circuits (Heisenberg-like scaling).  

Finally, we note that arriving at a definite list of circuits (an experiment design) required information based on the experimental circumstances.  The sets of fiducials and germs were based on the available native gates (specified by a target gate set), and the maximum depth of the circuits was based on a tradeoff between resources and accuracy (longer experiments are more resource-intensive, because they require more and deeper circuits, but they give more accuracy -- although only up to a maximum useful depth that is roughly the inverse of the decoherence rate).  GST experiments are tailored to a hardware's capabilities.

\subsection{Estimating gate set parameters in long-sequence GST\label{sec:ParameterEstimation}}

A gate set is a parameterized statistical model.  The parameterization may be simple (e.g., every gate element is a parameter) or nontrivial (see Section \ref{sec:Parameterization}).  Fitting a gate set model to data from the experiments described in Section \ref{sec:ExperimentSelection} is an example of \emph{parameter estimation}.  There are many ways to do this (see Section \ref{sec:LGSTMLE}).  In this paper, we focus on maximum likelihood estimation (MLE), which requires varying the gate set's parameters to maximize the probability of the data (or, for practical simplicity, its logarithm).  

The loglikelihood function can be constructed as follows.  Let $\mathcal{G}$ denote the model, $s$ index a circuit of the experiment design, and $N_s$ be the total number of times circuit $s$ was repeated in the GST experiment.  Furthermore, let $m_s$ be the number of outcomes of $s$, and $N_{s,\beta_s}$ denote the number of times outcome $\beta_s$ was observed.  The contribution to the total loglikelihood from $s$ is simply the multinomial likelihood function for an $m_s$-outcome Bernoulli scheme,
\begin{equation}
\logL_s = N_s\sum_{\beta_s}{f_{s,\beta_s} \log(p_{s,\beta_s})},\label{eq:perExperimentLogL}
\end{equation}
where $p_{s,\beta_s}$ is the \emph{probability} predicted by $\mathcal{G}$ of getting outcome $\beta_s$ from circuit $s$, and $f_{s,\beta_s} = N_{s,\beta_s}/N_s$ is the corresponding observed \emph{frequency}.  The total loglikelihood for the entire GST experiment is just the sum
\begin{equation}
\logL = \sum_{s}{ \logL_s} = \sum_{s,\beta_s} N_s f_{s,\beta_s} \log(p_{s,\beta_s}),\label{eq:DatasetLogLTP}
\end{equation}
where $s$ ranges over all of the circuits in the experiment design.  This derivation assumes that $\mathcal{G}$ is trace-preserving (TP), so that $\forall s \, \sum_{\beta_s} p_{s,\beta_s} = 1$.  An extension to non-TP gate sets requires some technical tricks and is explained in Appendix \ref{sec:NonTPLogLEstimation}.


Maximizing $\logL$ is made nontrivial by the structure of the problem.  This contrasts with state and process tomography, where the parameterized model is a density matrix $\rho$ or a superoperator $G$, each observable probability is a \emph{linear} function of the parameter, and the loglikelihood is a sum of logarithms of linear functions.  This means MLE state/process tomography is a straightforward convex optimization problem.

In contrast, the GST likelihood function (Eq.~\ref{eq:DatasetLogLTP}) is extremely non-convex.  Because gates appear repeatedly in circuits, the probabilities $p_{s,\beta_s}$ are nonlinear functions of gate elements (and, more generally, the parameters of $\mathcal{G}$).  The construction outlined previously causes the $p_{s,\beta_s}$ to \emph{oscillate}, creating a loglikelihood function that looks like an egg crate or optical lattice, with many local maxima.  Finding global maxima of such functions is generally hard.  

The gauge freedom creates more problems, by turning unique maxima into ``ridges'' that trace out gauge orbits.  Constraining the optimization to CP gate sets creates complexity, because the CP condition does not respect gauge symmetry and can create local maxima.  Ignoring the CP constraint makes optimization easier, but allows unphysical gate sets that can have zero or negative likelihood.

We have developed a particular pipeline of MLE and related estimators that reliably maximize Eq.~\ref{eq:DatasetLogLTP} when $\mathcal{G}$ is parameterized in one of several common ways (including with TP and CPTP constraints).  We present this method, not as the \emph{only} way to perform the parameter-estimation step of long-sequence GST, but as \emph{a} way of implementing this crucial step of the protocol.

Our approach is outlined in Algorithm \ref{alg:LSGST}. Here $\vec{\theta}$ is the vector of $\mathcal{G}$'s parameters being optimized, $\mathcal{D}$ is a data set, $\mbox{Truncate}(\mathcal{D}, L)$ is the subset of $\mathcal{D}$'s data corresponding to circuits whose germ-power has depth $\le L$, and $\mbox{Argmin}(S, \mathcal{G}, \mathcal{D}, \vec{\theta}_1)$ yields the $\vec{\theta}$ at which statistic $S(\mathcal{G}(\vec{\theta}), \mathcal{D})$ is minimal using a local optimizer seeded at $\vec{\theta}_1$.  $\mathcal{D}_0$ is the full GST data set, and $\vec{\theta}_0$ is the initial vector of model parameters, often provided by LGST (Section \ref{sec:LGST}).  We have found $m=2$ to work well in practice.

\begin{algorithm}[H]
\caption{Long-sequence GST\label{alg:LSGST}}
\begin{algorithmic}
\STATE $\vec{\theta} \leftarrow \vec{\theta}_0$
\FOR{$L \in 1,m,m^2,m^3,\ldots$}
\bindent
\STATE $\mathcal{D} \leftarrow \mbox{Truncate}(\mathcal{D}_0, L)$
\STATE $\vec{\theta} \leftarrow \mbox{Argmin}(\chi^2, \mathcal{G}, \mathcal{D}, \vec{\theta})$
\eindent
\ENDFOR
\STATE $\vec{\theta} \leftarrow \mbox{Argmin}(-\logL, \mathcal{G}, \mathcal{D}_0, \vec{\theta})$ 
\end{algorithmic}
\end{algorithm}

We highlight three important elements to this approach that we believe are particularly important to its success:

\vspace{1em}
\noindent \textbf{Optimization Stages.}  Our approach proceeds in multiple ``stages''.  At each stage, we run a traditional local optimization method (we find the Levenberg-Marquardt technique to give good results) on a \emph{subset} of all the data.  Each stage sets a maximum base circuit depth $L$, and at that stage only data from base sequences with depth less than or equal to $L$ are incorporated into the likelihood function (Eq.~\ref{eq:DatasetLogLTP}).  We choose $L=1,m,m^2,m^3,\ldots$, so that the stage corresponding to $L$ contains the circuits whose base circuits are $g^{\lfloor l/|g| \rfloor}$ for $l = 1,m,m^2\ldots,L$.  Successive stages add deeper circuits, while keeping all the shorter circuits.  By using the best-fit output of a stage as the starting point of the next we create a daisy chain of estimations that avoids local minima.  As long as finite sample fluctuations are kept small enough (by repeating each circuit enough times) the previous stage's best-fit estimate lies with high probability in the correct basin of the next stage's objective function, rendering its oscillatory nature benign.

\vspace{1em}
\noindent \textbf{Use of $\chi^2$ as a $\logL$ proxy.}  At every stage except the last one, the $\chi^2$ statistic is optimized instead of loglikelihood.  The $\chi^2$ is a weighted-sum-of-squares function that (like likelihood) quantifies goodness-of-fit.  Using our definitions above, it can be written as
\begin{equation}
\chi^2 = \sum_{s,\beta_s}{ N_s \frac{(p_{s,\beta_s}-f_{s,\beta_s})^2}{p_{s,\beta_s}}} = \sum_{s}{ \chi^2_s},\label{eq:chi2}
\end{equation}
where
\begin{equation}
  \chi^2_s = \sum_{\beta_s}{ N_s \frac{(p_{s,\beta_s}-f_{s,\beta_s})^2}{p_{s,\beta_s}} }\label{eq:chi2term}
\end{equation}
is the contribution from a single circuit $s$.
The weights ($1/p_{s,\beta_s}$) ensure that $\chi^2$ is a local quadratic approximation to the negative loglikelihood.  Compared with the loglikelihood, $\chi^2$ can be computed faster, and is more well-behaved as an objective function.  It has one significant disadvantage:  its minimum is a slightly biased estimator (see Appendix \ref{sec:chi2bias}).  Optimizing the loglikelihood in the final stage renders this a non-issue, and we find minimizing $\chi^2$ to be more robust at seeding the final $\logL$ maximization than performing $\logL$ maximizations all the time.  One notable exception to this occurs when the number of circuit repetitions is low and $\chi^2$ becomes a poor proxy for $\logL$.

\vspace{1em}
\noindent \textbf{Regularization.} We increase the reliability of optimization by regularizing both $\logL$ and $\chi^2$.  Both functions have poles when probabilities are zero, which can lead to numerical instabilities when probabilities are near zero.  By slightly altering each function, we make them more amenable to optimization.  A simple and effective way to regularize Eq.~\ref{eq:chi2} is to limit the least-squares weights to a maximum cutoff, e.g. $1/p_{min}$.  Since $\chi^2$ is already just a proxy for the negative loglikelihood, this has no effect on the final fit.  The $\logL$ function needs to be tweaked more carefully, by replacing Eq.~\ref{eq:perExperimentLogL} with its second-order Taylor series when $p_{s,\beta_s} < p_{min}$, where $p_{min}$ is chosen to be much less than the smallest possible non-zero frequency (e.g., $10^{-4}$ if each circuit is repeated 1000 times).  Because this alters the $\logL$ function only when a probability is much different than its observed frequency, it distorts the objective's value (and thereby the rigor of its interpretation) only for particularly bad fits.  We have evidence that these small adjustments to the standard $\logL$ and $\chi^2$ cause local optimization methods to be more robust, presumably because they avoid regions with very large gradient and widen basins of convergence.

\vspace{1em}
The method outlined above is not guaranteed to find a global maximum of $\logL$, but does so impressively often in practice.  Improvements to the algorithm are a deserving area for future work, but the algorithm described above demonstrates that the parameter estimation problem lying at the heart of long-sequence GST is tractable, enabling its practical use.

The long-sequence GST protocol described above -- designing an experiment, running that experiment to produce data, and finally analyzing the data via multi-stage MLE -- produces a best-fit gate set in an unknown gauge.  The gauge is irrelevant when predicting circuit outcomes, so the estimated gate set can be used immediately for this purpose.  (For example, it is possible to predict the success probabilities of RB circuits and extract the RB number.)   However, computing standard metrics such as fidelity and diamond-distance requires an additional step called \emph{gauge optimization}.  We discuss this, and other common post-processing, in Section \ref{sec:Interpretation}. Readers should feel free to jump to that section, if desired.  Section \ref{sec:AdvancedLongSequenceGST} covers several advanced topics that were omitted from our presentation thus far of long-sequence GST.


\section{Advanced long-sequence GST\label{sec:AdvancedLongSequenceGST}}
This section discusses two additional topics related to long-sequence GST.  Although they are not essential to the protocol as a whole, they infuse it with significant additional capability.  First, we formalize the notion of a gate set model, and suggest a natural path to extending long-sequence GST by creating additional models.  Second, we describe how the number of circuits in the GST experiment design (Section \ref{sec:ExperimentSelection}) can be sizeably reduced by taking advantage of overcompleteness present in the standard design.  The material here is not required by any of the remaining main text.


\subsection{Gate set models\label{sec:Parameterization}}
Gate sets, as we have defined them, contain general state, measurements, and superoperators.  A gate set $\mathcal{G}$ (Eq.~\ref{eq:GatesetDef}) represents each gate as a $d^2 \times d^2$ real-valued matrix, and each state preparation or measurement effect as a $d^2$-dimensional real-valued vector or dual vector, respectively.  Let us define \emph{matrix space} as $\mathcal{M}\ \sim \reals^{N_{\mathrm{e}}}$, where 
\begin{equation}
  N_{\mathrm{e}} = d^4 N_{\mathrm{G}} + d^2 \left(N_\rho + \sum_{m=1}^{N_{\mathrm{M}}} N^{(m)}_{\mathrm{E}}\right)\label{eq:numGatesetElements}
\end{equation}
is the total number of (real) elements in a gate set.  $\mathcal{M}$ is thus isomorphic to the direct sum of the vector spaces containing each gate set operation's matrix, vector, or dual vector.  Any gate set (as defined by Eq.~\ref{eq:GatesetDef}) is a point in matrix space, $\mathcal{G}\in\mathcal{M}$, with coordinates given by the elements of each gate set operation.


A gate set is a statistical model that assigns probabilities to the outcomes of quantum circuits.\footnote{Technically, a gate set is a \emph{factory}, which can produce a statistical model for any set of circuits. Since such a set of circuits is always implied by context in which gate sets are used, we allow ourselves to abuse terminology slightly.}

We use the term \emph{gate set model} to mean a mapping between a parameter space $\mathcal{P} \sim \reals^{N_{\mathrm{p}}}$ and $\mathcal{M}$.  The dimension of parameter space, $N_{\mathrm{p}}$, is typically less than $N_{\mathrm{e}}$.  Formally, a gate set model corresponds to a choice of $\mathcal{P}$ and a map 
\begin{equation}
W: \mathcal{P} \to \mathcal{M}.
\end{equation}
A gate set model is a parameterized statistical model that associates with every point in parameter space a gate set.  Long-sequence GST finds an optimal gate set by searching over this parameter space, optimizing over $\vec{\theta} \in \mathcal{P}$.  So by using different gate set models, we can constrain this optimization to subsets of the entire matrix space.  Informally, a basis for $\mathcal{P}$ defines ``knobs'' that can be adjusted -- e.g., by the MLE optimization, but also by hand if so desired -- to vary a gate set and make it more consistent with the observed data.

If we set $\mathcal{P} = \mathcal{M}$ and $W(x)=x$ then each element of every operation in the gate set is an independent parameter.  We call this the \emph{fully parameterized gate set model}, and we can use it for GST (as in Section \ref{sec:LongSequenceGST}).  However, it includes \emph{all} gate sets, even wildly nonphysical ones that violate complete positivity and/or trace preservation.  So it is useful to define \emph{smaller} gate set models that parameterize strict subsets of $\mathcal{M}$ (e.g., only CPTP gate sets).

\subsubsection{Gauge freedom in constrained gate sets}  

Gate set models may have gauge freedoms.  A gauge freedom exists if there is a transformation that can be applied to $\vec\theta\in\mathcal{P}$ that changes $\vec\theta$ but does not change any observable probability that can be computed from the corresponding gate set $W(\vec\theta)$.  We have already discussed the gauge freedoms of the fully parameterized gate set model; they correspond to similarity transformations by invertible matrices $M$ that act on each gate matrix as $G_k \to M G_k M^{-1}$ (see Eq.~\ref{eq:GaugeTransform}).  For any given $\mathcal{G}$, the action of \emph{all} possible $M$ on $\mathcal{G}$ traces out a \emph{gauge orbit} containing all the gate sets $\mathcal{G'} \in \mathcal{M}$ that are equivalent to $\mathcal{G}$.  For general gate set models, one must consider the pullback of Eq.~\ref{eq:GaugeTransform} to $\mathcal{P}$.  As we discuss below, this can make the analysis more complicated.  In $\mathcal{M}$, gauge freedoms correspond to gauge orbits -- sets of equivalent gate sets.  But in $\mathcal{P}$, these freedoms can map onto a different construct.  Still, it is helpful to picture the parameter space as being foliated into gauge orbits, like an onion or layered pastry.  Appendix \ref{sec:GaugeDOF} describes gauge transformations in more detail.

It can be useful to count the gauge degrees of freedom for a given gate set model.  All the gauge orbits, except for a set of measure zero, are manifolds of this fixed dimension.  (If one or more gates in a gate set have degenerate spectra, then they remain invariant under certain gauge transformations; the corresponding orbits have lower dimension, like the center of an onion).  Since $\mathcal{P}$ has dimension $N_{\mathrm{p}}$, then any point $\vec\theta$ can be varied in $N_{\mathrm{p}}$ linearly independent directions.  These define a local tangent space at $\theta$, which we can partition into orthogonal subspaces of $N_{\mathrm{p}}^{\mathrm{gauge}}$ \emph{gauge directions} that are tangent to the gauge orbit on which $\vec\theta$ lies, and $N_{\mathrm{p}}^{\mathrm{nongauge}}$ \emph{non-gauge directions} that are normal to the gauge orbit (see Appendix \ref{sec:GaugeDOF}).  It follows trivially that $N_{\mathrm{p}}^{\mathrm{gauge}} + N_{\mathrm{p}}^{\mathrm{nongauge}} = N_{\mathrm{p}}$.  This partition of $N_{\mathrm{p}}$ is same at almost all points $\vec\theta$ (the exceptions are the singular points corresponding to gate sets with degenerate gates), allowing us to view $\mathcal{P}$ mathematically as a \emph{fiber bundle}.

The fully parameterized gate set model has a parameter space with dimension
\begin{equation}
  N_{\mathrm{p}}^{\mathrm{full}} = N_{\mathrm{G}} d^4 + \left(N_\rho+\sum_{m=1}^{N_{\mathrm{M}}} N^{(m)}_{\mathrm{E}}\right) d^2,
\end{equation}
and it generally has $d^2$ gauge degrees of freedom exceptions occur when one or more operators commute with every operation in the gate set.

\subsubsection{TP and CPTP constrained models}

Quantum theory requires density matrices to have unit trace, and the operations acting on them to be trace-preserving (TP).  These are constraints on $\mathcal{G}$ which can be used to define a smaller gate set model.  The TP constraint corresponds to locking the first row of every superoperator to be $[1, 0, \ldots 0]$ and the first element of every state preparation vector to equal $1/\sqrt{d}$ (since $\Tr(B_0) = d/\sqrt{d} = \sqrt{d}$, see Section \ref{sec:HSSpace}).  Enforcing these constraints defines a \emph{TP parameterized gate set model} with $N_{\mathrm{p}} = N_{\mathrm{p}}^{\mathrm{TP}}$ parameters, where
\begin{equation}
  N_{\mathrm{p}}^{\mathrm{TP}} = N_{\mathrm{G}} d^2(d^2-1) + N_\rho (d^2-1) + \left( \sum_{m=1}^{N_{\mathrm{M}}} N^{(m)}_{\mathrm{E}} \right) d^2.
\end{equation}
The mapping $W$ is trivial to construct in this case: it leaves some of the elements of $\mathcal{G}$ fixed.

When running long-sequence GST, the TP parameterized gate set model is generally superior to the fully parameterized model, and presents no extra complications.  The gauge freedoms for the TP parameterized model are easy to derive; gauge transformations correspond exactly to matrices $M$ that are also TP, so instead of $d^2$ gauge parameters, the TP parameterized model has $d^2-d$ gauge parameters.

Quantum theory also requires operations to be completely positive (CP).  Imposing this constraint defines a \emph{CP parameterized gate set model}.  However, the CP constraint is harder to define (and impose) than the TP constraint.  Whereas the TP constraint is a constant equality and defines a subspace of the full parameter space, the CP constraint is an nonlinear inequality.

A mapping function $W$ whose range is constrained to CP gate sets can be constructed in several ways.  One is to represent each gate by the Cholesky factorization $T$ of its Choi matrix.  In this way, each gate is represented by a lower-triangular matrix $T_k$ with real diagonals, and the gate (transfer) matrix $G_k$ is obtained by applying the Choi-Jamiolkowski isomorphism to the Choi matrix $\chi_k = T_k^\dag T_k$.  It is usually desirable to apply the TP constraint as well, to get a \emph{CPTP parameterized gate set model}; this is done by additional constraints on each $T_k$ (unfortunately, the TP constraint is not as simple in the Choi representation).

An alternative way of constraining a gate to be CP is to write it in terms of an \emph{error generator}, which we denote $\xi$.  Let $G_k$ be the transfer matrix of a gate, and the corresponding ideal (error-free) CPTP operation be $G_k^0$. We define $G_k$'s error generator, $\xi$, to be the logarithm of the quotient $G_k(G_k^0)^{-1}$, so that
\begin{equation}
  G_k = e^\xi G_k^0.\label{eq:gate_errorgen_expression}
\end{equation}
The error generator $\xi$ is itself a superoperator, and when $\xi=0$, $G_k=G_k^0$ and the gate has no error.  By restricting $\xi$ to be a Lindbladian \cite{Lindblad_1976}, we can guarantee that $e^\xi$ is CPTP. Because CPTP maps form a semigroup, $G_k$ is CPTP as well.  The Lindblad form that ensures $e^\xi$ is CPTP is
\begin{equation}
\xi = \sum_{i=1}^{d^2} \alpha_i H_i + \sum_{j,k=2}^{d^2} \beta_{jk} S_{jk},\label{eq:errorgen}
\end{equation}
where operators $H_i$ and $S_{jk}$ act as
\begin{eqnarray}
  H_i & : & \rho \rightarrow i[\rho,B_i] \, \mbox{and}\\
  S_{jk} & : & \rho \rightarrow B_j \rho B_k - \frac{1}{2}\left( B_k B_j \rho + \rho B_k B_j \right)
\end{eqnarray}
on density matrices $\rho$, $\alpha_i$ is real, $\beta$ is a positive semidefinite (Hermitian) matrix. Here, $\{B_i\}$ is a basis for $\mathcal{B}(\mathcal{H})$ with the properties given in Section \ref{sec:HSSpace}.  Indices $j$ and $k$ sum only over the \emph{non-identity} elements, so they begin at 2.  The condition on $\beta$ can be implemented by parameterizing $\beta$'s Cholesky factorization as described above.  We have found that this parameterization facilitates optimization better than the Choi-matrix parameterization.  (This practical advantage makes this the CPTP parameterization of choice in \texttt{pyGSTi}.)  A downside to the error generator approach is that not all CPTP maps can be put into the form given by Eq.~\ref{eq:gate_errorgen_expression} - only those that can be infinitesimally generated.  We have not observed this restriction to have significant consequences in any experiment to date.

Using either parameterization, the number of parameters $N_{\mathrm{p}}$ for the CP and CPTP gate set models are the same as those for the full and TP models, respectively.

In the literature on state and process tomography, estimators that respect CP are common, and generally regarded as superior to unconstrained estimators.  The argument is simple:  nature and quantum theory only permit CP processes, so we should not consider non-CP estimates.  Furthermore, the CP constraint constitutes universally valid prior information, and must therefore improve estimation accuracy (at the cost of some bias).

The value of CP is more ambiguous in the context of GST.  The CP constraint \emph{can} be imposed, as described above, but doing so has some practical consequences.  It makes MLE estimation more complex \cite{ScholtenMLE2018}, and complicates the construction of error bars (see Section \ref{sec:ErrorBars}).  It also removes a valuable diagnostic:  the data \emph{should} be consistent with a CP gate set whether or not the constraint is imposed, so when GST returns a non-CP gate set (possible when the constraint \emph{isn't} imposed), this is a sign that something else is wrong -- usually some form of non-Markovian dynamics in the experiment.

Most importantly, the CP constraint interacts badly with the gauge freedom.  Every gate set on a gauge orbit is equivalent -- gauge transformations do not change any physically observable property.  But non-unitary gauge transformations \emph{do} change complete positivity!  This may seem paradoxical, since non-CP gates can produce negative circuit outcome probabilities.  But this operational interpretation \emph{requires} the ability to create arbitrary input states, including states entangled with a reference frame, and apply the gate to them.  Such states constitute an absolute reference frame, and the gauge freedom arises precisely because no such reference frame is available.  A gate set may be CP in one gauge, and non-CP in another gauge, because those two gauges imply different sets of allowable (non-negative) input states to the gates.  We explore these issues further in Appendix \ref{sec:GaugeDOF}.

In practice, we have found it useful to perform long-sequence GST using \emph{both} the CPTP and TP gate set models, and compare the results.  If a significantly better non-CP fit is found, then the two estimates should be examined carefully: either the CPTP fit got trapped in a local maximum, or significant non-Markovian dynamics occurred during the experiment.  If difference between the two fits is not statistically significant, then the CPTP fit is often preferable for later analysis because it is better behaved within post-processing (e.g., each gate's eigenvalues \emph{cannot} be greater than 1.0).

We conclude this discussion of gate set models by noting that we have only scratched the surface of possible models here.  The long-sequence GST framework (Section \ref{sec:LongSequenceGST}) is agnostic to which gate set model is used.  We have experimented with models ranging from the fully parameterized model to radically simplified models like a single-parameter model where each gate experiences depolarizing error with the same rate.  In this paper we do not dig more deeply into these possible variations or their applications: we generally assume that one of the fully-, TP-, or CPTP-parameterized models is being used.

\subsection{Fiducial-pair reduction\label{sec:FPR}}

%

In the ``standard long-sequence GST'' experiment design of Section \ref{sec:ExperimentSelection}, every base circuit is sandwiched between \emph{every} possible pair of preparation and measurement fiducial circuits to generate the full experiment design.  This enables complete tomography of each base circuit, which is clearly sufficient to estimate the gate set.  But it also yields very large GST experiments.  Information completeness requires $O(d^2)$ preparation fiducial circuits and $O(d)$ measurement fiducial circuits (because each one provides $d-1$ independent outcomes), giving $O(d^3)$ fiducial pairs.  In the 2-qubit case this means $\approx4^3=64$ circuits are required for each base circuit.)

However, many of these circuits are redundant.  Eliminating redundant circuits yields much smaller and more efficient experiments.  The redundancy stems from the fact that (as observed previously) each germ $g$ only amplifies certain ``directions'' in error space.  Because this subspace corresponds to \emph{commuting} deviations in $\seqaction{g}$ (the operation produced by $g$), its dimension is that of $\seqaction{g}$'s commutant.  This can be much smaller than the space defining $\seqaction{g}$ itself -- e.g., if $\seqaction{g}$ is a nondegenerate $d^2 \times d^2$ matrix, then although perturbations to $\seqaction{g}$ form an $d^4$-dimensional space, the commutant is only $d^2$-dimensional.  So measuring \emph{all} the fiducial pairs is redundant.  We only need to ensure that each base circuit is probed so that every amplified perturbation impacts the observed outcome probabilities.  Each amplified direction will affect the outcome probabilities corresponding to at least \emph{one} fiducial pair.  So if a given germ amplifies $m$ directions in parameter space, it is sufficient to select $m$ fiducial pairs that measure linearly independent probabilities sensitive to those $m$ variations.  Step 4 in Figure \ref{fig:GSTExperimentDesign} depicts how FPR functions during the construction of a GST experiment design.

The \texttt{pyGSTi} implementation performs ``fiducial pair reduction'' via an algorithm, described in Appendix \ref{sec:PyGSTiExperimentSelection}, that starts by constructing an informationally complete set of effective SPAM pairs, then selects germs, and finally eliminates redundant fiducial pairs for each base circuit, one by one, until no further fiducial pairs can be eliminated without losing sensensitivity to some amplified perturbation of the gate set's parameters.

A systematic study of the robustness of fiducial pair reduction is left as a topic for future work.  Anecdotal evidence suggests that fiducial pair reduction can reduce the size of GST experiments significantly, by 50-90\% in many cases.  However, eliminating redundancies in the data also tends to increase the risk that GST's parameter estimation step will get stuck at a local maximum, and we have observed this particularly in cases where the GST model does a poor job at explaining the data.  At this point we see the fiducial pair reduction technique as useful and sometimes necessary, but not one that should be applied in every case.

\section{Analyzing GST estimates\label{sec:Interpretation}}

We now turn from describing the GST protocols to interpreting their output.  GST is a characterization protocol, and its purpose is not just to provide overall performance estimation (like benchmarking) but to reveal \emph{how} an as-built component (e.g., gate or gate set) differs from its ideal.  This requires nontrivial analysis.  In this section, we give several ways to analyze or post-process the results of a GST experiment.  We will assume throughout that the GST optimization successfully finds a global optimum, i.e., it doesn't get stuck at a local maximum.  This assumption is motivated by the experiment design and iterative optimizaiton algorithm given in previous sections, as well as by our experience.  We first discuss how to assess the validity of the GST estimate (Section \ref{sec:GoodnessOfFit}).  If a GST estimate is valid, it may be helpful to gauge-optimize the estimate in order to compute common gauge-dependent metrics like process fidelity or diamond-distance (Section \ref{sec:GaugeOpt}).  Finally, we describe how to put error bars on quantities derived from a GST estimate (Section \ref{sec:ErrorBars}).

\subsection{Goodness-of-fit and Markovianity\label{sec:GoodnessOfFit}}

The first diagnostic from a GST experiment has nothing to do with the estimated gate set itself.  It is the \emph{goodness of fit} -- how well does that estimate (whatever it is) match the data?  GST's gate set model is capable of describing any set of Markovian gates in $d$-dimensional Hilbert space, along with state preparations and POVM effects (modulo optional TP and/or CP constraints).  So if it fails to fit the data, this is strong evidence that \emph{some} assumption of the model was violated.  We describe all such violations as ``non-Markovianity'', meaning that the observed behavior was influenced by some internal or external \emph{context} variable that was not included in the model \cite{Rudinger-PRX2019, veitia2020testing}.  Common phenomena of this type include slow drift~\cite{Proctor2019-oi}, leakage~\cite{Wood2018-wi,Wallman2016-kx}, persistent environments~\cite{Jing2018-ze,Chekhovich2013-fq}, pulse spillover~\cite{Mundada2019-pf,Gambetta2012-yu,Piltz2014-bk,Debnath2016-ny}, and gate-induced heating~\cite{Webb2018-rb,Rebentrost2009-ql}.  We do not attempt to diagnose specific phenomena here.  Instead we focus on how to quantitatively detect when the GST model has ``failed to fit the data'' as well as it should.  In cases where such pheomena are known to dominate the noise, it would be more appropriate to use techniques tailored to the corresponding type of non-Markovian noise, e.g.~Ref.~\cite{Proctor2019-oi}.

Consider a GST experiment containing $N_{\mathrm{exp}}$ distinct circuits.  It will produce a dataset described by $N_{\mathrm{o}}$ free parameters, where $N_{\mathrm{o}}$ is simply the number of independent circuit outcomes that can be observed.  In the typical case where there is just one native measurement ($N_{\mathrm{M}}=1$) that has $N^{(1)}_{\mathrm{E}}$ outcomes,
\begin{equation}
N_{\mathrm{o}} = N_{\mathrm{exp}} \left( N^{(1)}_{\mathrm{E}} - 1 \right).
\end{equation}
For a single qubit supporting just one native 2-outcome measurement, $N^{(1)}_{\mathrm{E}}=2$ and each circuit's data has $2-1=1$ independent degree of freedom.

\emph{Any} data set like this can be explained perfectly by a trivial ``maximal model'' with $N_{\mathrm{o}}$ parameters -- the one that assigns one probability for each independent outcome.  Fitting this maximal model to the data achieves $\logL_{\mathrm{max}}$ (see Section \ref{sec:ParameterEstimation}).

If GST's estimate has a likelihood of $\mathcal{L}$, then we can measure how well GST fit the data by comparing $\logL$ to the maximal model's $\logL_{\mathrm{max}}$.  If the data were produced by some Markovian gate set, then both the GST model and the maximal model are valid, and both should fit the data equally well \emph{if} we account for extra free parameters in the maximal model.  Wilks' theorem \cite{wilks1938} tells us how to do this accounting, and we can use it to quantify the fit quality of the GST estimate relative to that of the maximal model.  Wilks' theorem states that \emph{if} the gate set model is valid, then the loglikelihood ratio statistic between them is a $\chi^2_k$ random variable:
\begin{equation}
  2(\logL_{\mathrm{max}} - \logL) \sim \chi^2_k,
\end{equation}
where the maximal model has $k = N_{\mathrm{o}} - N_{\mathrm{p}}^{\mathrm{nongauge}}$ \emph{more} parameters than the gate set model has non-gauge parameters.  

If the loglikelihood ratio appears to have been sampled from a $\chi^2_k$ distribution, then the GST model fits the data as well as can be expected, and captures the behavior of the device exposed by this data.  On the other hand, if the observed loglikelihood ratio is so high that a $\chi^2_k$ distribution would be very unlikely to have produced it, then we have evidence that the data were generated by a non-Markovian process.  The $\chi^2_k$ distribution has mean $k$ and standard deviation $\sqrt{2k}$.  We quantify the observed \emph{model violation} by the number of standard deviations by which the loglikelihood ratio exceeds its expected value under the $\chi^2_k$ hypothesis:
\begin{equation}
N_\sigma \equiv \frac{2(\logL_{\mathrm{max}} - \logL) - k}{2\sqrt{k}}
\end{equation}
$N_\sigma \le 1$ indicates an extremely good fit that appears completely trustworthy.  (We have rarely seen this except on artificially simulated data).  Conversely, $N_\sigma \gg 1$ indicates significant model violation, meaning that no gate set can describe all of the data.\footnote{This statement relies on the stated assumption that the global likelihood optimization was successful}  Since the gate set only assumes Markovianity (and sometimes the physical TP and CP constraints), model violation indicates the presence of some kind of non-Markovian noise (as defined above).

We emphasize, however, that a large $N_\sigma$ value does not necessarily mean that the observed behavior is strongly non-Markovian.  It indicates high \emph{confidence} in the conclusion ``the Markovian model was violated''.  This does not quantify how much the model would have to be expanded to fit the data well, nor the probability of observing a surprising event, and it doesn't imply that the GST estimate is useless or untrustworthy.  If there is any model violating behavior at all, then $N_\sigma$ generally increases linearly with the number of times each circuit is repeated.  So more \emph{sensitive} experiments will yield higher $N_\sigma$, even with exactly the same physics.  Quantifying model violation in more operational and useful ways is a compelling topic for future research.

The $N_\sigma$ statistic is sensitive to the \emph{total} model violation, added up across all circuits.  It's also useful to identify individual circuits whose behavior is inconsistent with the GST estimate.  We can do this by examining each circuit's contribution to the loglikelihood, $\logL_s$ (Eq.~\ref{eq:perExperimentLogL}), and comparing \emph{it} to the maximal model.  Wilks' theorem predicts $2(\logL_{\mathrm{max},s}-\logL_s)$ should be $\chi^2_k$-distributed, with $k$ approximately equal to the number of independent outcomes of a single circuit.  \footnote{Technically it would be more accurate to subtract from this value the number of gate set parameters ``per circuit'', but this is usually much less than one and has virtually no impact.}  These per-circuit tests are almost independent from the ``total'' test described above -- either test may reveal model violation when the other does not, although most forms of model violation trigger both.

If many circuits are tested simultaneously for model violation (as is usually the case), it is important to raise the threshold for detecting a violation using extreme value theory.  If $K$ tests are performed in parallel, then to keep the probability of \emph{any} false positive below $\alpha$, each test should be performed at the $\alpha^{1/K}$ confidence level.

\begin{figure}
  \begin{center}
    \includegraphics[width=3.5in]{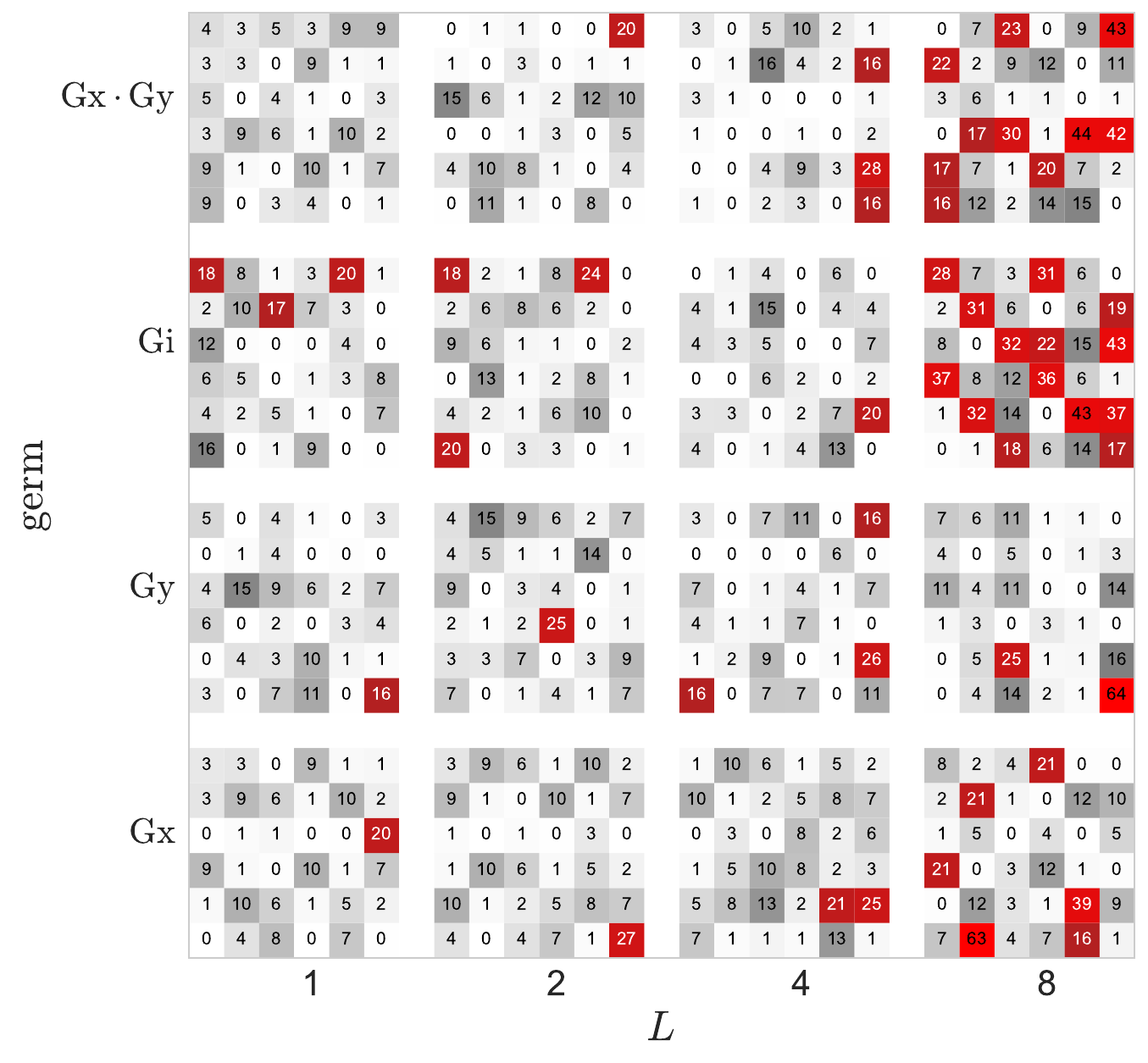}
    \caption{A sample model violation plot displaying the values of $2(\logL_{\mathrm{max},s}-\logL_s)$ for circuits $s$.  Each $6\times 6$ plaquette of colored squares represents a set of circuits based on the same base circuit $g^p$ with (maximum) depth $L$ increasing along the horizontal axis and the germ $g$ varying along the vertical axis.  The 36 distinct boxes \emph{within} a plaquette represent distinct circuits, formed by sandwiching $g^p$ between 6 different preparation fiducials (indexed by column) and 6 measurement fiducials (indexed by row).  In this particular case the single-qubit gate set ideally comprised of $X(\pi/2)$, $Y(\pi/2)$ and $I$ (the identity) was used, and both the preparation and measurement fiducials are the set $\{ \emptyset, X, Y, XX, XXX, YYY \}$, where $X$ and $Y$ are shorthand for $X(\pi/2)$ and $Y(\pi/2)$.\label{fig:exampleColorBoxPlot}}
  \end{center}
\end{figure}

We have found a plot of the per-circuit goodness-of-fit values, represented as a grid of colored boxes arranged by germ and base circuit depth (see Figure \ref{fig:exampleColorBoxPlot}), to be a useful diagnostic tool.  The color scale in Figure \ref{fig:exampleColorBoxPlot} changes from a linear grayscale to a logarithmic red-scale when the $\logL_s$ value of a box exceeds the $95$th percentile of the expected $\chi^2$ distribution (but see note above about adjusting for multiple tests).  Deeper circuits are more sensitive to most forms of non-Markovian noise.  If model violation (red boxes) appear preferentially in circuits containing a particular germ, this usually indicates that a gate appearing in that germ is the cause.

\subsection{Gauge optimization\label{sec:GaugeOpt}}

We now turn to analysis of estimated gate sets.  The output of ``doing GST'' on a quantum processor -- constructing an experiment, performing the circuits, and fitting a model to the data -- is a gate set that represents how that processor's gates act.  The catch, as noted repeatedly above, is that this estimate is only unique up to gauge transformations.  And while gauge transformations have no effect on outcome probabilities of circuits, they can have a huge effect on the individual transfer matrices representing the gates, and derived quantities of interest like fidelity.

The ideal solution to this problem would be to only compute and report \emph{gauge-invariant} properties.  But most of the metrics commonly used to assess quantum operations are \emph{not} gauge-invariant.  They vary -- usually quite a lot -- over gauge orbits.  This means they don't correspond to physically observable quantities \cite{Proctor2017-ru}.  These metrics originated as ways to quantify the performance of a single gate \emph{in the context of other, perfect operations} that form a reference frame.  Gauge freedom emerges from the absence of a reference frame, and the motivation for GST is that, despite common assumptions, experiments on quantum processors usually do not feature an absolutely reliable reference frame.  But even if they are flawed, gauge-variant metrics like fidelity, diamond distance, or entropy are not currently dispensable; no gauge-invariant replacements exist.  So in order to compute well-studied gauge-variant metrics such as process fidelity and diamond distance for GST estimates, we choose a particular gauge using what we call \emph{gauge optimization}.

Gauge optimization means reporting the GST estimate, $\hat{\mathcal{G}}$, in a gauge that optimizes some gauge-variant metric of ``closeness'' to the ideal target gate set.  In other words, we choose the gauge to make the gates look as good as possible.  The best metric is not unique or obvious, and can be highly situation-dependent.  The implementation of GST in \texttt{pyGSTi} \cite{Nielsen2020-lu} supports gauge optimization using a variety of metrics.  Actually performing the gauge optimization, once a metric is chosen, is straightforward and uninteresting (\texttt{pyGSTi} uses local gradient minimization, with occasional long random jumps to avoid rare local minimum traps).  We focus here instead on (1) the rationale and justification for gauge optimization, and (2) the rationale for choosing a metric to optimize.

\subsubsection{Why gauge optimization?}

A common and fundamental objection to GST and gauge optimization goes something like this: ``Intentionally seeking out and constructing the gate set \emph{closest} to the desired target seems like cheating, if not actively circular.''

No truly satisfactory answer to this can exist, because in the absence of a privileged reference frame, gauge-variant metrics have limited meaning and shouldn't be computed.  Gauge optimization is fundamentally a hack.  But the quantum computing community has been reluctant to abandon metrics like process fidelity, diamond distance, and trace distance that are deeply rooted in the theory and literature of quantum information -- not least because no good alternatives are known yet.

But, with that disclaimer, there are good arguments (which we find compelling) that gauge optimization is at least a \emph{good} hack, contrary to the objection proposed above.

First:  Gauge optimization is not an all-powerful tool for reducing errors.  It cannot alter any predicted circuit outcome probabilities, and is therefore powerless to improve any circuit-outcome-based performance metric.  Gauge optimization \emph{does} seek out the gate set closest to the desired targets -- but it searches over a very limited set of candidates!  Each gate's eigenvalues are gauge-invariant.  Target gates are almost always unitary, so they have unit-modulus eigenvalues.  Any errors that change a gate's eigenvalues -- e.g., stochastic errors that shrink a gate's eigenvalues toward zero -- cannot be disguised or eliminated by gauge transformations.  Almost every known physical error in gates \emph{cannot} be eliminated by gauge transformations.  Even coherent tilt errors, which manifest as relational mismatches between the rotation axes of different gates, can only be pushed from one gate to the other by gauge transformations.  Physical errors that \emph{do} correspond to gauge transformations can originate from an error that is global to the gate set being analyzed but not to the entire lab.  For example, a misalignment of one qubit such that all its gates and SPAM operations are rotated corresponds to a gauge transformation of the 1-qubit system (but this would \emph{not} correspond to a gauge transformation in a 2-qubit system where the other qubit was not rotated).

Second:  While gauge transformations cannot remove most physically plausible errors, they can \emph{create} arbitrarily large unphysical errors.  Applying a large unitary gauge transformation to the target gate set itself will make every gate appear to have large coherent errors, and thus large infidelities and diamond distances.  But of course, this is an illusion -- this gate set has no errors at all, because it's just another representation of the target.  As long as we are stuck with gauge-variant metrics, this example demands gauge optimization or an equivalent gauge-fixing procedure.

Third:  It's tempting to seek a different gauge-fixing procedure that produces a standard form for gate sets, without explicitly trying to make them look like the target gate set.  But, as the previous example shows, any gauge-fixing procedure that \emph{works} actually has to do this anyway -- just in disguise.  If we construct a gate set that's equivalent to the target, and then feed it into a gauge-fixing procedure, it absolutely \emph{must} return the target gate set, regardless of what the target is.  If it doesn't, then it will return a gate set that appears to have errors, which is not true (by construction).

\subsubsection{Metrics for gauge optimization}

Given an estimated gate set $\hat{\mathcal{G}}$ and a target $\mathcal{G}_0$, gauge optimization finds the gauge that minimizes some function 
\begin{equation}
f(\hat{\mathcal{G}},\mathcal{G}_0).
\end{equation}
Some general guidelines for choosing $f$ include:  it should be non-negative, it should be uniquely zero if $\hat{\mathcal{G}} = \mathcal{G}_0$, and it should facilitate minimization (e.g., smoothness and convexity are desirable).  However, it does not need to satisfy all the properties of a mathematical metric (e.g., the triangle inequality).

We examined variations on three different metrics, all derived from summing up some well-known function of two matrices over all the logic operations in the gate set\footnote{SPAM operations are included in these ``logical operations'', and are incorporated in various ways}:
\begin{enumerate}
\item Infidelity,
\item Trace/diamond distance,
\item Squared Frobenius (Euclidian) distance.
\end{enumerate}
Infidelity and trace/diamond distance have operational interpretations in quantum information.  The Frobenius distance between transfer matrices and density matrices generally does not.

Infidelity is not a suitable metric for gauge optimization for an interesting reason:  when applied to general matrices (with no positivity constraint), it is neither strictly positive nor uniquely minimized when $\hat{\mathcal{G}}=\mathcal{G}_0$.  This is related to the fact that infidelity is not actually a metric in the mathematical sense.  If $G$ is a unitary gate, then although $G$ has fidelity 1 with itself -- $F(G,G)=1$ -- it's actually possible to achieve fidelity \emph{greater} than 1 by performing a nonunitary gauge transformation, so $F(G,MGM^{-1}) > 1$ for some $M$.  We emphasize that in these circumstances, $M$ is not unitary, and $MGM^{-1}$ is not completely positive.  But nonetheless, this means that infidelity only satisfies the required desiderata if the CP constraint is imposed during gauge optimization.  This is inelegant, and poses practical problems in gauge optimization.

We found trace/diamond distance to be expensive to compute (because there is no closed form for diamond distance), hard to work with (because it is not smooth), and to yield results not significantly different from squared Frobenius distance.  In principle, diamond distance is an appealing metric -- it has deep operationally meaning, and there is a nice elegance in saying ``We have chosen the gauge that minimizes the diamond distance to the target gates.'' However, we found it too inconvenient in practice to justify this advantage.

We find the squared Frobenius distance to be the most useful gauge optimization metric.  It is well-behaved and fast to compute, and we have found few disadvantages to its use.  In particular, we utilize the \emph{weighted} sum of squared Frobenius distances between the estimated and target gate set elements,
\begin{eqnarray}
  f(\hat{\mathcal{G}},\mathcal{G}) &=&
  \sum_{i=1}^{N_\rho} \alpha_i \vert \hat{\rho}^{(i)} - \rho^{(i)} \vert^2
    + \sum_{i=1}^{N_{\mathrm{G}}} \beta_i \vert \hat{G}_i - G_i \vert^2 \nonumber \\
  & &  + \sum_{m=1}^{N_{\mathrm{M}}} \sum_{i=1}^{N^{(m)}_{\mathrm{E}}} \gamma_{m,i} \vert \hat{E}^{(m)}_i - E^{(m)}_i \vert^2,\label{eq:FrobeniusGauge}
\end{eqnarray}
where the notation of Eq.~\ref{eq:GatesetDef2} is used to denote the elements of the estimated ($\hat{\mathcal{G}}$) and target ($\mathcal{G}$) gate sets, and $\vert\cdot\vert$ represents the Frobenius norm.  The weights $\alpha_i$, $\beta_i$ and $\gamma_{m,i}$ are real numbers.

The weighting turns out to be important, for the odd and interesting reason that SPAM errors cannot be amplified.  Long-sequence GST with very deep circuits can resolve errors in \emph{gates} to very high precision, potentially below $10^{-5}$.  But achieving this sort of accuracy in SPAM operations is virtually impossible.  An error that rotates the initial state relative to the measurement axis by a small angle $\theta$ cannot be detected using fewer than $O(1/\theta^2)$ shots.  In an experiment where the longest circuit is $L$ gates deep, and each circuit is repeated $N$ times, estimation error in the SPAM operations is typically $O(1/\sqrt{N})$, while estimation error in the gates is $O(1/L\sqrt{N})$, which can be vastly smaller.

But gauge transformations can ``slosh'' certain errors -- coherent errors, in particular -- between SPAM operations and gate operations.  As a specific example, consider a GST experiment with $N=10^4$ and $L=100$, performed on a \emph{perfect} experimental system.  In this idealized case, all estimated errors are due entirely to finite-sample noise in the experiment.  We would expect the estimated errors in the gates to be around $\pm 1/(L\sqrt{N}) \approx 10^{-4}$, but estimated errors in the SPAM to be around $\pm 1/\sqrt{N} \approx 10^{-2}$.  But if we perform gauge optimization and seek to minimize the \emph{equally weighted} sum of Eq.~\ref{eq:FrobeniusGauge}, the optimal solution will ``split the difference,''  finding a gauge in which \emph{both} the SPAM and gate operations have about $0.5\times 10^{-2}$ coherent error.  This is a poor and misleading choice of gauge, because the highly precise estimate of the gates is polluted by estimation error in the SPAM.

One solution is to adjust the weights in Eq.~\ref{eq:FrobeniusGauge}, assigning different weights to SPAM operations ($\alpha_i$ and $\gamma_{m,i}$) and gate operations ($\beta_i$).  We have found that a good rule of thumb is to assume that the finite-sample estimation error in each operation is proportional to the largest number of times it appears in any circuit -- typically $L$ for gates and $1$ for SPAM -- and then give the corresponding term in the metric a weight proportional to the inverse square of its estimation error.  So, in the example above, we would set $\alpha_i = \gamma_{m,i} = 1$ and $\beta_i = 10^4$ and minimize Eq.~\ref{eq:FrobeniusGauge}.  This would tend to split relational error between SPAM and gates unequally, assigning 99\% of the error to the SPAM.

Another heuristic for avoiding poor gauge choices, which we find works even better in practice, is to perform several sequential ``stages'' of gauge optimization.  Each stage uses the Frobenius distance  metric (Eq.~\ref{eq:FrobeniusGauge}) with different weights \emph{and} a different optimization domain.  In the case of a fully parameterized gate set these stages are:
\begin{enumerate}
\item Optimize over the \emph{full gauge group} (over all invertible $M$ in Eq.~\ref{eq:GaugeTransform}) using \emph{uniform weights}.  This finds a decent gauge choice - one that may need some tweaking (see following steps) but that doesn't assign objectively unnecessary error, e.g., by rotating all of the gates and SPAM with respect to their targets.  Starting from this point, we next:
\item Optimize over the \emph{unitary group} (over all unitary $M$) using \emph{100\% weight on the gates} ($\alpha_i = \gamma_{m,i} = 0$, $\beta_i = 1$).  This rotates the gates into the best versions of themselves possible, and is justified in the common case where we know the gates to far higher accuracy than the SPAM.  By restricting to the unitary group, we disallow transformations that make the gates slightly better by drastically stretching or skewing the notion of distance.
\item Optimize over the \emph{SPAM gauge group} using \emph{100\% weight on the SPAM operations}.  We define the ``SPAM gauge group'' as the 2-parameter family $\{\mathrm{diag}(\alpha, \beta, \ldots \beta)\,:\, \alpha, \beta \in \reals\}$, where $\mathrm{diag}(\ldots)$ is a diagonal matrix with the given values on its diagonal.  Matrices of this form slosh error between SPAM operations and the non-unital part of the gates (recall that the $0$-th basis element is the identity) in a gate set.  Thus, this optimization reduces errors on the SPAM operations at the expense of increasing non-unital gate errors.  Because the SPAM gauge group doesn't necessarily preserve complete positivity, we include this as an optimization constraint.
\end{enumerate}
In the case of a TP gate set, the initial stage is modified to optimize only over TP matrices $M$ and the SPAM gauge group of the final stage is restricted to the 1-parameter family of diagonal matrices $\{\mathrm{diag}(1, \beta, \ldots \beta)\,:\, \beta \in \reals\}$ whose top-left element is one.  We treat CPTP gate sets by simply omitting the first stage of the procedure followed in the TP case.  

Gauge optimization yields a gate set $\hat{\mathcal{G}}_{\mathrm{GO}}$ that is gauge-equivalent to $\hat{\mathcal{G}}$, but effectively in standard form.  That is, any distinct yet equivalent gate set would be brought to the same form by gauge optimization, provided that the same metric was used for gauge optimization.  From $\hat{\mathcal{G}}_{\mathrm{GO}}$, gauge-variant quantities can now be computed, and assigned meaning on the grounds that at least they've been computed in a fixed and well-motivated gauge.  Computing gate metrics such as the fidelity or diamond distance requires some type of informed gauge-fixing is required, and gauge optimization as described here is the best method we have found.

\subsection{Error bars\label{sec:ErrorBars}}

The GST analysis pipleine described so far produces a \emph{point estimate} of the gate set -- a single gate set that fits the data well.  But estimation is like throwing darts at a board:  no dart ever hits the exact center of the target.  Even if we assume there exists a ``true'' gate set, $\mathcal{G}_{\mathrm{MLE}}$ will at best be close to it.  We need to estimate \emph{how} close.  We need what physicists tend to call ``error bars'', and what statisticians refer to as a \emph{region} estimator.  

At least three kinds of region estimators are commonly used: confidence regions \cite{Blume-Kohout2012-fq}, credible regions \cite{Ferrie_ellipsoids_2014, Granade2017qinferstatistical}, and standard errors \cite{Efron_standarderror_1981}.  The details of what each means, and their relative merits, are rather technical, and good discussions relevant to tomography can be found in \cite{Blume-Kohout2012-fq,Christandl2012-am}.  Here, we mostly follow the confidence region approach, but we do not attempt rigor.  Our goal is to demonstrate that putting plausible, reasonable error bars around GST estimates, by adapting well-known techniques, is possible.  We welcome rigorous critiques and improvements.

We have successfully used two techniques to equip GST estimates with error bars.  The first is bootstrapping.  The second is likelihood-ratio confidence regions.  Our limited testing has found both to reliably produce reasonable error bars.  However, our testing also makes it clear that GST is a ``messy'' statistical problem, and further research and development are needed.  To put this more bluntly:  we would \emph{not} attempt to publish the error bar algorithms reported here as a standalone paper at this time, because they are insufficiently developed.  But, at the same time, we believe it is irresponsible to publish or deploy a QCVV protocol without \emph{any} mention of error bars.  Since the overall theory of GST is mature and overdue for publication, our imperfect solution to these frustrated constraints is to discuss the best available error bar technology for GST, including this clear disclaimer.

\subsubsection{General problems with region estimators}  

There is no universal agreement on the ``correct'' way to report uncertainty about any tomographic estimate.  However, the most principled approach appears to be to construct and report either a confidence region \cite{Blume-Kohout2012-fq, Christandl2012-am, Faist2016-pb} or credible region \cite{Shang2013-ao, Granade2016-qy}.  In either approach, the tomographer describes their uncertainty by (1) choosing a probability $\alpha$ (e.g., $\alpha = 95\%$), then (2) reporting a subset $R$ of the parameter space, of more or less arbitrary shape, for which a statement like ``$R$ probably contains the true parameter, with probability $\alpha$.'' can be made.\footnote{We emphasize that this statement's precise phrasing and meaning are tricky, and vary between the two approaches.}

Both approaches pose some significant practical challenges.
\begin{enumerate}
\item There is no standard or efficient way to describe an arbitrary region in a high-dimensional space.
\item This construction requires the tomographer to pick a particular $\alpha$ in advance.  But this value is rarely meaningful to end users, and there is absolutely no guarantee that a region for $\alpha'\neq\alpha$ can be extrapolated from the $\alpha$ region.
\item Researchers and end users are almost always more concerned with some particular scalar function $f(\vec\theta)$ of the high-dimensional tomographic parameter $\vec\theta$ than with $\vec\theta$ itself.  Extracting an \emph{interval estimate} for $f(\vec\theta)$ from a \emph{region estimate} for $\vec\theta$ is nontrivial, and if done crudely can dramatically overestimate the uncertainty in $f$.
\end{enumerate}
None of these problems invalidate the theoretical validity of optimal confidence or credible regions.  But because of them, most experiments either ignore error bars, or simplify theoretically rigorous techniques so ruthlessly that their provable properties (optimality or rigorous coverage) are lost.  For example, the first issue can be mitigated by specializing to regions of a particular shape -- e.g., ellipsoids, or spheres in a particular metric -- but if coverage probability is maintained, this usually requires regions that are far from optimal and overestimate uncertainty.

The two methods we discuss here can both be motivate by a fairly common ansatz:  \emph{we assume that local asymptotic normality (LAN) applies}.  When LAN holds,
\begin{enumerate}
\item the likelihood function for the \emph{current} data is Gaussian, and
\item the likelihood function for \emph{other hypothetical data sets} that could have been observed (but weren't) is almost surely Gaussian with almost the same covariance matrix.
\end{enumerate}
These consequences mean that we can treat the underlying statistical problem like a Gaussian shift model.  The MLE will be $\mathcal{N}(\vec\theta_{\mathrm{true}},\sigma)$ distributed around the true parameter value, and the likelihood function will be a Gaussian whose covariance matrix is that same $\sigma$.

Inasmuch as this assumption holds, it resolves all of the practical problems above:
\begin{enumerate}
\item Every optimal confidence region is an ellipsoid centered at the MLE with covariance $\sigma$.
\item Changing $\alpha$ just requires scaling that ellipsoid up or down.
\item Confidence intervals for any linear function $f(\vec\theta)$ can be extracted very straightforwardly from the confidence region (by marginalizing the confidence ellipsoid and applying a Bonferroni correction), and the same procedure can be used when $f(\vec\theta)$ is nonlinear as long as it's approximately linear over the extent of the confidence ellipsoid.
\end{enumerate}
Of course, LAN will never hold exactly, and we can construct GST examples where it is a pretty bad approximation.  In these cases, ellipsoid confidence regions will (usually) overestimate uncertainty, and (occasionally) have insufficient coverage probability.  More research is required to identify and map out the regimes where the LAN ansatz used here is unreliable.

\subsubsection{Bootstrapping}

The basic idea of the bootstrap is very simple:  repeatedly simulate the experiment that generated the real data to generate a large number of ``fake'' datasets, analyze each of them to produce a point estimate, and use their scatter (variance) to construct a region.  This works very well for Gaussian shift models, as long as a large enough sample of fake datasets is constructed.  It is \emph{known} to break down when the estimator is biased, when it is significantly heteroskedastic, or if the bootstrap distribution is undersampled.  Both bias and heteroskedasticity are known problems for state and process tomography due to positivity constraints.  We find these to be less important for GST for two reasons.  First, we do not usually impose CPTP constraints explicitly.  Second, long-sequence GST can achieve such high accuracy that the data bound the estimate comfortably away from the zero-probability boundaries where heteroskedasticity becomes large.  When we have cross-validated bootstrapping against the alternative method described below, we have observed consistent results.

The two basic types of bootstrap are \emph{parametric} (in which the ``fake'' datasets are simulated from scratch using the MLE gate set) and \emph{non-parametric} (in which each circuit's outcomes are resampled directly from the data, with replacement).  We have performed both types, and have not seen significant differences.  However, on theory grounds, the parametric version should be more reliable because it is less affected by constraint-induced bias, and by certain small-sample-size effects.

Applying standard bootstrap methods to GST encounters two specific problems.  First, the GST model has a lot of parameters.  One qubit gate sets usually have at least $n=30$ free parameters, while 2-qubit gate sets usually have at least $n=1000$.  Unless at least $n^2$ samples are generated, the sample covariance will necessarily be rank-deficient, so estimating the population covariance by bootstrapping requires \emph{at least} $O(n^2)$ samples.  Since each bootstrap sample requires running an MLE analysis on a fake dataset, which can take minutes or hours, the time required to generate $10^6$ (or even 1000) samples is excessive.

So, instead of trying to construct a confidence \emph{region} for the entire gate set (which, as noted above, is almost never directly useful), we use the bootstrap to construct confidence intervals for scalar parameters directly.  Estimating the uncertainty in a single parameter this way requires only $O(1)$ samples, and $k$ distinct parameters (e.g., all the matrix elements of the transfer matrices in the gate set) can be reliably bootstrapped with only $O(\log k )$ samples.

The second issue is more subtle, and stems from the gauge freedom in gate sets.  We discuss this in Section \ref{sec:GaugeFixingBootstrap} below.

\subsubsection{Confidence regions based on the Hessian of the likelihood}

Likelihood-ratio (LR) confidence regions are an alternative and logically independent approach to quantifying uncertainty.  They are based on the intuition that the true $\vec\theta$ will probably have a high likelihood -- not much less than the maximum -- and so the set of all $\vec\theta$ with likelihood above some threshold constitutes an efficient confidence region.  The basic theory for LR confidence regions in tomography can be found in Ref.~\cite{Blume-Kohout2012-fq}.

If we assume that LAN applies, then the likelihood function is Gaussian, its logarithm is quadratic, and so it is completely determined by its second derivatives or \emph{Hessian} evaluated at the MLE.   Under these assumptions, a LR confidence region can be computed simply by
\begin{enumerate}
\item Calculating the Hessian $H$ of the loglikelihood at the MLE, which is somewhat tedious but straightforward.
\item Inverting $H$ and multiplying $H^{-1}$ by an appropriate scaling factor to define an ellipsoid that coincides with the a specific contour of the likelihood function.  
\end{enumerate}
The desired contour is defined by $\chi^2$ theory, and corresponds to gate sets $\mathcal{G}$ whose loglikelihood satisfies
\begin{displaymath}
  -2\log\left(\frac{\mathcal{L}(\mathcal{G})}{\mathcal{L}(\hat{\mathcal{G}})}\right) = \lambda_{\mathrm{thresh}}(k,\alpha),
\end{displaymath}
where $k=N_{\mathrm{p}}^{\mathrm{nongauge}}$ is the number of non-gauge parameters in the gate set, $\mathcal{L}(\hat{\mathcal{G}})$ is the maximum likelihood for any gate set (e.g., GST's point estimate), $\alpha$ is the desired confidence level, and $\lambda_{\mathrm{thresh}}$ is the $\alpha$ quantile of a $\chi^2_k$ distribution.

Confidence intervals for linear functions $f(\vec\theta) = \vec{v}\cdot\vec\theta$ of the gate set can be derived easily from this procedure.  We simply project the covariance matrix $H^{-1}$ onto the subspace spanned by $\vec{v}$ to get the variance of a scalar, and then set $\lambda_{\mathrm{thresh}}$ based on $k=1$ free parameter.  However, we have to deal with the gauge freedom first.

\subsubsection{Dealing with gauge freedom}  

GST's gauge freedom creates problems for both the bootstrapped and the Hessian-based regions described above.  All the gate sets on a gauge orbit are completely equivalent.  Imagine that we could parameterize the set of all gate sets as $\mathcal{G} = (X,Y)$, where $X$ describes which gauge orbit $\mathcal{G}$ is on and $Y$ specifies its location on the gauge orbit.  $X$ would include all the gauge-invariant parameters, and $Y$ would constitute pure gauge parameters.  In this scenario, how would we describe our uncertainty about (i.e., put error bars on) $Y$?  On one hand, the data provide absolutely no information about $Y$, suggesting total uncertainty and large error bars.  But, on the other hand, we may simply choose a gauge by fiat, suggesting absolute certainty and vanishing error bars.

The answer is made clear by considering how ``gauge parameters'' ($Y$, above) affect the variables that scientists wish to learn from GST -- e.g., superoperator matrix elements, fidelities, or diamond norms.  These are all functions of the gate set, but they are not gauge invariant.  So if we know $X$ exactly, but allow $Y$ to vary wildly, then these desirable variables will also vary.  Thus, if we assign large error bars to gauge parameters, we necessarily get large uncertainty about fidelities and other gate properties -- even if we have learned everything that can be learned about the gate set ($X$) to high precision!  

So we can only assign consistent error bars if we \emph{fix} the gauge.  Every point estimate necessarily has a fixed gauge, by its very nature -- $\hat{\mathcal{G}} = (\hat{X},\hat{Y})$, so $\hat{\mathcal{G}}$ has a fixed value of $Y$ (gauge).  But a region is a \emph{set} of points $\{\mathcal{G}\}$.  It is easy to construct a confidence region containing gate sets with different gauges -- say, $\mathcal{G}_1 = (X_1,Y_1)$ and $\mathcal{G}_2 = (X_2,Y_2)$, where $Y_1 \neq Y_2$.  If we allow this to happen, then the error bars on interesting but non-gauge-invariant quantities will be artificially inflated.  So fixing the gauge for a \emph{region} requires ensuring that every element of the region has the same $Y$ coordinates.

This is harder than it sounds, and is actually something of a category mistake.  Above, we said ``Imagine that we could'' separate $G$ into gauge-invariant parameters and pure gauge parameters.  This is not generally possible in gauge theories, because it would require a global definition of parallel transport that does not generally exist.  Different gauge orbits are typically not the same, or even isomorphic.  For example, consider two gate sets $\mathcal{G}_1,\mathcal{G}_2$ that each contain just a single gate $G_i$ -- but in $\mathcal{G}_1$, $G_i$ is the identity matrix, while in $\mathcal{G}_2$, $G_i$ is nondegenerate.  One $G_i$ is invariant under all gauge transformations, while the other is not.  So the gauge orbit of $\mathcal{G}_1$ is a zero-dimensional point (the entire gate set $\mathcal{G}_1 = \{G_i\}$ is invariant under gauge transformations), while that of $\mathcal{G}_2$ is much larger and more complex.  No coordinate system ($Y$) can describe them both, and if a confidence region includes both $\mathcal{G}_1$ and $\mathcal{G}_2$ there is no way to demand that they have ``the same $Y$ coordinate''.

Instead, we gauge-fix a region by first identifying each \emph{orbit} (equivalence class of gate sets) in the region, assigning it a single representative gate set, and selecting those representatives to minimize the size (variance) of the resulting set of gate sets.  This approach can be used for both bootstrapped and Hessian-based regions, but plays out differently in practice for the two constructions.

\subsubsection{Gauge-fixing bootstrap regions\label{sec:GaugeFixingBootstrap}}

We have found the following procedure effective at eliminating effects of gauge freedom from boostrapped error bars.  Each gate set that is sampled via the bootstrap is ``gauge optimized'' using a fixed procedure, which sets its gauge to make it as close as possible to the target gate set.  This is a good operational approximation to requiring every element of the region to be ``in the same gauge''. By making every representative as close as possible to a fixed target, it minimizes the radius and variance of the entire region.

\subsubsection{Gauge-fixing Hessian-based regions\label{sec:GaugeFixingHessian}}

This is a bit trickier, because in normal statistical scenarios, the Hessian $H$ of the loglikelihood is full rank.  But each gauge degree of freedom creates a direction in parameter space, at the MLE, along which the loglikelihood is locally constant.  It has no curvature in these directions, and thus its second derivative vanishes, giving $H$ a kernel.  Before we can invert $H$, we must get rid of these zero eigenvalues by projecting $H$ onto its support.  This projection corresponds to ``slicing'' an uncertainty ellipsoid that is infinitely extended along the tangent plane of the gauge orbit at the MLE, which yields a finite and bounded ellipse.  However, minimizing the size of the region requires choosing the right plane in which to slice.  This is somewhat technical, and details are given in Appendix \ref{sec:GaugeInErrorBars}.  The resulting projected Hessian has full rank and can be inverted and analyzed as given above.  It defines a covariance tensor in (non-gauge) gate set space that, when scaled by an appropriate factor, defines an ellipsoid that is a valid $\alpha$ confidence region.  Writing down this ellipsoid explicitly (as a tensor), while possible, is not very practical.  Instead, we use it to define error bars (confidence intervals) for any and all interesting \emph{scalar} quantities (e.g., fidelities, diamond norms, superoperator matrix elements, etc).

We compute such confidence intervals as follows.  Let $f(\mathcal{G})$ be a scalar function of a gate set parameters with linearization $f(\mathcal{G}) \approx f_0 + \nabla f \cdot (\mathcal{G}-\mathcal{G}_{\mathrm{MLE}})$.  We define an $\alpha$ confidence interval around the best-estimate value $f_0 = f(\mathcal{G}_{\mathrm{MLE}})$ by computing
\begin{equation}
\delta f = \sqrt{ (P(\nabla f))^{\dagger} \cdot (P(H)/C_1)^{-1} \cdot P(\nabla f)  } \label{eq:HessianErrorBars}
\end{equation}
where $P$ indicates projection onto the (correctly chosen) support of $H$, $P(H)$ is the projected Hessian and $P(\nabla f)$ is the similarly projected gradient of $f$.  $C_1$ is a scalar constant which satisfies $\mathrm{CDF}_1(C_1) = \alpha$, where $\mathrm{CDF}_k$ is the cumulative density function of the $\chi^2_k$ probability distribution.  With $\delta f$ so defined, $f_0 \pm \delta f$ specifies a good $\alpha$ confidence interval for $f$.  Within the linear approximation to $f$, which is valid for small $\delta f$, this interval corresponds to minimizing and maximizing the value of $f$ over the contour of the loglikelihood corresponding to a $\alpha$ confidence interval \emph{if} the loglikelihood had a single parameter.

We emphasize that this does not construct a $\alpha$ confidence \emph{region}.  Equation \ref{eq:HessianErrorBars} can be used to construct $\alpha$ confidence \emph{intervals} for each of the $N_{\mathrm{p}}^{\mathrm{nongauge}}$ parameters, but the region formed by taking the product of all these intervals contains the truth only if every one of the intervals contains its parameter.  This occurs with probability at least $(\alpha)^{N_{\mathrm{p}}^{\mathrm{nongauge}}}$, not $(\alpha)$.  A true $\alpha$ region can be readily constructed by simply replacing $C_1$ with $C_k$ in Eq.~\ref{eq:HessianErrorBars}, where $k = N_{\mathrm{p}}^{\mathrm{nongauge}}$.

When error bars on a scalar quantity are needed, we believe the confidence intervals constructed above  are more meaningful than projections of the full $\alpha$ confidence region, for two reasons.  First, these intervals are consistent with the error bars reported by the parametric bootstrap, which yields standard errors for each parameter independently.  Those intervals would have to be expanded significantly to construct a joint confidence region.  Secondly, we are often interested in the uncertainty of single scalar quantity (e.g., diamond norms), independently of others.  The confidence intervals constructed this way correctly measure this uncertainty.

\section{Summary}
We have presented gate set tomography as a method for characterizing a quantum logic device.  GST is in some ways a successor to existing tomographic methods such as quantum state and process tomography, but it is also fundamentally different:  a gate set is a entity unto itself, \emph{not} just a collection of unrelated components.  So GST is really a \emph{new} type of tomography that probes a different object (a gate set) with unique properties.  There is still considerable overlap between GST and other tomographic methods, so we conclude by recalling and summarizing GST's primary distinguishing (and novel) features.
\begin{itemize}
\item \textbf{Calibration free}: GST neither requires nor even considers a pre-calibrated reference frame, nor relies on any priori assumptions about the accuracy of any operations.
\item \textbf{Self consistent}: GST produces self-consistent descriptions of all the operations (state preparations, gates, and measurements) exposed by a quantum processor.  This is different from estimating each quantum state and superoperator independently, both because it requires no reference frame, and because (as a result) it exposes gauge degrees of freedom that make some errors \emph{relational} between operations, rather than endemic to a specific gate.
\item \textbf{Hyperaccurate}: The use of deep, periodic circuits enables GST estimates' precision to achieve Heisenberg-like scaling (modulo gauge freedom; see Appendix \ref{sec:NumericalVerification}).
\item \textbf{Constraint-friendly}:  GST is not restricted to estimating arbitrary quantum transfer matrices, but can naturally incorporate almost any constraint or physics-inspired model, by treating a gate set as a parameterized model.  Immediately useful examples includ restricting the estimate to be trace preserving (TP) or completely positive and trace preserving (CPTP).
\item \textbf{Model validation}:  It is very natural within GST to \emph{detect and quantify} violation of the underlying model that is being fit, using standard statistical techniques.  This is a useful byproduct of the overcomplete experiments used by GST, and can be used as a built-in warning system to identify systematic errors in characterization.  
\end{itemize}

Although this paper is intended as the first \emph{comprehensive} description of gate set tomography, including the details of the long-sequence variant, GST has been used in a variety of published experiments \cite{GST2013,GST2015, GST2016, Blume-Kohout2017-kn, Rol2017-wn, Mavadia2018-al, Ware2018-cq, Proctor2019-oi, Song2019-fg, White2019-ls, Hong2020-vc, Joshi2020-wo, Zhang2020-ux} in several qubit technologies.  Like all protocols, it has limitations, which we've tried to expose and discuss here.  But GST is also a mature protocol with extensive real-world validation in experiments.  A reference implementation of GST is available in the free, open-source Python package \texttt{pyGSTi} \cite{pygsti,Nielsen2020-lu}.  Extensions of GST to include additional constraints and to analyze time-dependent data are currently under investigation, as are variants of the protocol which scale more favorably with the number of qubits.  But even if these extensions should fail to produce useful practical tools, the development of GST to date has contributed something we believe is of lasting value: it demonstrated the principle of self-consistently characterizing all the operations on a quantum logic device, and revealed that this problem is both more and less than the sum of its parts, thanks to the unexpected role of gauge freedom.

\begin{small}
\vspace{0.2cm}
\noindent \textbf{Acknowledgements.} This work was supported by the U.S. Department of Energy, Office of Science, Office of Advanced Scientific Computing Research Quantum Testbed Program, Intelligence Advanced Research Projects Activity, and the Laboratory Directed Research and Development program at Sandia National Laboratories. Sandia National Laboratories is a multi-program laboratory managed and operated by National Technology and Engineering Solutions of Sandia, LLC., a wholly owned subsidiary of Honeywell International, Inc., for the U.S. Department of Energy's National Nuclear Security Administration under contract DE-NA-0003525. All statements of fact, opinion or conclusions contained herein are those of the authors and should not be construed as representing the official views or policies of the U.S. Department of Energy, or the U.S. Government.
\end{small}

\appendix
\addcontentsline{toc}{section}{Appendices}

\section{Gauge\label{sec:Gauge}}
\subsection{Gauge degrees of freedom\label{sec:GaugeDOF}}
GST estimates gate set parameters by varying these parameters to better fit a given set of data. A naive goal would be to estimate \emph{every} parameter with high accuracy -- i.e., to estimate a best-fit \emph{point} within the parameter space $\mathcal{P}$.  However, it is usually the case (in GST analyses) that at every point in parameter space there are directions along which none of the predicted probabilities change. Variations along these directions do not affect the goodness-of-fit (likelihood) at all.  We call them \emph{gauge directions}.  

Gauge directions arise from the class of gauge transformations $T_M: \mathcal{M} \to \mathcal{M}$ acting on the space of explicitly represented gate sets.  They were defined by Eq.~\ref{eq:GaugeTransform} and we repeat them here:
\begin{eqnarray}
\sbra{E^{(m)}_i} &\to& \sbra{E^{(m)}_i} M^{-1} \nonumber \\
\sket{\rho^{(i)}} &\to& M \sket{\rho^{(i)}} \nonumber \\
G_i &\to& M G_i M^{-1},
\end{eqnarray}
where $M$ is any invertible superoperator.  In this appendix, we refer to these as \emph{matrix-space gauge transformations}.  Such transformations map one gate set in $\mathcal{M}$ to another without changing any observable probability (see Eq.\ref{eq:generalFreqToProb}).  The set of transformations $T_M$ forms a group isomorphic to the Lie group $\mathrm{GL}(d^2)$, as $T_M \cdot T_{M'} = T_{MM'}$ and $\left(T_M\right)^{-1} = T_{M^{-1}}$.  We call $M \in \mathrm{GL}(d^2)$ an element of the ``gauge group'' $\mathrm{GL}(d^2)$.  The action of $T_M$ can be pulled back by the gate set model's mapping ($W$) to act on the parameter space $\mathcal{P}$.  This pullback is not always a well-defined function as $T_M$ does not, in general, respect the constraints on $\mathcal{M}$ imposed by $W$.   When it \emph{is} well-defined, we call the resulting pulled back transformation on parameter space a \emph{parameter-space gauge transformation}.  In this appendix, the unqualified term ``gauge transformation'' will always refer to a parameter-space transformation and $T_M: \mathcal{M} \to \mathcal{M}$ will be referred to as a ``matrix-space gauge transformation''.

In the case of the fully parameterized model, $\mathcal{P}$ is isomorphic to $\mathcal{M}$, and the gauge transformations are identical to the matrix-space gauge transformations of Eq.~\ref{eq:GaugeTransform}.

In the case of a TP parameterized model, the gauge transformations are given by restricting Eq.~\ref{eq:GaugeTransform} to the parameterized elements of $\mathcal{M}$ (i.e.~excluding the first row of $G_i$ and the first element of $\sket{\rho^{(i)}}$) and constraining $M$ to be TP (i.e., its first row should be $[1,0,0,\ldots 0]$).  The gauge transformations of the TP model form a subgroup of the matrix-space gauge transformations, which is isomorphic to the subgroup of $\mathrm{GL}(d^2)$ comprised of all $d \times d$ matrices whose first row is $[1,0,0,\ldots 0]$.  This nice structure results in a well-defined gauge group on the parameter space.

When dealing with a CP (or CPTP) parameterized model, there is no such structure and the set of gauge transformations that respect the model's constraints varies from point to point in parameter space.  The CP constraint does \emph{not} correspond to a constraint of $M \in \mathrm{GL}(d^2)$ to one of its subgroups, and so there is no well-defined gauge group.  In this sense, we say the CP constraint ``does not play nicely with the gauge''.  In this scenario, it may be possible to identify a gauge \emph{subgroup} that does respect the constraints.  For example, consider restricting the set of all matrix-space gauge transformations to those for which $M$ corresponds to a \emph{unitary} action on $\mathcal{H}$.  Since these transformations respect the CP and TP constraints and have a group structure isomorphic to the unitary group, then form a proper gauge subgroup.

Each gauge direction at point $p\in\mathcal{P}$ corresponds to a degree of freedom within the class of gauge transformations.  For example, in a single-qubit fully parameterized gate set model, the gauge transformations are in one-to-one correspondence with the invertible $4\times 4$ matrix $M$ of Eq.~\ref{eq:GaugeTransform}.  So there are 16 independent gauge directions at every point in parameter space, defined by varying each of the elements of $M$.  In a single-qubit TP-parameterized model there are only 12 such gauge directions because of the restriction that $M$ be TP.  In the CP (CPTP) case there are 16 (12) gauge directions at every point in parameter space \emph{except} those corresponding to gate sets where at least one gate lies on the boundary of CP space.  At such points, the gauge freedom is constrained, and there are fewer gauge directions.\footnote{There may fewer than 16 and 12 gauge parameters in special circumstances when there are classes of would-be gauge transformations which commute with all of the gates and SPAM elements.}

As we stated in the main text, the \emph{number} of gauge directions is fixed almost everywhere by the gate set model, but the gauge directions themselves vary from one parameter-space point to another.  The gauge freedom thus gives rise to curved ``gauge manifolds'' (or orbits) within parameter space whose points all yield \emph{identical} probabilities and thus identical fits to any data.  If $S: \mathcal{P} \to \mathcal{P}$ is a gauge transformation and $\vec{\theta}$ is a vector of gate set parameters, then the gate sets $W(\vec{\theta})$ and $W(S(\vec{\theta}))$ are \emph{physically indistinguishable} ($W: \mathcal{P} \rightarrow \mathcal{M}$ is the gate set model's mapping).  This situation is similar to that found in electromagnetism and quantum field theory, where the natural mathematical description of an entity is over-complete or redundant.  

\begin{figure}
    \begin{center}
    \includegraphics[width=3.5in]{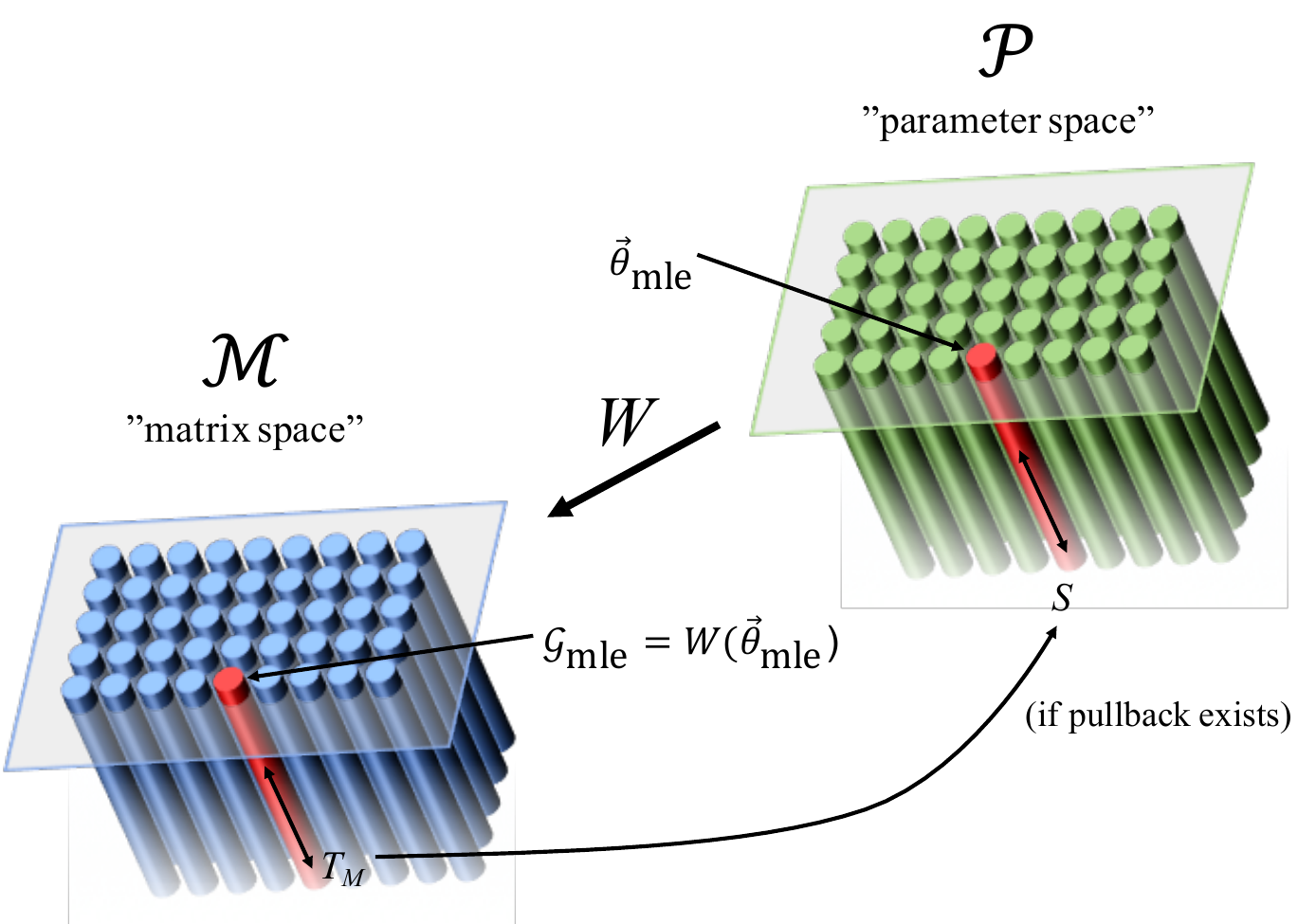}
    \caption{\textbf{Matrix and Parameter spaces.}  The matrix space $\mathcal{M}$ of gate sets has an inherent fiber bundle structure as described in the text.  The gate set model's mapping $W:\mathcal{P} \to \mathcal{M}$ can be used to pull back this structure onto the parameter space $\mathcal{P}$.  Each fiber in $\mathcal{M}$ consists of a gauge-equivalent set of gate sets, and movement within these fibers is accomplished by the matrix-space gauge transformations $T_M$ (Eq.~\ref{eq:GaugeTransform}).  Movement within the fibers of $\mathcal{P}$ are accomplished by ``gauge transformations'' as referred to by the text.  Since the goodness-of-fit between a gate set and data is invariant under gauge transformations, the ``best-fit'' gate set, $\mathcal{G}_{MLE}$ is really a best-fit fiber.\label{fig:spaces}}
    \end{center}
\end{figure}

The parameter space, because it is foliated into gauge manifolds, takes the structure of a \emph{fiber bundle}.  Fibers are given by the local gauge directions, and the base manifold given by the perpendicular non-gauge directions.  Importantly, the base manifold depends on the metric chosen for $\mathcal{P}$.  In essence (see Figure \ref{fig:spaces}), $\mathcal{P}$ can be viewed as a space of physically distinguishable fibers of equivalent gate sets (or, more precisely, their pre-images).  In this picture, ``fixing a gauge'' means choosing a representative point from each of the fibers to define a set (which may be a manifold, if the choice is performed in a smooth way) of physically distinguishable gate sets.  When the gate set model yields gauge transformations with a group structure -- i.e., when there is a well-defined gauge group -- the fibers correspond to orbits of the gauge group.

Let us consider the tangent space $\mathcal{T}_q$ at a given point $q \in \mathcal{M}$.  Recall that the dimension of $\mathcal{M}$  is given by $N_{\mathrm{e}}$ (Eq.~\ref{eq:numGatesetElements}).  $\mathcal{T}_q$ is spanned by the derivatives with respect to some basis for $\mathcal{M}$.  The derivatives of the infinitesimal gauge transformations will span a subspace of $\mathcal{T}_q$, which we call the \emph{gauge subspace} at $q$.  Let us write an element of the gauge group $\mathrm{GL}(d^2)$ (see Eq.~\ref{eq:GaugeTransform}) as
\begin{equation}
  M = \exp(K),\label{eq:gaugeGroupElAsExp}
\end{equation}
where $K$ is a $d^2 \times d^2$ real matrix.  The space of infinitesimal gauge transformations is spanned by the operations
\begin{eqnarray}
\sbra{E^{(m)}_i} &\to& \sbra{E^{(m)}_i}(I-K) \nonumber \\
\sket{\rho^{(i)}} &\to& (I+K)\sket{\rho^{(i)}} \nonumber \\
G_i &\to& G_i + [K,G_i] \label{eq:InfGaugeTransform},  
\end{eqnarray}
which can be found by inserting Eq.~\ref{eq:gaugeGroupElAsExp} into Eq.~\ref{eq:GaugeTransform} and keeping only first order terms in $K$.  Square brackets in Eq.~\ref{eq:InfGaugeTransform} denote the commutator.  Taking a derivative with respect to each element $K_{jk}$ yields the gauge subspace directions
\begin{eqnarray}
& & -\sbra{E^{(m)}_i} u_{jk}, \nonumber \\
& & u_{jk} \sket{\rho^{(i)}}, \,\mbox{and} \nonumber \\
& & \left[ u_{jk},G_i \right] \label{eq:GaugeDirections},
\end{eqnarray}
where $u_{ij}$ is the matrix whose $i,j$-th element equals $1$ and whose other elements are zero.  These gauge directions can be viewed as length-$N_{\mathrm{e}}$ vectors, $\vec{dQ}_{ij} = \frac{\mathrm{d}T_M}{\mathrm{d}K_{ij}} \in \mathcal{T}_q$, where $i$ and $j$ range between $1$ and $d^2$.  Let $dQ$ be the $N_{\mathrm{e}} \times d^4$ matrix with the $\vec{dQ}_{ij}$ as its columns, and let $dP$ be the $N_{\mathrm{e}} \times N_{\mathrm{p}}$ Jacobian of $W$ at point $p\in\mathcal{P}$.  The first $N_{\mathrm{p}}$ components of each vector in a basis for the nullspace of $\left[ dP | dQ \right]$ give a basis for the pullback of the gauge subspace to $\mathcal{T}_{\vec{\theta}}$, the tangent space at $\vec{\theta} \in \mathcal{P}$.  That is, the columns of $\mathbb{P}$, where
\begin{equation}
  \left[ \begin{array}{c} \mathbb{P} \\ \hline \mathbb{Q} \end{array} \right] \equiv \mathrm{nullspace}(\left[ dP | dQ \right]),
\end{equation}
form a basis for the gauge space at $\vec{\theta}$ and projection onto this space is given by the action of the projector
\begin{equation}
\mathbb{P}_{\mathrm{gauge}} = \mathbb{P} \left(\mathbb{P}^T \mathbb{P}\right)^{-1} \mathbb{P}^T.\label{eq:gaugeSpaceProjector}
\end{equation}

We call the number of gauge directions at each point (equivalently the dimension of the fibers) the ``number of gauge parameters'', and denote it $N_{\mathrm{p}}^{\mathrm{gauge}}$.  The number of remaining dimensions in parameter space,  $N_{\mathrm{p}}^{\mathrm{nongauge}} = N_{\mathrm{p}}-N_{\mathrm{p}}^{\mathrm{gauge}}$ (also the dimension of the base manifold) is called the ``number of non-gauge parameters''.  Note that while the gauge directions generally vary from point to point, the number of total ($N_{\mathrm{p}}$), gauge ($N_{\mathrm{p}}^{\mathrm{gauge}}$), and non-gauge ($N_{\mathrm{p}}^{\mathrm{nongauge}}$) parameters are globally fixed almost everywhere by our common gate set models (full, TP, and CP) and can be treated as constants.

Gauge degrees of freedom are very relevant to the central task of parameter estimation because at any point in parameter space there is no difference in the goodness of fit along the gauge directions (i.e. along the gauge fiber).  This means that though we naively desire a best-fit \emph{point} in parameter space (i.e. knowledge of all the parameters), the data can at most determine a best-fit \emph{fiber} (a point of the base manifold).  GST data analysis (e.g., MLE) seeks to find a gate set whose parameters mark a point within the best-fit fiber: the data cannot distinguish between points within a fiber.  Any specific best-fit gate set necessarily assigns values to all $N_{\mathrm{p}}$ parameters, but only $N_{\mathrm{p}}^{\mathrm{nongauge}}$ linear combinations of those parameters are actually physically meaningful, or can have finite error bars, as discussed in Section \ref{sec:GaugeInErrorBars}.  Gauge transformations, which act as automorphisms of each gauge fiber, may be used to map the initial best-fit gate set to another physically identical one.  We say that two gate sets mapped to each other under gauge transformations are \emph{gauge-equivalent}.

This gauge ambiguity -- the inability to pin down $N_{\mathrm{p}}^{\mathrm{gauge}}$ parameters of a gate set -- is unavoidable and stems from a relativism intrinsic to gate sets (as discussed in the main text).  For example, there is no unique matrix representation for a $X(\pi/2)$ gate.  Instead an $X(\pi/2)$ gate is defined by its behavior \emph{relative} to the other gates, state preparations, and effects which comprise the gate set.  This is related to the standard freedom to choose a basis in linear algebra; the fundamental issue here is that the data do not and cannot choose a single preferred basis (i.e.~reference frame).

Gauge ambiguity makes it difficult to compare two gate sets, since two distinct gate sets corresponding to different points in $\mathcal{P}$ may actually be physically equivalent.

\textbf{Example.} Consider the gate set $\mathcal{G} = \{\rho=\proj{0},E=\proj{1},G_1=G_X\}$, where $G_X: \rho \to \sigma_x\rho\sigma_x$.  Any unitary change of basis is a valid gauge transformation, so $\mathcal{G}$ is gauge-equivalent to $\mathcal{G}' = \{\rho'=\proj{+},E=\proj{-},G_1 = G_Z\}$, where $G_Z: \rho \to \sigma_z\rho\sigma_z$.  If an experimentalist \emph{intended} to produce $\mathcal{G}$, but tomography reports the (equivalent) gate set $\mathcal{G}'$, then by all the obvious metrics the device is wildly out of spec.  Of course, this is merely a misunderstanding resulting from different choices of reference frame.  To avoid such communication failures, it is necessary that all parties agree on a shared gauge.

\subsection{Gauge considerations in error bars\label{sec:GaugeInErrorBars}}
Following Section \ref{sec:GaugeFixingHessian}, suppose we compute the Hessian of the loglikelihood at a given point in parameter space.  In the absence of gauge freedoms the Hessian would be full rank.  In our case, however, the presence of $N_{\mathrm{p}}^{\mathrm{gauge}}$ gauge degrees of freedom mean that at the MLE there are $N_{\mathrm{p}}^{\mathrm{gauge}}$ directions in parameter space along which the loglikelihood has no (zero) curvature.  Stated equivalently, the $N_{\mathrm{p}} \times N_{\mathrm{p}}$ Hessian matrix will have $N_{\mathrm{p}}^{\mathrm{gauge}}$ zero eigenvalues.  If one were to proceed with the Hessian-based confidence region formalism, ignoring the gauge freedoms, trouble would arise when needing to invert a rank-deficient Hessian.

As usual, the gauge freedom gives us the opportunity to make a choice -- namely how to fix the gauge.  From the point of view of uncertainty analysis,  we don't need to have any uncertainty about the position along the gauge directions because we can fix it.  The data tell us nothing about the gauge directions because they are physically meaningless, yet because they are meaningless we may choose them however we want, and therefore know them completely.

Practically this means we may ignore the the Hessian components along the gauge directions and only consider those along the ``non-gauge'' directions, defined as those orthogonal to the gauge directions.  However, \emph{orthogonality is metric-dependent}.  Our choice of a parameter-space metric will determine which directions are considered to be the non-gauge directions, and therefore will influence the Hessian and consequently any region estimates and error bars.  Recognizing this, and dealing with it, is critical for assigning sensible error bars.

\begin{figure}
  \begin{center}
    \includegraphics[width=1.75in]{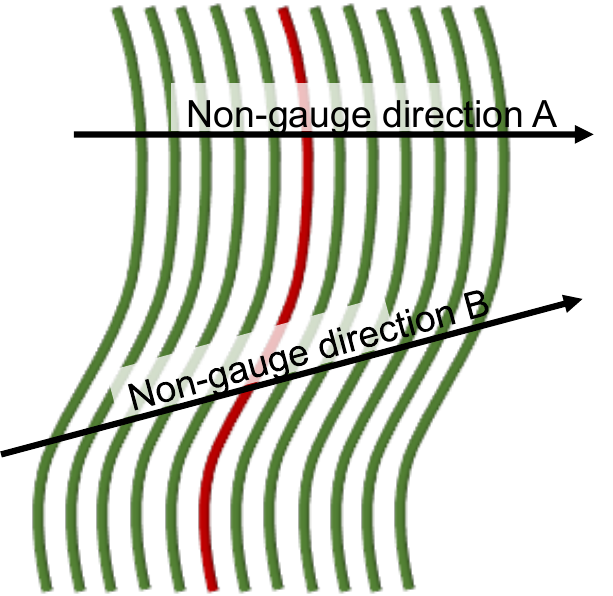}
    \caption{Diagram representing different choices of the non-gauge directions used to define the space on which error bars are computed.  Lines represent different gauge orbits (fibers), with the central red line indicating the maximum-likelihood (ML) orbit.  Computing Hessian-based error bars requires that we fix the gauge not just within the ML orbit but in a \emph{region} around this orbit (so we can project the Hessian and remove its zero eigenvalues).  Performing this regional gauge fixing can be done in many ways, as it amounts to selecting a point within each fiber; the arrows give two such examples.\label{fig:gaugeSlicing}}
  \end{center}
\end{figure}

To visualize how error bars are affected by the choice of a metric, let us return to the fiber-bundle picture of parameter space presented in Section \ref{sec:GaugeDOF} and Figure \ref{fig:spaces} above.  The MLE, a single point in parameter space, is an arbitrary point within the fiber corresponding to the best-fit equivalence class of gate sets.  The gauge directions are well-defined: they are the directions given by Eq.~\ref{eq:GaugeDirections} along which the loglikelihood is invariant, and they form a $N_{\mathrm{p}}^{\mathrm{gauge}}$-dimensional space.  However, there are many different $N_{\mathrm{p}}^{\mathrm{nongauge}}$-dimensional \emph{non-gauge} spaces which are linearly independent from the gauge space, and would be considered orthogonal to the gauge space in \emph{some} metric.  As Figure \ref{fig:gaugeSlicing} illustrates, choosing this space is like choosing a specific angle/way of cutting through the fiber bundle.  by choosing a specific non-gauge space, we are essentially choosing how much movement along the fibers (the gauge directions) should be coupled to movement between the fibers.

Almost any set of $N_{\mathrm{p}}^{\mathrm{nongauge}}$ independent directions, or equivalently almost any metric, will yield a valid non-gauge space.  Moreover, there is no ``correct'' metric: we are free to choose whichever we please.  But our choice will affect the error bars.  So, to choose \emph{wisely}, we may perform an optimization over metrics, minimizing a weighted sum of the size of the error bars on 1) the matrix elements of each superoperator representation, and 2) the vector elements of each state preparation and measurement effect.

An alternative way of choosing a metric involves partitioning the parameters into ``gate'' and ``SPAM'' groups.  We then define each group's ``intrinsic error'' as the minimum average size of the error bars on parameters in that group when other groups' error bars are ignored (allowed to be large).  We then choose the metric which weights the parameters of the different partitions according to the ratios of their intrinsic error rates.

Once a metric is selected, the Hessian can be projected onto the corresponding non-gauge space by constructing a projector using an expression identical to Eq.~\ref{eq:gaugeSpaceProjector}, but where the columns of $\mathbb{P}$ are \emph{non-gauge} rather than gauge directions.

\section{The extended LGST (eLGST) protocol\label{sec:eLGST}}


The extended-LGST (eLGST) protocol was an important milestone on the way to the long-sequence GST protocol described in the main text.  As we state in Section \ref{sec:HistoricalLSGST}, it utilizes structured circuits based on repeated germs (similar to long-sequence GST), but uses LGST in combination with a least-squares objective function to arrive at its final estimate (instead of MLE used in long-sequence GST).  As such, it constitutes an interesting fossil - an extinct intermediate step - that, despite being of little practical use, is worth outlining here.

The eLGST analysis proceeds as follows.  Let $\{g_i\}$ be the set of germ circuits, and let the data constitute LGST experiments on base circuits of the form $g_i^{l_j}$ -- that is, $l$ repetitions of each germ circuit, for some set of integers $\{l_j\}$.  Each circuit is repeated $N$ times to yield data.  

First, LGST is used for each base circuit, to obtain a reasonably accurate estimate of $\seqaction(g_i)^{l_j}$ for that base circuit.  We denote this estimate by $\widehat{\seqaction(g_i)^{l_j}}$, and assume that
\begin{equation}
  \widehat{\seqaction(g_i)^{l_j}} = \seqaction(g_i)^{l_j} \pm \frac{O(1)}{\sqrt{N}}.
\end{equation}
Now, if the $g_i$ were just numbers (not matrices), then we could get a high-precision estimate by simply taking the low-precision estimate $\widehat{\seqaction(g_i)^l}$, for some very large $l$, and computing its $l$th root,
\begin{equation}
  \widehat{\seqaction(g_i)} = \left(\widehat{\seqaction(g_i)^l}\right)^{1/l} = \seqaction(g_i) \pm \frac{O(1)}{l\sqrt{N}}.\label{eq:elgstScaling}
\end{equation}
Getting this same result when $g_i$ is a matrix is a bit more complicated, because matrix roots are very unstable.  To make this concept work for matrices, we need to 1) iteratively increase $l$ and 2) avoid explicitly taking $l$th roots of matrices.  These technical points are discussed below.  The following iterative algorithm is used in eLGST to get a high-precision estimate of $g_i$:

\begin{enumerate}
\item Perform LGST on base circuits of the form $g_i^l$ for $l=1,2,4,8,16\ldots l_{\mathrm{max}}$ to obtain approximate estimates $\widehat{\seqaction(g_i)^{l_j}}$, where $i$ indexes the germ circuit.
\item For each $g_i$, set the \emph{initial} $\widehat{\seqaction(g_i)}$ equal to $\widehat{\seqaction(g_i)^1}$.
\item Use least-squares optimization to find the $\widehat{\seqaction(g_i)}$ that minimizes
\begin{equation}
\delta_2 = |\widehat{\seqaction(g_i)} - \widehat{\seqaction(g_i)^1}|_2^2 + |\widehat{\seqaction(g_i)}^2 - \widehat{\seqaction(g_i)^2}|_2^2.
\end{equation}
\item Repeat Step 3 with a new cost function $\delta_4 = \delta_2 + |\widehat{\seqaction(g_i)}^4 - \widehat{\seqaction(g_i)^4}|_2^2$.
\item Continue repeating Step 3 with cost functions $\delta_8,\delta_{16},\ldots \delta_{l_{\mathrm{max}}}$ until all data has been incorporated.
\item Use the same approach to extract high-precision estimates of the gates ($G_k$) from the estimates of the germs ($\seqaction(g_i)$).
\end{enumerate}
This algorithm was found to be reliable and accurate when tested on simulated data, and thus inspired the current stable algorithm used for long-sequence GST.

The problems with matrix roots mentioned around Eq.~\ref{eq:elgstScaling} are worth a brief discussion.  The first problem is demonstrated even by the simple case where $\seqaction(g_i)$ is a complex scalar, $\seqaction(g_i) = e^{i\theta}$. Now $\seqaction(g_i)^l = e^{i(l\theta\ \mathrm{mod} 2\pi)}$, and so the $l$th root is multi-valued (i.e., $\theta \to \theta + n\frac{2\pi}{l}$ leaves all observable probabilities invariant).  We have to choose the right \emph{branch}.  This is impossible without further information.

We solve this problem by iteratively increasing $l$ from 1 to its final value logarithmically.  For example, if initially $l=32$, we would include the base circuits for $l=1,2,4,8,16,32$.  The additional cost is only logarithmic in maximum $l$, and as long as $N$ is large enough, it completely solves the problem of choosing the correct branch.  We begin by using the $l=1$ data to get a decent estimate of $\seqaction(g_i) \pm O(1)/\sqrt{N}$.  Then, we use the $l=2$ data to deduce that
\begin{equation}
  \seqaction(g_i) \approx \pm\sqrt{\widehat{g_i^2}},
\end{equation}
and use the $l=1$ estimate to identify unambiguously which of the two branches indicated by the $\pm$ symbol is correct.  We repeat this process recursively for each successively larger $l$.

The second problem is peculiar to matrices.  The procedure given above works very well for scalar $\seqaction(g_i) = e^{i\theta}$, but not for matrix-valued $\seqaction(g_i)$.  To see this, consider the example of
\begin{equation}
  \seqaction(g_i) = \sigma_Z.
\end{equation}
Clearly, $\seqaction(g_i)^2 = \Id$.  If we perform LGST on $\seqaction(g_i)^2$, we will generally obtain $\widehat{g_i^2} = \Id \pm O(1)/\sqrt{N}$, where the error term is a small, random matrix.  Suppose (without loss of generality) that we get $\widehat{g_i^2} = \Id + \epsilon \sigma_x = \epsilon\sigma_x$, where $\sigma_\cdot$ denotes a standard Pauli matrix.  There are multiple square roots of $\widehat{g_i^2}$, but this isn't the main problem; the main problem is that none of them are even \emph{close} to the true value of $\seqaction(g_i) = \sigma_z$!  Instead, every square root of $\epsilon\sigma_x$ is diagonal in the $\sigma_x$ basis.

The root of the problem is that the matrix square-root function is highly non-smooth, and can rip apart the topology of matrices.  To fix it, we observe that we should be looking for a $\widehat{g_i}$ such that $\widehat{g_i}^2 \approx \widehat{g_i^2}$ -- \emph{not} such that $\widehat{g_i} \approx \sqrt{\widehat{g_i^2}}$.  These two equations are not equivalent due to the topological violence that can be hiding in the matrix root.

These tricks are implemented in the final version of eLGST given above, and were pivotal to the success of eLGST as a protocol.

\section{Addressing overcompleteness in LGST with pseudo-inverses \label{sec:LGSTovercomplete}}

Although using exactly $N_{f1} = N_{f2} = d^2$ fiducial states and effects makes the mathematics of LGST convenient, there are practical advantages to using more than $d^2$ fiducials.  They include
\begin{itemize}
\item Redundancy: overcomplete fiducials provide more information and allow self-consistency checking, 
\item Precision: with some gate sets, such as Clifford gates, optimal fiducial sets (2-designs) can only be generated with more than $d^2$ fiducials,
\item Feasibility: if only projective measurements with $d$ effects are used, then at least $d+1$ of them are required for informational completeness, which yields $d^2+d$ effects.
\end{itemize}

Overcomplete fiducials yield vector spaces of observable probabilities whose dimension exceeds that of Hilbert-Schmidt space.  The $A$, $B$, and (generally) $P_k$ matrices are not square, and therefore not invertible.  However, all the steps from the previous derivation still hold -- they just require slightly more complicated linear algebra.

Let us return to the derivation at Eq.~\ref{eq:PeqAGB}.  We have measured all the $N_{f1}\times N_{f2}$ elements of $P_k$ and $\tilde{\Id}$, and we want to solve for $G_k$ in these two equations:
\begin{eqnarray*}
P_k &=& AG_kB \\
\tilde{\Id} &=& AB.
\end{eqnarray*}
The elements of $P_k$ and $\tilde{\Id}$ will experience small random perturbations because of finite sample errors in estimates of probabilities.  So in general, these equations will have no exact solutions -- the rank of $P_k$ and $\Id$ will be higher than $d^2$, but $AB$ can only have rank $d^2$ (because we demand reconstructed gates of this dimension).  Therefore, we seek the best \emph{approximate} solution.  A simple and elegant criterion for ``best'' (though not actually \emph{the} best criterion; see ``Maximum likelihood estimation'' below) is least-squares.  We seek the $G_k$ (and also $A$ and $B$) that minimize $|P_k - AG_kB|^2$ and $|\tilde{\Id}-AB|^2$.

To find $A$ and $B$, we start with the identity
\begin{equation}
|\tilde{\Id}-AB|^2 = \Tr\left[\tilde{\Id}^T\tilde{\Id} - \tilde{\Id}^TAB - B^TA^T\tilde{\Id} + B^TA^TAB\right],
\end{equation}
differentiate with respect to $A$ (using the rule that $\nabla_A\Tr[AX] = X^T$ and $\nabla_A\Tr[A^TX] = X$), set the resulting gradient to zero and solve.  This yields
\begin{equation}
A = \tilde{\Id}B^T(BB^T)^{-1}.\label{eq:LSQ_A}
\end{equation}
Some useful observations about this expression:
\begin{itemize}
\item This is a valid generalization of the expression derived previously ($A = \tilde{\Id}B^{-1}$), because $B^T(BB^T)^{-1}$ is a right inverse of $B$ -- i.e., multiplying it on the left by $B$ yields $\Id$.
\item Furthermore, if $B$ is square and invertible, then this inverse is unique, so this expression reduces to $A = \tilde{\Id}B^{-1}$ in that case.
\item This inverse is well-conditioned. Although $B$ is generally singular, $BB^T$ is square and full-rank (by informational completeness).
\end{itemize}

Next, we write out $|P_k - AG_kB|^2$ the same way, take the gradient with respect to $G_k$, set it to zero, and solve for $G_k$.  This yields
\begin{equation}
G_k = (A^TA)^{-1}A^TP_k B^T(BB^T)^{-1},
\end{equation}
and substituting $A$ from Eq.~\ref{eq:LSQ_A} yields
\begin{equation}
G_k = B\left[B^T\left(B\tilde{\Id}^T\tilde{\Id}\right)^{-1}B\right]\tilde{\Id}^T P_k 
 \left[B^T(BB^T)^{-1}\right]. \label{eq:LSQG1}
\end{equation}
This, again, is a valid generalization of the expression derived earlier ($G_k = B\tilde{\Id}^{-1}P_k B^{-1}$).  If $B$ is invertible, then this entire expression reduces to the earlier one.  But this expression \emph{only} requires inverting $d^2\times d^2$ matrices that have full rank if the fiducials are informationally complete.

One problem remains; $B$ (which is still unknown) appears in the estimate.  In the previous derivation, where $B$ was assumed to be square and invertible, we showed that although the final expression for $G_k$ involved $B$, it \emph{only} determined the gauge, and had no impact on any predicted probability.

In Eq.~\ref{eq:LSQG1}, this is not \emph{quite} true.  $B$ does affect the estimate\ldots but \emph{only through its support}.  Recall that $B$'s dimensions are now $d^2 \times N_{f1}$.  If the fiducial states are informationally complete, then $B$ has rank $d^2$.  Its columns span the entire $d^2$-dimensional state space, but its rows only span a $d^2$-dimensional subspace of the $N_{f1}$-dimensional space of observable probabilities.

We can write $B$ as $B = B_0 \Pi$, where $\Pi$ is a $d^2\times N_{f1}$ projector onto \emph{some} $d^2$-dimensional subspace, and $B_0$ is an arbitrary full-rank $d^2\times d^2$ matrix.  Substituting this into Eq.~\ref{eq:LSQG1} yields a remarkable simplification:
\begin{equation}
G_k = B_0 \left(\Pi \tilde{\Id}^T\tilde{\Id}\Pi^T\right)^{-1}\left(\Pi\tilde{\Id}^T P_k \Pi^T\right) B_0^{-1}. \label{eq:LSQG}
\end{equation}
Here, $B_0$ clearly determines the gauge and \emph{only} the gauge.  $\Pi$, on the other hand, has a real effect.  What's happening here is simple:  the high-dimensional information contained in the data ($\tilde{\Id}$ and $P_k$) has to be compressed into the lower-dimensional space of the estimate $G_k$.  The estimator is a linear map, so the only way to do this is by projection, and $\Pi$ (the projector onto the support of $B$) is that projection.  So we are free to choose $B$ more or less arbitrarily (as long as it has rank $d^2$) -- but its support \emph{will} affect the estimate, while its form on that support affects only the gauge.

To choose $\Pi$ optimally, we recall that $A$ and $B$ are chosen to minimize $|\tilde{\Id}-AB|^2$.  If we write out $AB$ using Eq.~\ref{eq:LSQ_A} and the identity $B = B_0\Pi$, we get
\begin{equation}
AB = \tilde{\Id}B^T(BB^T)^{-1}B = \tilde{\Id}\Pi^T\Pi,
\end{equation}
and if we define the complement projector $\Pi_c = \Id_{N_{f1}\times N_{f1}} - \Pi^T\Pi$, then
\begin{equation}
|\tilde{\Id}-AB|^2 = \Tr\left[\Pi_c \tilde{\Id}^T\tilde{\Id}\Pi_c\right].
\end{equation}
This is uniquely minimized by choosing $\Pi$ to be the projector onto the $d^2$ right singular vectors of $\tilde{\Id}$ with the largest singular values.  With this specification for $\Pi$, Eq.~\ref{eq:LSQG} is a completely specified LGST estimator for $G_k$.

A more or less identical (but easier) calculation yields linear inversion estimates of the native states and effects, in terms of the directly observable probability vectors $\vec{R}^{(l)}$ and $\vec{Q}^{(m)}_l$:
\begin{eqnarray}
\sket{\rho^{(l)}} &=& B_0 \left(\Pi\tilde{\Id}^T\tilde{\Id}\Pi^T\right)^{-1}\Pi\tilde{\Id}^T \vec{R}^{(l)},\\
\sbra{E^{(m)}_l} &=& \left[\vec{Q}^{(m)}_l\right]^T \Pi^T B_0^{-1}.
\end{eqnarray}

\section{Implementations in \texttt{pyGSTi}\label{sec:PyGSTiImplementation}}
This appendix provides descriptions of several algorithms used in \texttt{pyGSTi}'s implementation of gate set tomography.  These are not included in the main text because they describe implementation choices that are not fundamental to GST as a protocol.  These implementation details demonstrate that various steps of the GST protocol are tractable in practice, and may be helpful to readers seeking to implement GST in computer code.

\subsection{Circuit selection\label{sec:PyGSTiExperimentSelection}}

Recall that the primary goal when selecting circuits for long-sequence GST is to amplify all possible gate errors.  More specifically, the germs must amplify all of a gate set's parameters with the exception of those linear combinations that correspond to gauge freedoms -- we want to amplify $N_{\mathrm{p}}^{\mathrm{nongauge}}$ independent parameter directions as defined in Appendix \ref{sec:GaugeDOF}.  By imposing a periodic structure on circuits (see Section \ref{sec:ExperimentSelection}), the choice of circuits becomes one of choosing (together) a set of \emph{germs} $g_i$, germ-powers $p$, and effective SPAM pairs (when fiducial-pair reduction is used, see \ref{sec:FPR}).  The native set of SPAM operations is a property of the target gate set (input by the user).  The selection of (effective-preparation, effective-measurement) pairs means selecting appropriate fiducial circuits to precede or follow the some or all of the native state preparations or measurments.  In the typical case when there is just a single native state preparation and single native POVM, this pair selection amounts to just selecting pairs of fiducial circuits.

In \texttt{pyGSTi}'s most tested (and trusted) implementation of circuit selection, the tasks of selecting fiducials, germs, and germ-powers are separated for simplicity.  After these tasks complete, one may optionally perform fiducial pair reduction.  We consider each task in turn below.

Throughout this appendix, $\mathcal{G}$ refers to a gate set model for the quantum processor at hand. So, in addition to using the notation of Eq.~\ref{eq:GatesetDef} to describe members of $\mathcal{G}$, we may also speak of $\mathcal{G}$'s parameters.  $\mathcal{G}$ should approximate the true gate set (i.e. the to-be-determined best-fit estimate) so that circuits that amplify $\mathcal{G}$'s parameters also amplify the parameters of the best-fit estimate, and likewise for informationally complete fiducial circuits.  Initially, we set $\mathcal{G}$ to be a TP-parameterized gate set containing the ideal (target) gates.  Usually the best-fit gate set is close enough to the ideal one that the experiment design resulting from this initial choice of $\mathcal{G}$ is all that is ever needed.  However it is possible, if necessary, to iterate this process, using, e.g., the initial GST estimate to generate a new experiment design which gives rise to a second GST estimate, etc.

\tocless\subsubsection{Fiducial selection}
Fiducial selection refers to the process of (independently) choosing informationally complete (IC) sets of effective state preparations and effective measurements.  That is, fiducial selection is the process that determines $\left\{H_k\right\}$ and $\left\{F_k\right\}$ in Eqs.~\ref{eq:effectiveEffects} and \ref{eq:effectivePreps}, respectively.

A set of matrices is IC if and only if it spans the vector space $\mathcal{B}(H)$.  This requires at least $d^2$ linearly independent elements.  While choosing $d^2$ random circuits would almost certainly result in an IC set, elements could likely be chosen that are ``more'' linearly independent.  This notion of an amount of linear independence is quantified by the spectrum of the Gram matrix $\tilde{\Id}$, defined (as in  Eq.~\ref{eq:Itilde}) by
\begin{equation}
\tilde{\Id}_{i,j} = \sbraket{E_i}{\rho_j}.
\end{equation}
If either $\left\{\sket{\rho'_j}\right\}$ or $\left\{\sbra{E'_i}\right\}$ fails to be IC, the Gram matrix will fail to have $d^2$ non-zero (to machine precision) singular values.  As the $d^2$-th largest magnitude singular value approaches zero this indicates that one of the sets is close to being linearly dependent, which is to be avoided.  Thus, an optimal set of fiducial circuits (one whose elements span $\mathcal{B}(H)$ as uniformly as possible) can be found in practice by maximizing the smallest of the top $d^2$ singular values of the Gram matrix over many candidate sets.  We typically consider as a candidate set all the circuits with depth below some cutoff, and have only a single native state preparation and measurement ($N_\rho = N_{\mathrm{M}} = 1$).  The precise details of how fiducial selection is done are non-essential to achieving the desired asymptotic accuracy scaling, as this only requires informational completeness.  More uniformly IC sets are beneficial because they decrease the prefactor in the the accuracy scaling (i.e.~the y-offset of the diamond-distance vs. depth series in Figure \ref{fig:LongSeqScaling}).   

For fiducial selection to succeed, one must compute the Gram matrix using good estimates of the actual gates.  When the true (or best-fit) gate set is only slightly different from $\mathcal{G}$, then the actual gates will prepare states (and effects) that are close to the expected ones -- and therefore close to uniformly IC. 
If the true gates are sufficiently far from those used in fiducial selection, this is readily detected in the singular values of the empirical Gram matrix (Eq.~\ref{eq:Itilde}), and remedied by iterating over multiple experiment-generation steps as described at the beginning of this section.  From this point forward, let us assume fiducial selection has succeeded in identifying sets of $N_{f1}$ effective preparations and $N_{f2}$ effective POVM effects.

\tocless\subsubsection{Germ Selection\label{sec:pyGSTi_germselection}}
Germ selection seeks to identify a set of low-depth circuits whose powers make useful ``operations of interest'' for tomography.  Let us begin by considering more generally a completely arbitrary set of ``operations of interest'', $\mathcal{O}$, and asking: ``What needs to be in this set to amplify all possible gate set parameters?''.

By performing the circuits that sandwich each $O \in \mathcal{O}$ between all pairs of effective state preparations and measurements, we are able to estimate the probabilities
\begin{equation}
\left\{ \sbraopket{E_j}{\seqaction(O)}{\rho_i} \right\} \,\,\mbox{for}\,\, i=1\ldots N_{f1},\,j=1\ldots N_{f2}.
\end{equation}
Naively we could simply take $\mathcal{O}$ to be the set of single gates (each a depth-1 circuit) in the gate set, and repeatedly measure the prescribed circuits many times.  These are the same circuit used by process tomography, and are, up to some additional short circuits (see Eqs.~~\ref{eq:Itilde}-\ref{eq:LGST_Q}), sufficient to characterize the gates using LGST.  However, by repeatedly measuring each circuit $N$ times we are only able estimate each gate's elements (and thereby the gate set parameters, assuming these parameters are linear in the elements) to an accuracy proportional to $1/\sqrt{N}$ (see Section \ref{sec:LongSequenceGST}).  This is just the standard stochastic error scaling arising from the $1/\sqrt{N}$ scaling of the standard deviation of the mean in a Gaussian or binomial distribution.  To achieve $1/N$ scaling, we must include deeper circuits.  Intuitively, we want to take each parameter $\theta$ (an element of $\vec{\theta}$ after fixing a basis for parameter space) in the gate set model and map it to a probability that depends on it as $p\approx p\theta$, where $p$ is (or is proportional to) the germ-power.

Consider the case of a single-qubit gate $G_{\mathrm{x}}$ that is intended to be an $x$-rotation by $\pi/2$ radians but is actually an $x$-rotation by $\theta = \pi/2 + \epsilon$, where $\epsilon$ is small.  Let us suppose that every circuit is measured $N$ times.  If $\seqaction(O) = G_{\mathrm{x}}$ then we would estimate $\theta\mod 2\pi = \pi/2 + \epsilon \pm \alpha/\sqrt{N}$ for some constant $\alpha$.  (The $\mod 2\pi$ arises because rotations by $2\pi$ are undetectable.)  If $\seqaction(O) = G_{\mathrm{x}}^2$, which is a rotation by $2\theta = \pi + 2\epsilon$, then we would estimate $2\theta \mod 2\pi = \pi + 2\epsilon \pm \alpha/\sqrt{N}$ and thus $\theta \mod \pi = \pi/2 + \epsilon \pm \alpha/(2\sqrt{N})$.  More generally, if $\seqaction(O) = G_{\mathrm{x}}^p$, then we would estimate $p\theta \mod 2\pi = p\pi/2 + p\epsilon \pm \alpha/\sqrt{N}$ and thus $\theta \mod 2\pi/p = \pi/2 + \epsilon \pm \alpha/(p\sqrt{N})$.  While it may be initially concerning that as $p$ increases the ``branch'' ambiguity in $\theta$ increases (as one is unable to discriminate between angles separated by $2\pi/p$), this issue can be circumvented by probing $G_{\mathrm{x}}^p$ for \emph{multiple} values of $p$ by choosing logarithmically-spaced depths $l$ and letting $p=\lfloor l/|g| \rfloor$ (see Section \ref{sec:SelectingGermPowers})).  One only needs logarithmically-spaced $p$ (or $l$) values to determine the correct ``branch'' for $\theta$, as each $p$ can be used to rule out a constant factor of the possible branches.  The same reasoning lies behind the Robust Phase Estimation protocol \cite{RudingerRPE_PRL2017}.

In the end, we see that by increasing $l$ the uncertainty in $\theta$ decreases as $1/l$.  Since the total number of circuits scales at most linearly with $l$ (and more often than not \emph{logarithmically} since $l$-values are logarithmically spaced), this is at worst a Heisenberg-like scaling in the total circuit number and markedly better than the $1/\sqrt{N}$ scaling that would be obtained by just increasing $N$ for the single $\seqaction(O) = G_{\mathrm{x}}$.  


This simple example suggests that we should include in $\mathcal{O}$ logarithmically spaced powers of each of the gates.  These circuits amplify \emph{some} but not \emph{all} types of gate errors (i.e.~parameters of the gate set).  Let us return to our single-qubit example, but now suppose $G_{\mathrm{x}}$ is an exact $\pi/2$ rotation around a slightly wrong axis.  This corresponds to the unitary map
\begin{equation}
G_{\mathrm{x}} = e^{-i(\pi/4)(\cos\epsilon \sigma_x + \sin\epsilon \sigma_y)},
\end{equation}
where $\sigma_x$ and $\sigma_y$ are the Pauli operators and $\epsilon$ is assumed small.  This ``tilt error'', is \emph{not} amplified by $G_{\mathrm{x}}^p$, which is easily seen by observing that $G_{\mathrm{x}}^4 = \Id$, so the error cancels, or ``echos'', itself out after just four repetitions.  More sophisticated circuits are needed to amplify tilt errors.  In this example, it is sufficient to probe $G_{\mathrm{x}}G_{\mathrm{y}}$, where $G_{\mathrm{y}}$ is a perfect single-qubit $\pi/2$ rotation around the $y$-axis.  $G_{\mathrm{x}}G_{\mathrm{y}}$ is then a rotation by $2\pi/3 + \epsilon/\sqrt{3}$.  Therefore, performing $(G_{\mathrm{x}}G_{\mathrm{y}})^p$ amplifies the deviation $\epsilon$ by a factor of $p$ and, as before, allows estimation of $\epsilon$ to an accuracy scaling as $1/(p\sqrt{N})$.  This illustrates the need for circuits other than the gates themselves which must be repeated to amplify all of a gate set's parameters.  We call each of the circuits which is repeated, $O\in\mathcal{O}$, a ``germ'' (consistent with Section \ref{sec:ExperimentSelection}).

The general situation gets rapidly complicated -- e.g., if $G_{\mathrm{y}}$ were not perfect in the above example, then $G_{\mathrm{x}}G_{\mathrm{y}}$ alone could not distinguish between $y$-axis tilt in $G_{\mathrm{x}}$ and $x$-axis tilt in $G_{\mathrm{y}}$.  In general, every circuit is sensitive to some nontrivial linear combination of gate set parameters.  To choose a set of germs, we create a list of candidate germs (e.g., every circuit shorter than some cutoff depth), and for each candidate germ $g$ identify what linear combination of parameters it amplifies.  We do this by computing the Jacobian,
\begin{equation}\nabla_g^{(p)} \equiv \frac{1}{p}\left.\frac{\partial \left[\seqaction(g)^p\right]}{\partial \vec{\theta}}\right|_{W(\vec{\theta})=\mathcal{G}}, \label{eq:Jacobian}
\end{equation}
where $\seqaction(g)$ is the $\mathcal{B}(\mathcal{H}) \to \mathcal{B}(\mathcal{H})$ map formed by composing the elements of $g$ (see Eq.~\ref{eq:sigmaDefinition}), and $\vec{\theta} \in \mathcal{P}$ is a vector of gate set parameters.  Note that we are justified in assuming we have access to this entire Jacobian -- i.e.~the derivatives of \emph{all} the elements of $\seqaction(g)^p$ -- because we have committed to sandwiching the base circuit between \emph{informationally complete sets of fiducials}.  As we will see in Section \ref{sec:SequenceReduction}, inclusion of all of these fiducial pairs is overkill and we can remove many of them.  Note also that, as in the main text, since the germs are only able to amplify errors in the gates and not in the SPAM, we will assume a gate set model whose parameters only vary its gates.

In general, $\seqaction(g)$ is a $d^2 \times d^2$ matrix and the gate set has $N_{\mathrm{p}}$ parameters ($N_{\mathrm{p}} = N_{\mathrm{G}} d^4$ in the case of fully parameterized gate set model).  This means $\nabla_g^{(p)}$ is a $d^4\times N_{\mathrm{p}}$ matrix.  Its $d^4$ right singular vectors indicate linear combinations of model parameters that $\seqaction(g)$ amplifies, and the corresponding singular values quantify how much they are amplified.  A zero singular value indicates a parameter combination that is not amplified at all (like the tilt error discussed above).  We expect each germ to amplify at least the $d^2$ linear combinations of parameters given by its eigenvectors.  It may amplify more than $d^2$ if its spectrum is degenerate and remains so for all unitary perturbations (e.g., a single-qubit unitary gate will always have two unit eigenvalues).  The amplification properties of a set of $N_{\mathrm{g}}$ germs $\left\{g_1\ldots g_{N_{\mathrm{g}}}\right\}$ is described by the $N_{\mathrm{e}} \times N_{\mathrm{p}}$ Jacobian
\begin{equation}
J^{(p)} = \left(\begin{array}{c} \nabla_{g_1}^{(p)} \\ \nabla_{g_2}^{(p)} \\ \vdots \\ \nabla_{g_{N_{\mathrm{g}}}}^{(p)} \end{array}\right).\label{eq:germJacobian}
\end{equation}

Our goal is to choose germs that provide high sensitivity at ``large'' values of $p$ (or equivalently $l$, if $p=\lfloor l/|g| \rfloor$).  In practice, it is not useful to make $l$ larger than $1/(|g|\eta)$, where $\eta$ is the rate of stochastic or depolarizing noise. To select germs, however, we ignore this effect and make the simplifying assumption that the gates (and therefore $\seqaction(g)$) are reversible.  In practice, we project the gates of $\mathcal{G}$ onto their closest unitary and consider random unitary variations of $\mathcal{G}$ to avoid the effects of accidental degeneracies.  Under the assumption of unitary gates, it is possible to define the $l\to\infty$ limit of the Jacobian in Eq. \ref{eq:Jacobian}.  Using the product rule and that $\seqaction(g)^{-1} = \seqaction(g)^\dagger$,
\begin{eqnarray}
\nabla_g^{(p)} &=& \frac{1}{p}\sum_{n=0}^{p-1} \seqaction(g)^n \frac{\partial \seqaction(g)}{\partial \vec{\theta}} \seqaction(g)^{p-1-n} \\
&=& \left[ \frac{1}{p} \sum_{n=0}^{p-1} \seqaction(g)^n \nabla_g^{(1)} (\seqaction(g)^\dagger)^n \right]\seqaction(g)^{-(p-1)}
\end{eqnarray}
As $p\to\infty$, the average over all powers $n$ of $\seqaction(g)$ \emph{twirls} $\nabla_g^{(1)}$.  By Schur's lemma, the effect of twirling is to project $\nabla_g^{(1)}$ onto the commutant of $\seqaction(g)$ -- i.e., onto the subspace of matrices that commute with $\seqaction(g)$.  Furthermore, multiplication by the unitary $\seqaction(g)^{-(p-1)}$ is merely a change of basis, and has no effect on the right singular vectors or the singular values of $\nabla_g$.  So, up to an irrelevant change of basis:
\begin{equation}
\lim_{p\to\infty}{\nabla_g^{(p)}} = \Pi_{\seqaction(g)}\left[ \nabla_g^{(1)} \right],
\end{equation}
where $\Pi_{\seqaction(g)}$ is the projection onto the commutant of $\seqaction(g)$.

This framework defines a notion of completeness for germs.  The set of germs $\left\{g_i\right\}_{i=1}^{N_{\mathrm{g}}}$ is \emph{amplificationally complete} (AC) if and only if the right singular rank of its Jacobian, the $p \rightarrow \infty$ limit of Eq.~\ref{eq:germJacobian}, equals the total number of physically accessible parameters in the gate set, $N_{\mathrm{p}}^{\mathrm{nongauge}}$ (see Appendix \ref{sec:GaugeDOF}).  To build an AC set of germs, it is sufficient to add germs, one by one, to a set until its Jacobian has rank equal to $N_{\mathrm{p}}^{\mathrm{nongauge}}$.  One may continue to optimize the set of germs numerically by adding and removing germs (taken from some master candidate list), and only keeping modifications that lower a certain score function.  The primary score function used in \texttt{pyGSTi} is

\begin{equation}
f(\left\{g_1\ldots g_{N_{\mathrm{g}}}\right\}) = \frac{\Tr\left[(J^\dagger J)^{-1}\right]}{N_{\mathrm{g}}}.
\end{equation}
This score estimates the mean squared error of estimation if a fixed number of counts are spread over the $N_{\mathrm{g}}$ distinct germs.

\tocless\subsubsection{Base circuit selection}
It remains to specify the number of repetitions (i.e.~the values of $l$ and therefore $p$) for each germ used to construct the base circuits of a GST experiment design.  We have motivated the use of logarithmically spaced germ powers (above, and in Appendix \ref{sec:eLGST}), and as described in the main text (Section \ref{sec:SelectingGermPowers}) the maximum number of repetitions should be chosen so that the \emph{depth} of the repeated germ does not exceed $L_\eta$, a rough maximum circuit depth before the output probabilities cease to yield useful information.

The main idea is to repeat each germ a logarithmically increasing number of times until the depth of the repeated germ circuit reaches $L_\eta$.  We define the set of ``maximum depths'', $\left\{l_i = m^{i-1}\right\}_{i=1}^{N_\mathrm{l}}$ (with $m=2$ usually) such that $l_{N_\mathrm{l}} \le L_\eta$, and define $l_{\mathrm{max}} = l_{N_\mathrm{l}} = \max_i l_i$ for later convenience.  Let us write the germ power $p$ formally as a function of germ and max-depth:
\begin{equation}
  p(g,l) = \lfloor l/|g| \rfloor,
\end{equation}
so that $p(g,l)$ is the maximum number of times $g$ may be repeated without exceeding depth $l$ (if $|g| > l$ then $p(g,l)=0$).  Germ $g$ is then repeated $p(g,l_i)$ times for each $i = 1 \ldots N_\mathrm{l}$.   Note that this yields slightly different base circuits from just choosing logarithmically-spaced powers ($p$) - see Figure~\ref{fig:GSTExperimentDesign} step 2.  In particular, when $|g| > 1$ we may end up including additional circuits that we would not have if $p$ was itself logarithmically spaced.  These additional circuits add to the over-completeness of the data, which we discuss later.  The reason for choosing logarithmically-spaced maximum depths ($l$) rather than exponents ($p$) is primarily so that we can, in most cases, use a single set $\{l_i\}$ for \emph{all} germs.  If a common set of powers ($p$) was used for all the germs, this would either result in wasted circuits of depth greater than $l_{\mathrm{max}}$ or the failure to repeat short germs as many times as one could.  A more flexible (but more complicated) approach is to vary $l_{\mathrm{max}}$ from germ to germ based on prior knowledge of the device's stochastic noise behavior (e.g., when two-qubit gates have more stochastic error than one-qubit gates).  \texttt{pyGSTi} allows users to specify per-germ maximum depths, but this is not the default behavior.


In the end, the complete set of \texttt{pyGSTi}-generated GST circuits is as follows.  With all the repeated germs given by $\mathcal{O} = \left\{g_i^{p(g_i,l_j)}\right\}_{i,j}$, where $\left\{l_j\right\}_{j=1}^{N_\mathrm{l}}$ is the set of $N_\mathrm{l}$ maximum depths, the probabilities corresponding to Eq.~\ref{eq:probabilities4GSTcondensed} are given by (using $p_{i,j} = p(g_i,l_j)$):
\begin{equation}
  \sbraopket{E_a}{\seqaction(g_i)^{p(g_i,l_j)}}{\rho_b},\label{eq:pyGSTi_probabilities4GSTcondensed}
\end{equation}
where $a=1\ldots N_{f1}$, $b=1\ldots N_{f2}$,  $i=1\ldots N_{\mathrm{g}}$, and $j=1\ldots N_\mathrm{l}$.  Expanding the fiducial circuits gives the probabilities in terms of just native operations (elements of the gate set),
\begin{equation}
  \sbra{E^{m(a)}_{t(a)}}\seqaction\left(H_{h(a)}\right)  \seqaction\left(g_i\right)^{p(g_i,l_j)} \seqaction\left(F_{f(b)}\right)\sket{\rho^{r(b)}}.\label{eq:pyGSTi_probabilities4GST}
\end{equation}

\tocless\subsubsection{Fiducial pair reduction.\label{sec:SequenceReduction}}
While the set of circuits corresponding to Eq.~\ref{eq:pyGSTi_probabilities4GST} achieves the goal of amplifying all of the gate set model's parameters, it isn't as efficient as it could be.  This is because we have included a complete set of fiducial pairs for each base circuit.  In this section we expand upon the description of fiducial pair reduction (FPR) in the main text (Section \ref{sec:FPR}).

FPR is especially needed when we consider scaling GST beyond a single qubit.  Moving to more qubits necessarily brings dramatic increases in the dimensionality of the parameter manifold, but the standard GST experiment design increases more rapidly than is required by the greater number of parameters.  An ambitious 1-qubit GST experiment uses 6 fiducial circuits, 11 germs, and up to 13 powers ($L=1,2,4,\ldots 8192$).  This adds up to approximately $6\times 6\times 11\times 13=5148$ circuits.  This sounds like a lot -- but a direct generalization of this method to 2 qubits would require 36 fiducial circuits and 80 germs, for roughly $1.3\times 10^6$ distinct circuits.  This is not practical.

Thankfully, this many circuits is not necessary.  As discussed in the main text, the standard experiment design contains redundancies stemming from the fact that we don't need \emph{full} tomographic information for each repeated germ ($g_i^{p(g_i,l_j)}$) circuit.  We only need to be able to extract the amplified linear combinations of parameters for each germ, which are typically much fewer ($\approx d^2$) than the $\approx d^3$ circuits that pinpoint the $\approx d^4$ elements in $\seqaction\left(g_i^{p(g_i,l_j)}\right)$.  As such, we roughly expect that we can reduce the number of circuits by a factor of $d^3/d^2 = d$ and still maintain the desired Heisenberg scaling.

One way of doing this is to simply randomly remove some fraction of the $N_{f1}\cdot N_{f2}$ circuits corresponding to each repeated-germ.  This technique seems to work in practice but has not been studied extensively.  An alternate technique is to remove a specific set of the $N_{f1}\cdot N_{f2}$ ``fiducial pairs'' (corresponding to $(a,b)$ pairs of indices in Eqs.~\ref{eq:probabilities4GSTcondensed}-\ref{eq:probabilities4GST}) for each germ.  The same set of fiducial pairs is removed for every power of the same germ.  To construct a reduced set of $M$ fiducial pairs for germ $g$ we construct the projectors $\left\{\mathbb{P}_k\right\}$ where $\mathbb{P}_k = J_k J_k^T$ and $J_k$ is the $M \times N_{\mathrm{p}}$ matrix whose $m$-th row is the derivative of $\sbraopket{E_{a_m}}{\seqaction(g)}{\rho_{b_m}}$ with respect to $k$-th amplified parameter of $g$.  The ``amplified parameters'' of $g$ are the non-zero elements of $\seqaction(g)$'s Jordan normal form, as these are the generalization of $\seqaction(g)$'s eigenvalues which include off-diagonal elements of degenerate blocks.  If the rank of $\sum_k \mathbb{P}_k$ is equal to the maximal number of parameters $g$ can amplify then the set of fiducial pairs $\left\{(a_m,b_m)\right\}_{m=1}^M$ is sufficient and may be used as the reduced set of fiducial pairs for $g$.  (The maximal number of parameters $g$ can amplify can easily be found by computing the rank when including all possible fiducial pairs.)  For each germ we find a minimal set of fiducial pairs which amplifies the maximum number of parameters, and this collection of per-germ fiducial pairs dictates a reduced set of circuits which should maintain the same desired accuracy scaling.  We find this to be true in practice, but that the FPR process also makes the parameter estimation less robust to data that does not fit the gate set model.  The redundancy, as one might expect, serves to stabilize the numerical optimization used in the parameter estimation portion of GST.

It is also possible to consider different maximum-depth sets $\left\{l_i\right\}$ for different germs, so that the number of times each each germ is repeated is strictly a logarithmic sequence.  We expect that this would similarly lead to GST maintaining the desired scaling on data that can be fit well but that it could lead to an overall lack of robustness to data which cannot be fit well.  We have not performed simulations to verify this hypothesis, and leave this as a subject of future work. 

\subsection{Non-TP loglikelihood optimization\label{sec:NonTPLogLEstimation}}
If the gate set is not TP, then the expression for the loglikelihood in Eq.~\ref{eq:perExperimentLogL} above must be modified.  In doing so, we will make use of a general relationship between the Poisson and Multinomial distributions.
Consider a multinomial distribution with event probabilities $\left\{n_j\right\}$ and number of samples $N$.  Now suppose the number of samples is no longer fixed but instead is Poisson distributed with mean $\Lambda=N$.  This results in the previously multinomially-distributed $n_j$ becoming independently Poisson-distributed with rates $\lambda_j = n_j \Lambda = n_j N$.  This is easier to see in the reverse direction, starting from independent Poisson-distributed $n_j$, and post-selecting on $N = N_0$ where $N_0$ is a fixed number.  This introduces correlations among the $n_j$, and the $n_j$ become multinomially distributed. 

To see this, suppose we have $K$ Poisson-distributed random variables $n_{j}$, each with its own rate parameter $\lambda_{j}$:
\begin{equation}
  \mathrm{Pr}(n_{j}) = \frac{e^{-\lambda_{j}}\lambda_{j}^{n_{j}}}{n_{j}!}.
\end{equation}
In turn, the random variable $X \equiv \sum_{j}n_{j}$ is Poisson distributed, with rate parameter $\lambda \equiv \sum_{j}\lambda_{j}$:
\begin{equation}
  \mathrm{Pr}(X = N) = \frac{e^{-\lambda}\lambda^{N}}{N!}.\label{eq:PoissonXN}
\end{equation}
Further, since the $n_{j}$ are independent, the probability of observing a particular \emph{set} of values $\{n_{j}\}_{j=1}^{K}$ is simply the product of the individual probabilities:
\begin{equation}
  \mathrm{Pr}(\{n_{j}\}) = \prod_{j}\mathrm{Pr}(n_{j}) = e^{-\lambda}\prod_{j}\frac{\lambda_{j}^{n_{j}}}{n_{j}!}.
\end{equation}
What is the probability of observing this particular set of values, conditioned on $X = N_{0}$? By the definition of conditional probability, this is given by
\begin{equation}
  \mathrm{Pr}(\{n_{j}\}~|~X = N_{0}) =\frac{\mathrm{Pr}(\{n_{j}\} \cap \{X = N_{0}\})}{\mathrm{Pr}(X=N_{0})}.
\end{equation}
The denominator of the above expression has already been calculated (Eq.~\ref{eq:PoissonXN}); what remains is to calculate the numerator. Notice that if the sum of the $\{n_{j}\}$ is not $N_0$, then the numerator is zero, while if their sum \emph{is} $N_{0}$, then the numerator is $\mathrm{Pr}(\{n_{j}\})$:
\begin{equation}
  \mathrm{Pr}(\{n_{j}\} \cap \{X = N_{0}\}) =\begin{cases}\mathrm{Pr}(\{n_{j}\})~~~~~\sum_{j}n_{j} = N_{0}\cr 0~~~~~~~~~~~~~~~\sum_{j}n_{j} \neq N_{0}\end{cases}.
\end{equation}
Therefore, the conditional probability is
\begin{align*}
\mathrm{Pr}(\{n_{j}\}~|~X = N_{0}) &=\left(e^{-\lambda}\prod_{j}\frac{\lambda_{j}^{n_{j}}}{n_{j}!}\right)\left(\frac{N_{0}!}{e^{-\lambda}\lambda^{N_{0}}}\right)\\
&=\frac{N_{0}!}{n_{1}!n_{2}!\cdots n_{K}!}\left(\frac{\prod_{j}\lambda_{j}^{n_{j}}}{\lambda^{N_{0}}}\right)\\
&=\frac{N_{0}!}{n_{1}!n_{2}!\cdots n_{K}!}\prod_{j}\left(\frac{\lambda_{j}}{\lambda}\right)^{n_{j}}\\ 
&=\frac{N_{0}!}{n_{1}!n_{2}!\cdots n_{K}!}\prod_{j}\left(\frac{\lambda_{j}}{\sum_{k}\lambda_{k}}\right)^{n_{j}},
\end{align*}
which is a multinomial distribution over $N_{0}$ trials, and with event probabilities $p_{j} = \frac{\lambda_{j}}{\sum_{k}\lambda_{k}}$.  Note that when moving to the third line in this derivation we used the fact $N_{0} = \sum_{j}n_{j}$.

When a gate set is not constrained to be TP, it means that the predicted outcome probabilities for a circuit need not sum to one, and equivalently that the predicted total number of counts need not equal the actual (observed) number of counts.  Assuming the predicted number of counts will be (Poisson-) distributed around the actual number, we find ourselves in the situation just described with $\left\{n_j\right\}$ corresponding to the outcome probabilities $\left\{p_{s,\beta_s}\right\}_{\beta_s}$.  We may thus think of $\lambda_{s,\beta_s} \equiv N_s p_{s,\beta_s}$ as the \emph{rate} of seeing outcome $\beta_s$ after circuit $s$ and write the likelihood for a single circuit as the likelihood for independent Poisson distributions:
\begin{eqnarray}
\mathcal{L}_s &=& \prod_{\beta_s} \frac{\lambda_{s,\beta_s}^{N_{s,\beta_s}} e^{-\lambda_{s,\beta_s}}}{ N_{s,\beta_s}! }\nonumber\\
\logL_s &=& \sum_{\beta_s} N_{s,\beta_s} \log( N_s p_{s,\beta_s}) - N_s p_{s,\beta_s} - \log( N_{s,\beta_s}! )\nonumber\\
\logL'_s &=& \sum_{\beta_s} N_{s,\beta_s} \log( p_{s,\beta_s} ) - N_s p_{s,\beta_s}
\end{eqnarray}
where in the final line we have for convenience discarded the $p_{s,\beta_s}$-independent $N_{s,\beta_s}\log(N_s)$ and $\log(N_{s,\beta_s}!)$ terms and defined $\logL'_s$.  The total loglikelihood is the sum of each $\logL'_s$, giving
\begin{equation}
\logL = \sum_{s}{ \logL'_s} = \sum_{s,\beta_s N_s} N_{s,\beta_s} \log( p_{s,\beta_s} ) - N_s p_{s,\beta_s} \label{eq:DatasetLogL}
\end{equation}
The second term in Eq.~\ref{eq:DatasetLogL} can be viewed as a Lagrange multiplier (or penalty term) which softly enforces the TP constraint that $\sum_{\beta_s} p_{s,\beta_s} = 1$ for all $s$.  Furthermore, we note that this ``Poisson-picture'' form of the loglikelihood is identical (up to an unimportant constant shift) to Eq.~\ref{eq:DatasetLogLTP} for TP gate sets, and so Eq.~\ref{eq:DatasetLogL} can be interpreted as a more general formulation which allows for both TP and non-TP gate sets.

\texttt{pyGSTi} uses this Poisson-picture formulation of $\logL$ all the time, both because it is more general (can handle cases when the gates are not TP) and because it makes the function more amenable to least-squares optimization techniques such as the Levenberg-Marquardt method.  As noted in the text (Section \ref{sec:ParameterEstimation}), \texttt{pyGSTi} also regularizes the $\logL$ functions so that evaluating the objective function at negative probabilities (as can be predicted by non-CP-constrained gate set models) is possible.

\section{Numerical verification\label{sec:NumericalVerification}}
This appendix presents a numerical verification of the accuracy scaling claims derived in the main text. Here we present evidence that
\begin{itemize}
\item the accuracy with which LGST reconstructs a gate set scales as $O(1/\sqrt{N})$, where $N$ is the number of times each circuit is run on the device (so there are $N$ sampled outcomes per circuit), and
\item the accuracy with which long-sequence GST reconstructs a gate set scales as $O(1/L)$, where $L$ is the maximum depth of the base circuits used in long-sequence GST.
\end{itemize}


Both points can be handled similarly, as they both involve testing whether the slope of a particular plot is near a target value.  Our results consist of verifying the values of such slopes over many \emph{trials} on 1- and 2-qubit systems.  Each trial consists of the following steps:

\begin{enumerate}
\item \textbf{Choose a ``true'' gate set.}  This identifies the trial, and is a slightly perturbed ideal/target gate set.  The true gate set contains imperfect operations.
\item \textbf{Simulate circuits using the true gate set.}  Circuit outcomes are generated by sampling $N$ times the multinomial distribution corresponding to the outcome probabilities predicted by the true gate set.  For LGST trials, the number of circuits is fixed and $N$ is varied to form multiple data sets each corresponding to a different value of $N$.  For long-sequence GST trials, $N$ is fixed, and data sets corresponding to different maximum base-circuit depth $L$ are created.
\item \textbf{Perform LGST or GST on the simulated data sets.}  This leads to a series of estimated gate sets -- one for each value of $N$ or $L$, respectively.
\item \textbf{Plot the accuracy of the LGST (GST) estimates vs. $N$ ($L$) on a log-log plot, and compute its best-fit slope.}  Accuracy is computed as a distance to the true gate set (see below), and quantifies how well the estimate matches the true gate set.
\end{enumerate}
In the end, each trial leads to a slope that we compare with the target value of $-1/2$ (for LGST trials) or $-1$ (for GST trials).

Trials are constructed by first selecting a \emph{perfect} gate set that has unitary gates.  We consider 6 1-qubit perfect gate sets and 5 2-qubit perfect gate sets that we commonly see in hardware.  Each 1-qubit (2-qubit) perfect gate set is perturbed 100 (10) times by applying (a) independently random depolarization to each gate, with a maximum strength of $10^{-3}$, (b) a constant SPAM error of $10^{-2}$ (1\%), and (c) independently random rotations of up to $10^{-3}$ radians about each axis ($x$, $y$, and $z$ for 1-qubit and axes corresponding to the 15 nontrivial Pauli matrices in 2-qubit cases).  These noise strengths were chosen to be realistic values for current state-of-the-art devices.  Each perturbed gate set is then used to generate a series of data sets by varying $N$ or $L$ as described above.

Quantifying the accuracy of LGST/GST is done by computing the average diamond-norm distance, denoted $\diamond_{avg}(\mathcal{G},\mathcal{G}_0)$, between the gates of the true gate set $\mathcal{G}_0$ and those of the estimate $\mathcal{G}$, where
\begin{equation}
  \diamond_{\mathrm{avg}}(\mathcal{G}_1,\mathcal{G}_2) = \frac{1}{N_G} \sum_i || G_i^{(1)} - G_i^{(2)} ||_\diamond.\label{eq:diamondAvg}
\end{equation}
In Eq.~\ref{eq:diamondAvg}, $G_i^{(1)}$ are the gates belonging to $\mathcal{G}_1$, $G_i^{(2)}$ are the gates belonging to $\mathcal{G}_2$, and the diamond norm is given by
\begin{equation}\label{eq:diamondNorm}
|| G - G_0 || _\diamond = \sup_{\rho} || (G \otimes \mathbbm{1}_d ) [\rho] - (G_0 \otimes \mathbbm{1}_d ) [\rho] ||_1,
\end{equation}
where $\rho$ ranges over all valid quantum states.

Since the diamond distance is \emph{not} gauge invariant (see Appendix \ref{sec:GaugeDOF}) we gauge-optimize the estimated gate sets to the true gate set before computing it.  If we didn't, the results wouldn't be meaningful and we would not expect to observe $\diamond_{\mathrm{avg}}$ to decrease in the expected way.  The choice of average diamond distance here is somewhat arbitrary.  If we instead used the infidelity, the Frobenius distance, or another similar well-behaved quantity measuring the difference between two gate sets, we would expect to see the same scaling behavior.

Figure \ref{fig:LGSTScaling} shows the average diamond distance vs.~$N$ for estimates obtained from LGST, and confirms the expected $1/\sqrt{N}$ scaling (slopes are near -0.5).  This is true for all the single- and two-qubit gate sets we tested. In the figure legends, single-qubit rotation gates are given by their axis and angle, e.g., $X(\pi/2)$ denotes a x-axis rotation by $\pi/2$ radians.  The $N_{\phi}$ gate is a $\pi/2$-rotation gate about the tilted $(x,y,z) = (\sqrt{3}/2, 0, -1/2)$ axis.  Two-qubit gates are given as either two letters that indicate a tensor product of single-qubit $\pi/2$ rotation gates (e.g., $XI$ means a $X(\pi/2)$ gate on qubit 1 and an idle on qubit 2), or as a commonly known CNOT or CPHASE gate.

The $O(1/\sqrt{N})$ scaling leads to an overall $O(1/\sqrt{N'})$ scaling, where $N'$ is the \emph{total} number of samples, since the number of circuits required by LGST is fixed and $N$ is proportional to $N'$.

\begin{widetext}

\begin{figure}[H]
  \begin{center}
    \begin{tabular}[c]{cc}
      \includegraphics[width=3.5in]{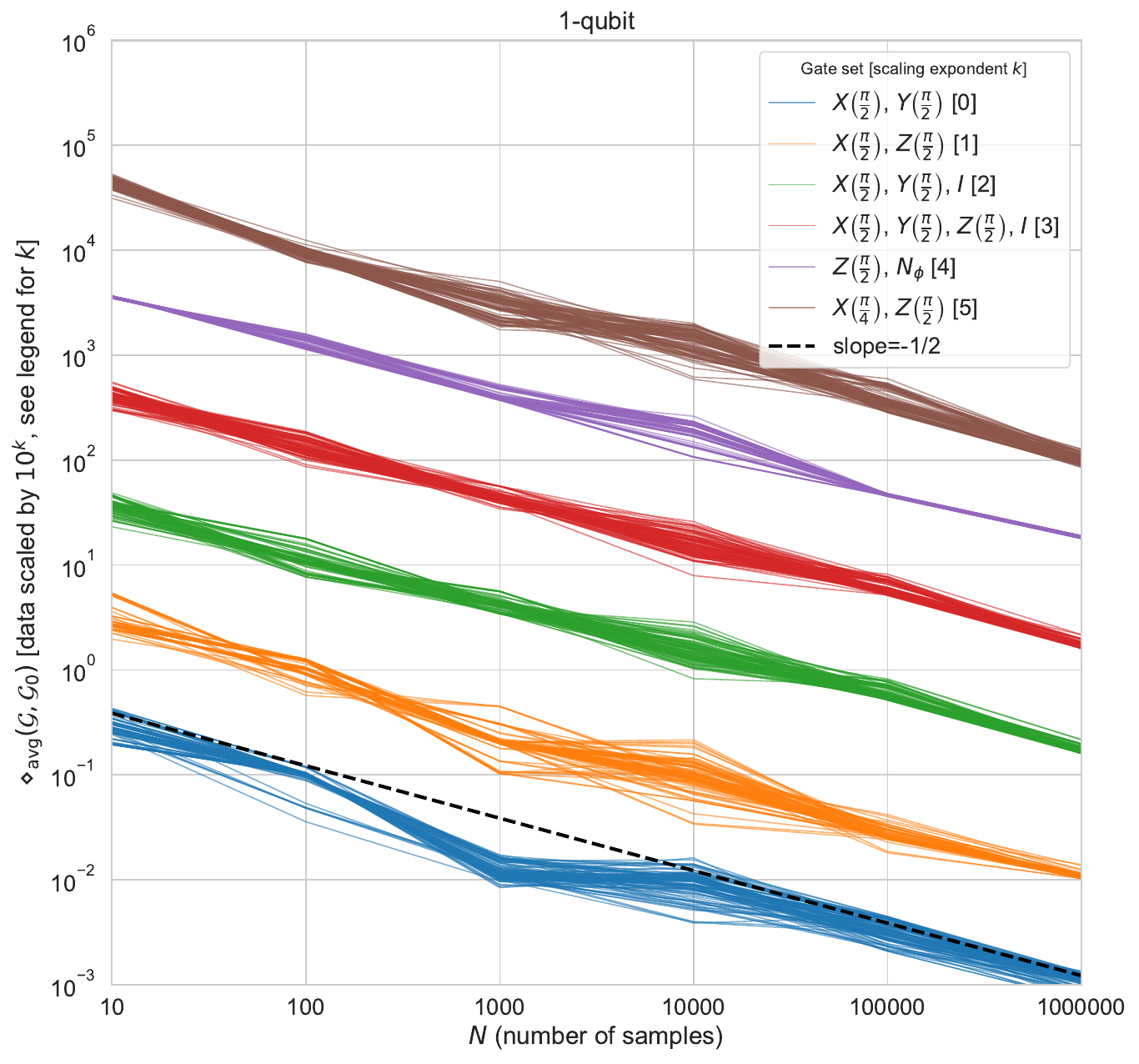} &
      \includegraphics[width=3.5in]{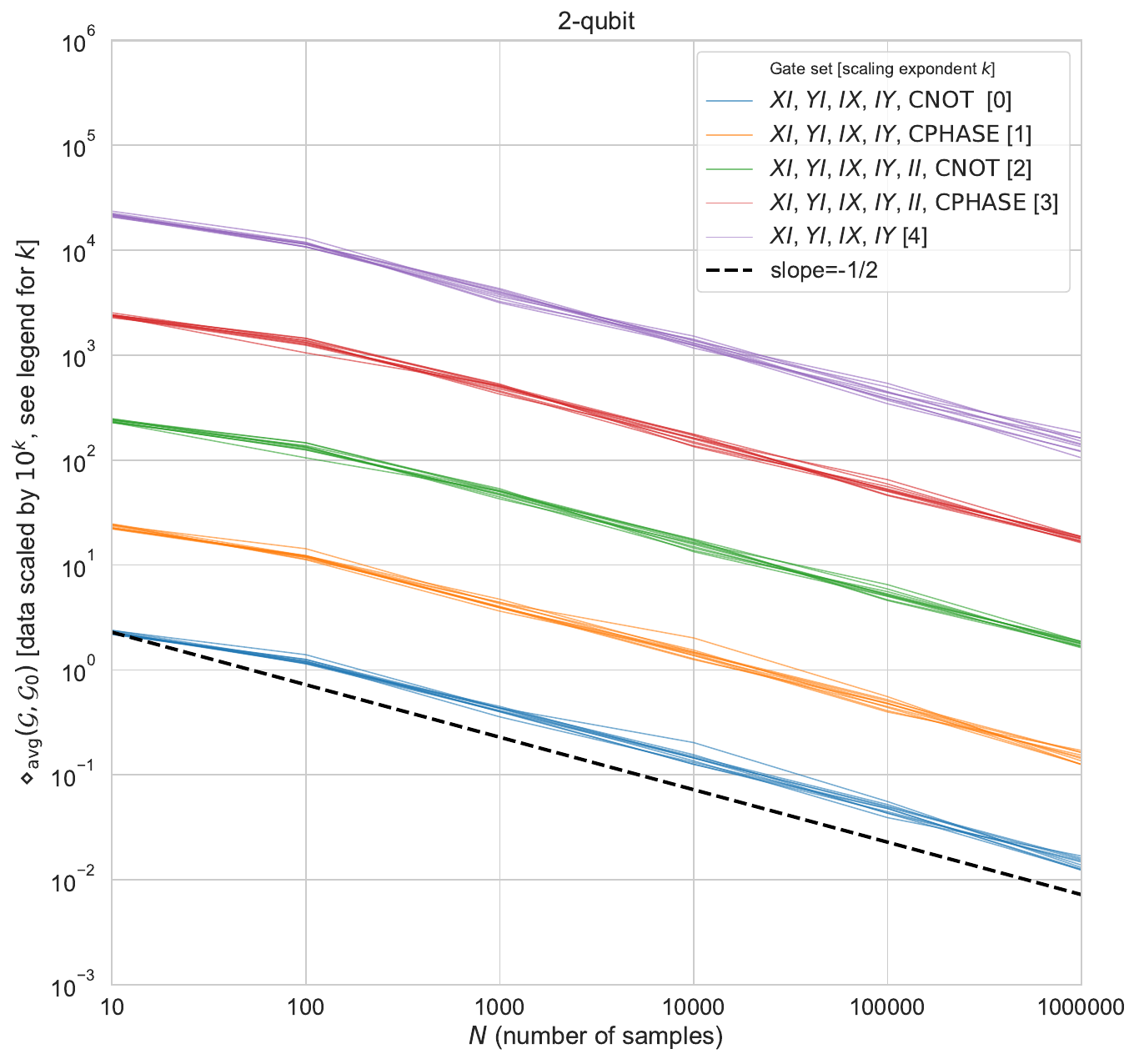} \\
      \includegraphics[width=3.5in]{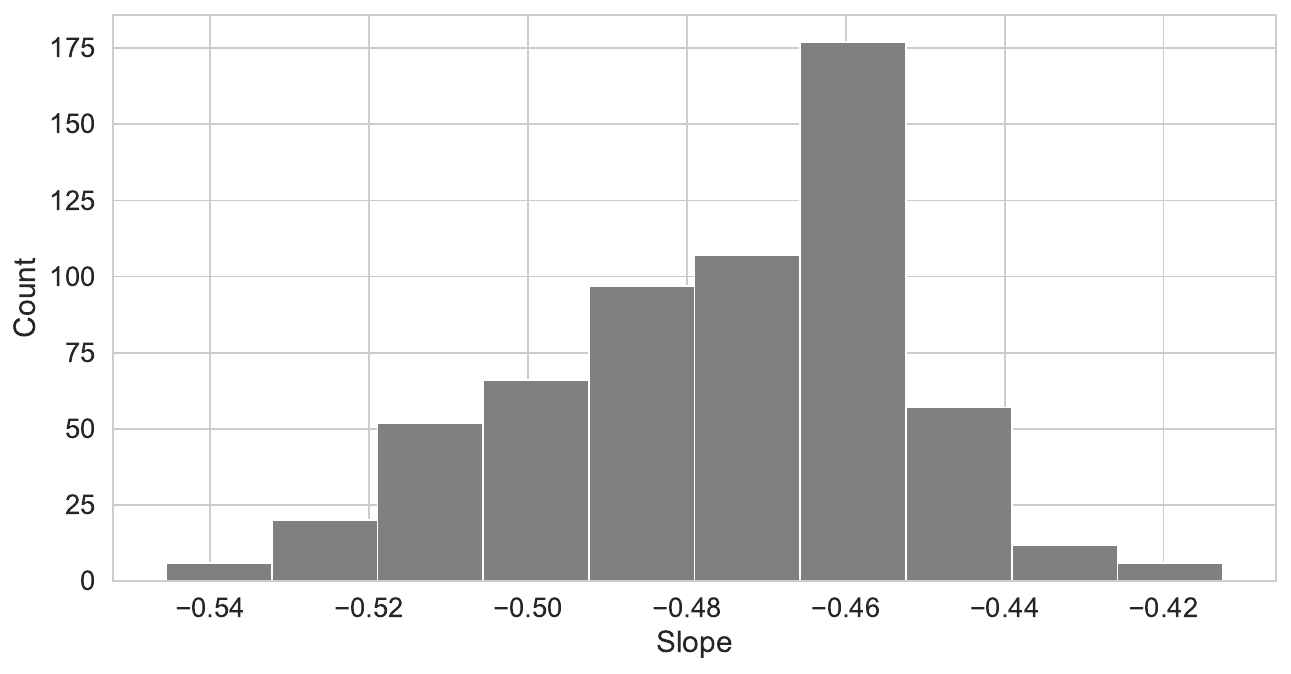} &
      \includegraphics[width=3.5in]{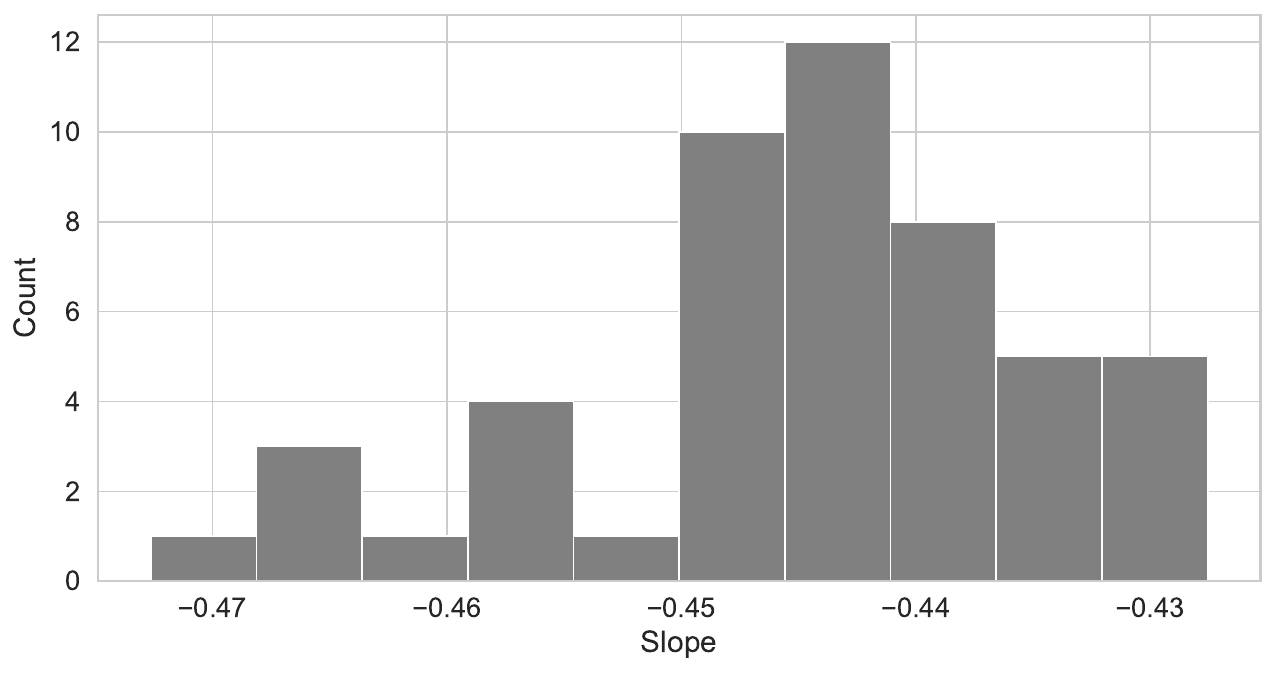}
    \end{tabular}

    \caption{\textbf{$1/\sqrt{N}$ dependence of LGST accuracy.}  The average diamond norm distance between the gates of a LGST estimate and the data-generating ``true'' gate set is computed as a function of the number of data samples per circuit, $N$. The dashed line with slope -0.5 indicates the expected $O(1/\sqrt{N})$ scaling.  The left and right columns of plots correspond to 1- and 2-qubit results, respectively.  At each value of $N$, 100 (1-qubit cases) or 10 (2-qubit cases) perturbations of the original perfect gate set were used (color-coded). A line is plotted for each perturbed gate set.  Y-axis values are multiplied $10^k$ to aid in visibility (this shifts lines upward by $k$ grid cells).  The integer $k$ is different for each perfect gate set, and is given in the legend.  The legend's $k$-value indicates the numbr of grid cells the data must be shifted \emph{down} to make the y-axis values are correct.  Below each plot is a histogram of the best-fit slopes of all the lines within it. \label{fig:LGSTScaling}}
  \end{center}
\end{figure}

\begin{figure}[H]
  \begin{center}
    \begin{tabular}[c]{cc}
      \includegraphics[width=3.5in]{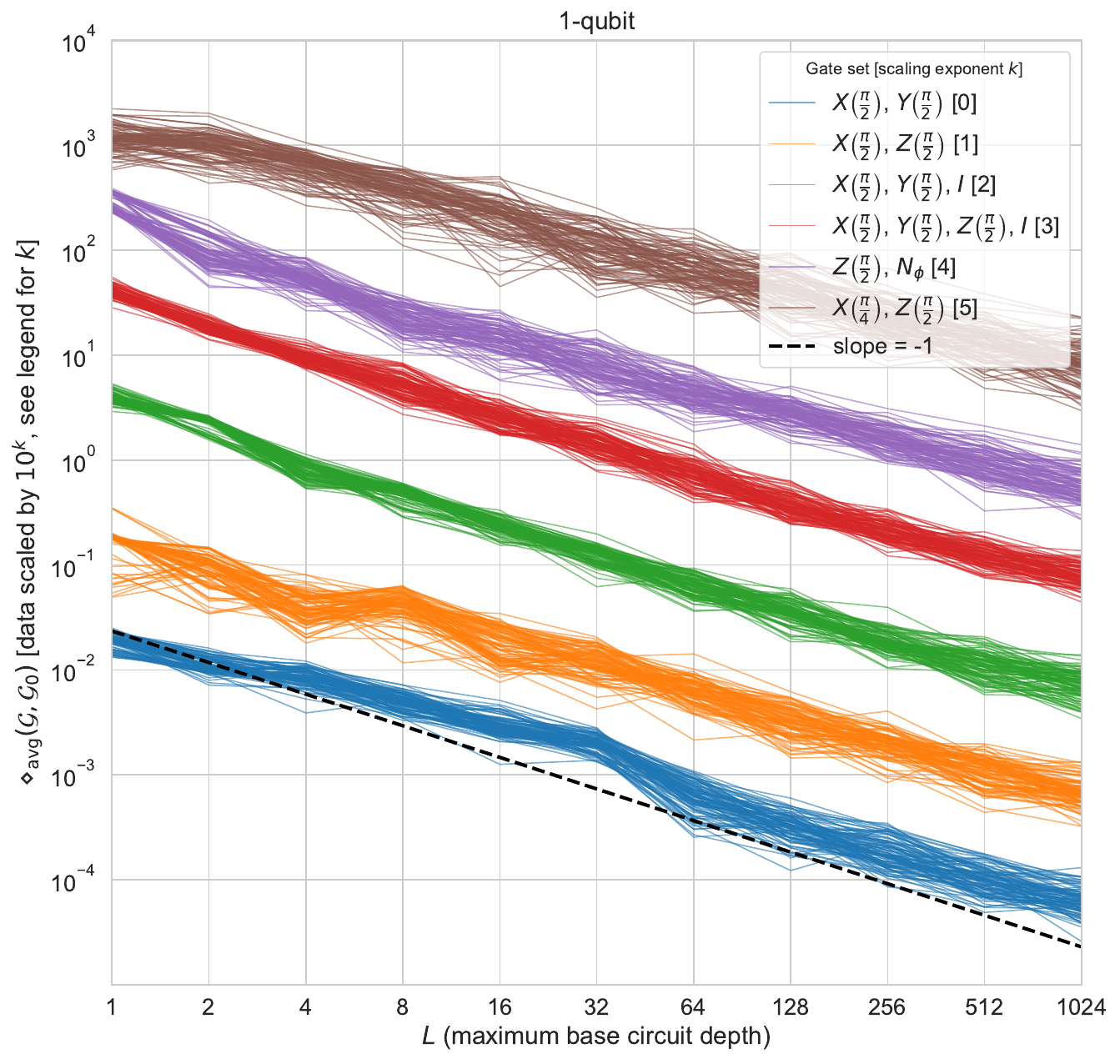} &
      \includegraphics[width=3.5in]{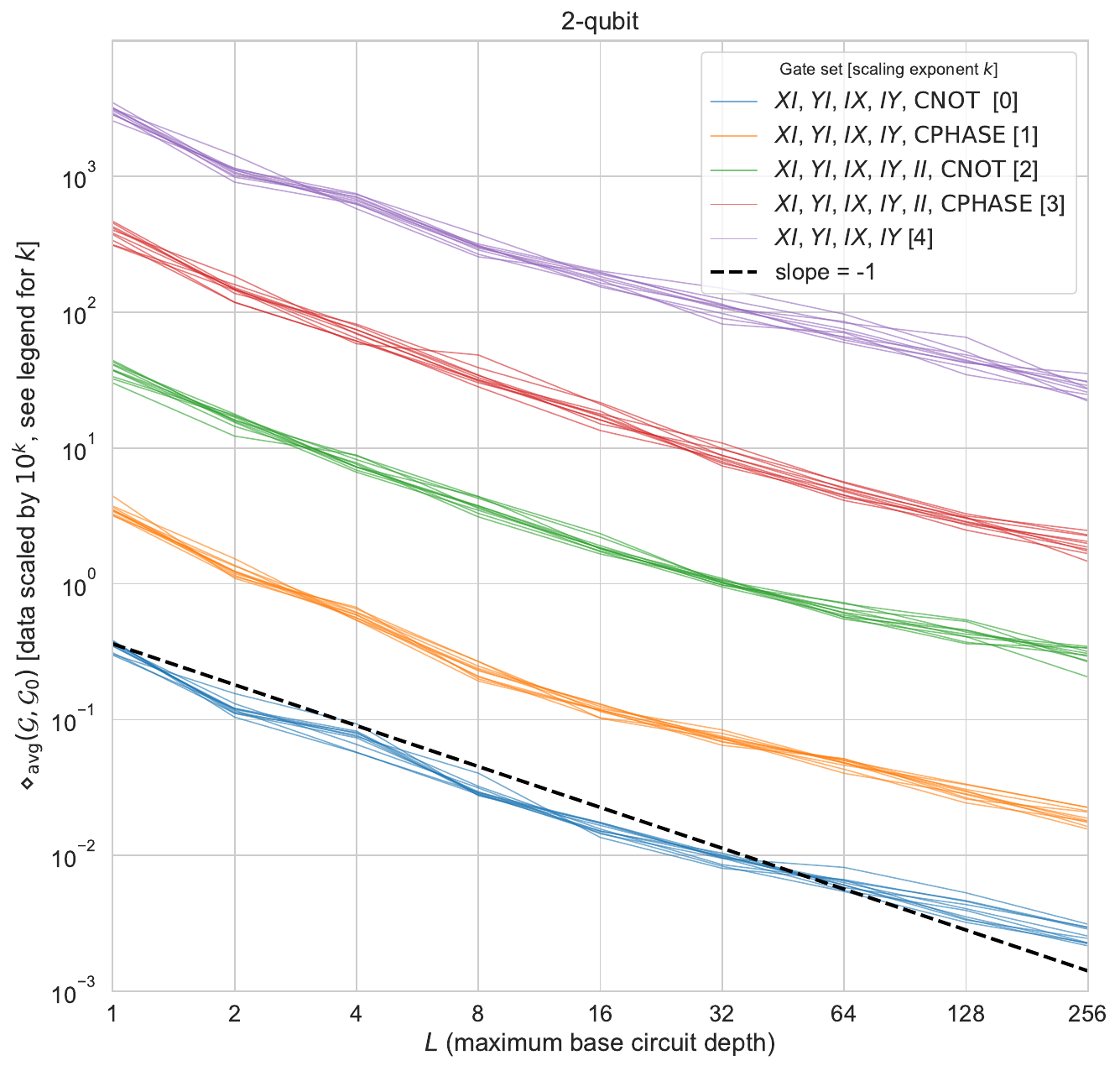} \\
      \includegraphics[width=3.5in]{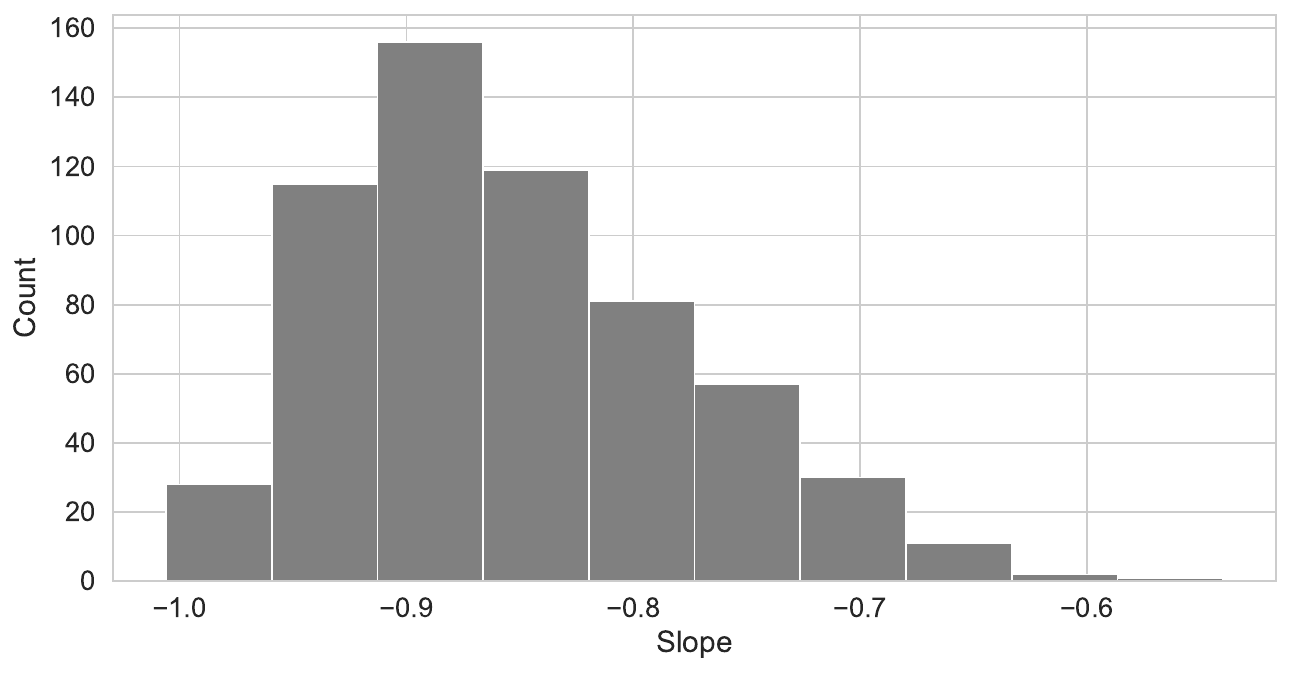} &
      \includegraphics[width=3.5in]{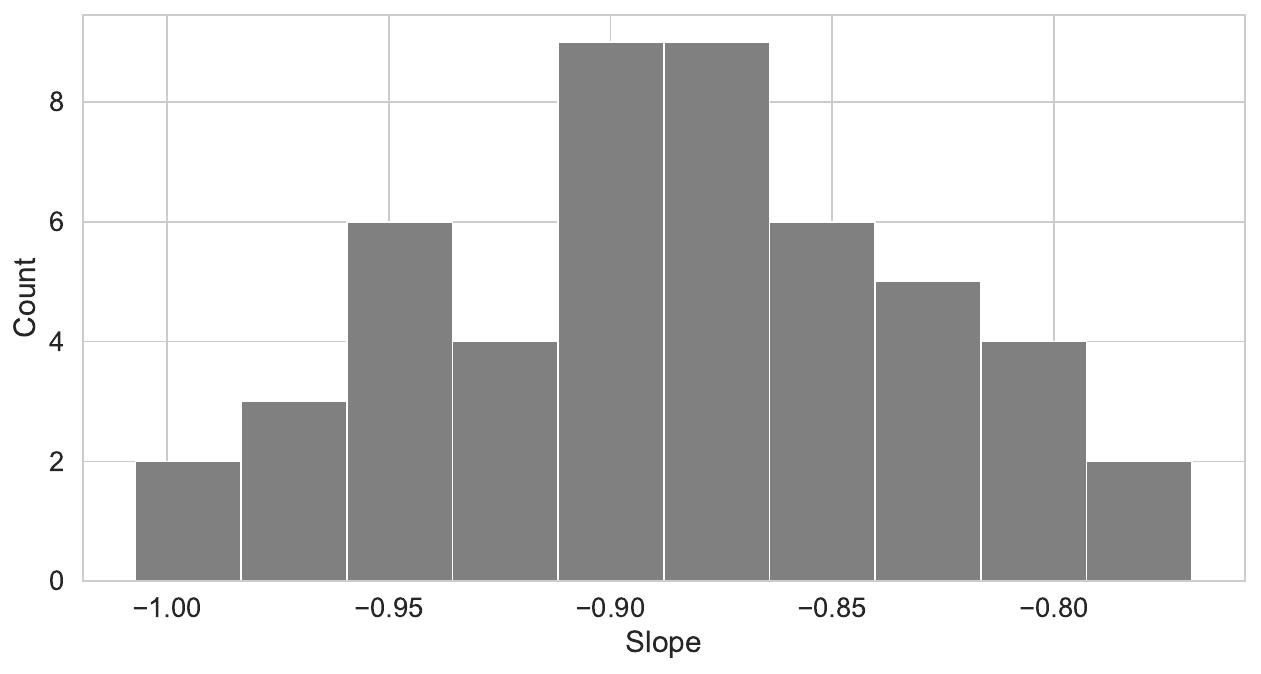}
    \end{tabular}

    \caption{\textbf{$1/L$ dependence of long-sequence GST accuracy.}  Upper plots show the average diamond norm distance, $\diamond_{\mathrm{avg}}(\mathcal{G},\mathcal{G}_0)$, between each long-sequence GST estimate $\mathcal{G}$ and the corresponding data-generating gate set $\mathcal{G}_0$ is displayed as a function of the maximum base-circuit depth $L$.  The dashed line has slope equal to -1.0, indicating the expected $O(1/L)$ scaling.  Left and right columns of plots correspond to 1- and 2-qubit results, respectively.  At each value of $L$, 100 (1-qubit cases) or 10 (2-qubit cases) perturbations of the original perfect gate set were used (color-coded). A line is plotted for each perturbed gate set.  Y-axis values are multiplied $10^k$ to aid in visibility (this shifts lines upward by $k$ grid cells).  The integer $k$ is different for each perfect gate set, and is given in the legend.  The legend's $k$-value indicates the numbr of grid cells the data must be shifted \emph{down} to make the y-axis values are correct.  Below each upper plot is a histogram of the best-fit slopes of all the lines within it.\label{fig:LongSeqScaling}}
  \end{center}
\end{figure}

\end{widetext}

Long-sequence GST utilizes deep circuits to amplify gate errors and thereby make them more visible.  By design, the sensitivity of GST to gate errors should scale as $1/L$, where $L$ is the maximum depth of the repeated germ circuits.  Thus, we expect GST to achieve a $O(1/L)$ accuracy scaling. Figure \ref{fig:LongSeqScaling} shows the average diamond distance vs.~$L$ for estimates obtained from LGST, and confirms this $1/L$ scaling (slopes are close to $-1$).

The long-sequence GST simulations were run with fixed $N=1000$ and the maximum base-circuit lengths chosen to be logarithmically-spaced powers of 2 from 1 to $L$.  There is some expected ``roll-off'' at larger values of $L$ that is believed to be due to $L$ approaching $L_\eta$ (the depth at which circuit outcomes no longer provide useful information because the quantum state is completely decohered).  This slight bend is more visible in the 2-qubit case.

This $O(1/L)$ scaling easily translates to an overall $O(1/N')$ scaling, where $N'$ is the \emph{total} number of samples, since by fixing $N$ and pessimistically assuming that there are $L$ different maximum-base-circuit-depth values (usually there are $log(L)$), $N'$ is proportional to $L$.

All of the analysis shown here used the LGST and long-sequence GST implementations in \texttt{pyGSTi} version 0.9.9.2.

\section{$\chi^2$ estimator SPAM bias\label{sec:chi2bias}}
In the course of our analysis, we discovered a subtle problem with the minimum-$\chi^2$ estimation: it can be significantly \emph{biased}.  To see this, consider a simple example: estimating a coin's bias.  Suppose that two circuits are done.  In each circuit the coin is flipped 100 times.  The first circuit yields 0 heads, and the second yields 1.  In this simple case the data could just be combined into a single data set with $N=200$ and $n=1$, in which case any reasonable estimator yields $\hat{p}=0.005$.  If, instead, we combine the \emph{circuits} into a single likelihood function,
$$L(p) = p^1(1-p)^{99} \times p^0(1-p)^{100},$$
 then its maximum is still at $\hat{p}_{\mathrm{MLE}}=0.005$.  But if we combine the circuit into a single $\chi^2$ function,
$$\chi^2(p) = \frac{(p-0.01)^2}{p(1-p)} + \frac{(p-0)^2}{p(1-p)},$$
then the minimum is at
$$\hat{p}_{\mathrm{m}\chi^2} = \frac{\sqrt{19801}-1}{19800} \approx 0.007.$$
This is not in itself bias, but when this estimator is averaged over all data sets, we find that minimum-$\chi^2$ estimation overestimates the probability of rare events.  If the true underlying probability were $0.01$, then $\expect{\hat{p}}$ would be $0.0118$.  As $p$ gets smaller, the bias gets worse:
$$\lim_{p\to0}\expect{\hat{p}} = \frac{\sqrt{19801}-1}{99}p \approx 1.411p.$$
This example is remarkably relevant, because the estimation of SPAM error in GST proceeds almost identically to this.  Many distinct circuits -- each typically with $N\approx 100$ counts -- are combined into a single $\chi^2$ or likelihood function, whose optimum gives the estimated SPAM error.

Indeed, we observe that the using the minimum-$\chi^2$ estimation can overestimate SPAM error by as much as 100\%.  While the bias usually doesn't have a direct effect on gate estimates (deep circuit data dominates them, and the deep circuit data is not dominated by rare events), it can indirectly induce a noticeable underestimate of the RB error rate (by causing the SPAM error rate to be overestimated).  Using the full MLE instead resolves these issues.  The ML estimate is generally quite accurate, and the ``residual likelihood'' (or, more precisely, the loglikelihood ratio statistic w/r.t. a maximal model) is $\chi^2$ distributed when the model is valid, and forms an excellent test statistic for model validation.  

\bibliographystyle{plainnat}
\bibliography{citations}

\end{document}